\documentclass[prx,superscriptaddress,aps, bibliography, longbibliography, twocolumn, reprint, showpacs, footnoteinbib]{revtex4-1}
\usepackage{bbm}

\usepackage[T1]{fontenc}
\usepackage{color}
\definecolor{note_fontcolor}{rgb}{1, 0, 1}
\usepackage{amsmath}
\usepackage{amssymb}
\usepackage{graphicx}

\makeatletter
\usepackage{bbm}
\usepackage[T1]{fontenc}
\usepackage[utf8]{inputenc}
\usepackage{color}
\definecolor{note_fontcolor}{rgb}{1, 0, 1}
\usepackage{amsmath}
\usepackage{amssymb}
\usepackage{graphicx}
\usepackage{tabularx}
\usepackage{makecell}
\usepackage{colortbl}
\makeatletter

    \def\CT@@do@color{%
      \global\let\CT@do@color\relax
            \@tempdima\wd\z@
            \advance\@tempdima\@tempdimb
            \advance\@tempdima\@tempdimc
    \advance\@tempdimb\tabcolsep
    \advance\@tempdimc\tabcolsep
    \advance\@tempdima2\tabcolsep
            \kern-\@tempdimb
            \leaders\vrule
                    \hskip\@tempdima\@plus  1fill
            \kern-\@tempdimc
            \hskip-\wd\z@ \@plus -1fill }
    \makeatother

\usepackage{dcolumn} 
\usepackage{xcolor}
\usepackage{boldline,multirow}

\usepackage{graphicx}

\newcommand{\revVK}[1]{{\color{black}#1}}
\newcommand{\rev}[1]{{\color{black}#1}}
\newcommand{\rvI}[1]{{\color{black}#1}}

\usepackage{soul}
\usepackage{ulem} 
\usepackage[breaklinks=true,colorlinks=true,linkcolor=blue,urlcolor=blue,citecolor=blue]{hyperref}

\newcommand{\be}{\begin{equation}}
\newcommand{\ee}{\end{equation}}
\def\ba{\begin{aligned}}
\def\ea{\end{aligned}}
\newcommand{\bea}{\begin{eqnarray}}
\newcommand{\eea}{\end{eqnarray}}
\newcommand{\bes}{\begin{subequations}}
\newcommand{\ees}{\end{subequations}}

\newcommand\mean[1]{\ensuremath{\langle#1\rangle}}

\newcommand\lrp[1]{\left(#1\right)}

\newcommand{\la}{\left\langle}
\newcommand{\ra}{\right\rangle}
\newcommand{\lb}{\left[}
\newcommand{\rb}{\right]}
\newcommand{\lp}{\left(}
\newcommand{\rp}{\right)}

\renewcommand{\Im}{{\rm \, Im\,}}

\newcommand{\theo}[1]{{\color{blue}$[ #1 ]$}}
\newcommand{\extr}[1]{{\color{red}$\{ #1 \}$}}
\newcommand{\blue}[1]{{\color{blue}$#1$}}
\newcommand{\red}[1]{{\color{red}$#1$}}

\begin{document}

\title{Fragile extended phases in logarithmically-normal Rosenzweig-Porter model. }

\author{I.~M.~Khaymovich}
 \affiliation{Max-Planck-Institut f\"ur Physik komplexer Systeme, N\"othnitzer Stra{\ss}e~38, 01187-Dresden, Germany }

\author{V.~E.~Kravtsov}
\affiliation{Abdus Salam International Center for Theoretical Physics - Strada Costiera~11, 34151 Trieste, Italy}
\affiliation{L. D. Landau Institute for Theoretical Physics - Chernogolovka, Russia}

\author{B.~L.~Altshuler}
\affiliation{Department of Physics, Columbia University, New York, NY 10027, USA}
\affiliation{Russian Quantum Center, Skolkovo, Moscow Region 143025, Russia}

\author{L.~B.~Ioffe}
\affiliation{Google Inc., Venice, CA 90291 USA}
\affiliation{National Research University HSE, Laboratory for Condensed Matter Physics,
  Myasnitskaya str., 20, 101978, Moscow, Russia }
\begin{abstract}
In this paper we suggest an extension of the Rosenzweig-Porter (RP) model, the LN-RP model,
in which the off-diagonal matrix elements have a wide, log-normal distribution.
We argue that this model is more suitable to describe a generic many body
localization problem.
In contrast to RP model, in LN-RP model a  \rev{fragile weakly ergodic phase appears that
 is characterized by broken basis-rotation symmetry
which the fully-ergodic phase, also present in this model,
strictly respects in the thermodynamic limit. }
Therefore, in addition to
the localization and ergodic transitions in LN-RP model there exists also the
 transition between
the two ergodic phases (FWE transition).
We suggest
new criteria of stability of the non-ergodic phases which give the points of localization and ergodic
 transitions and prove that
the Anderson localization transition in LN-RP model \rev{involves a jump in the fractal
dimension of the egenfunction support set}.
We also formulate the criterion of FWE transition and obtain the full phase diagram of the model.
We show that truncation of the log-normal tail shrinks
the region of weakly-ergodic phase and restores the multifractal and the fully-ergodic phases.
\end{abstract}

\maketitle
%
%

\section{Introduction}
\begin{figure}[h!]
\centering
\begin{minipage}{0.8\linewidth}
\includegraphics[width=1\linewidth]{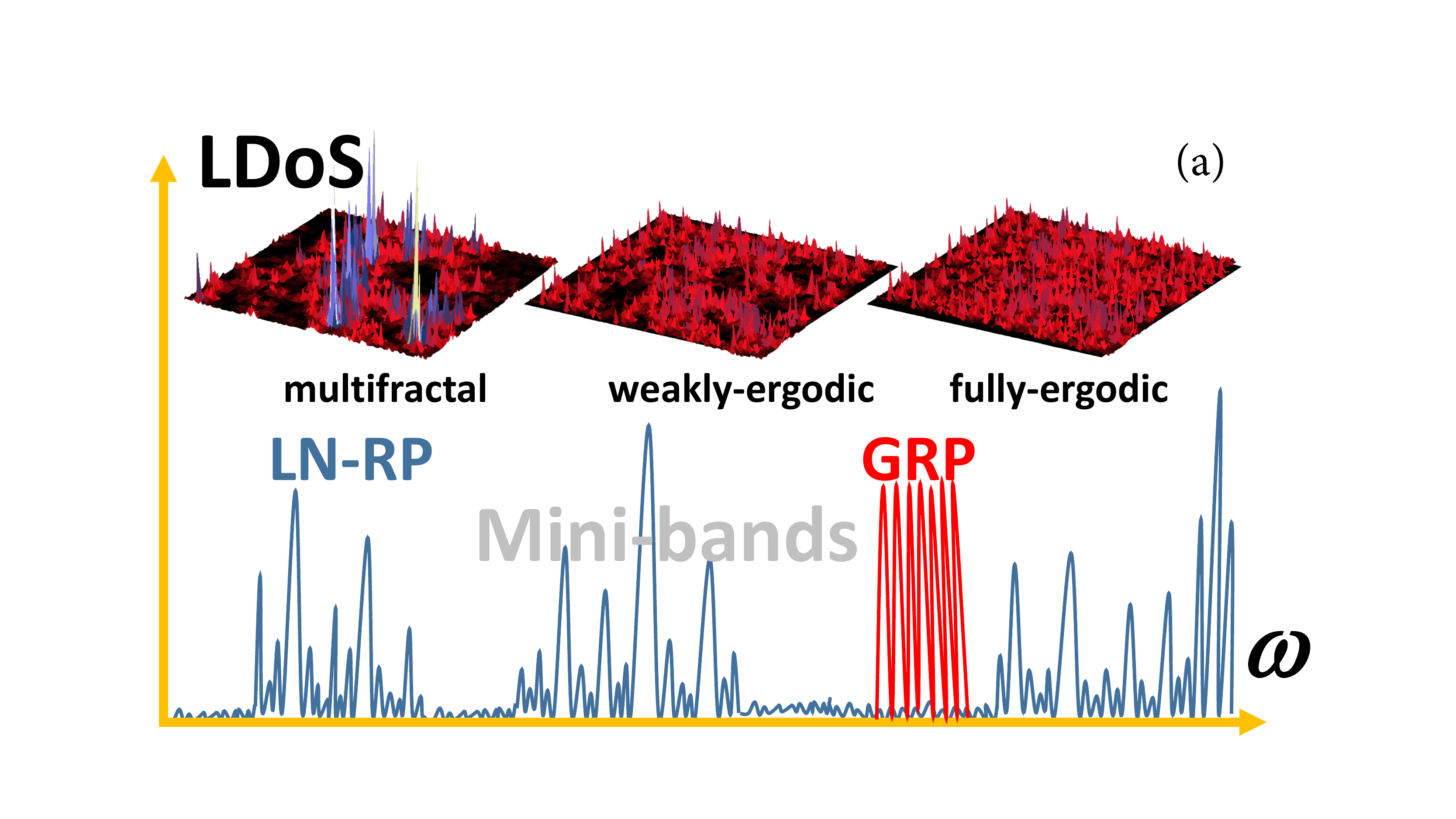}
\end{minipage}
\begin{minipage}{0.9\linewidth}
\centering
\includegraphics[width=0.46\linewidth]{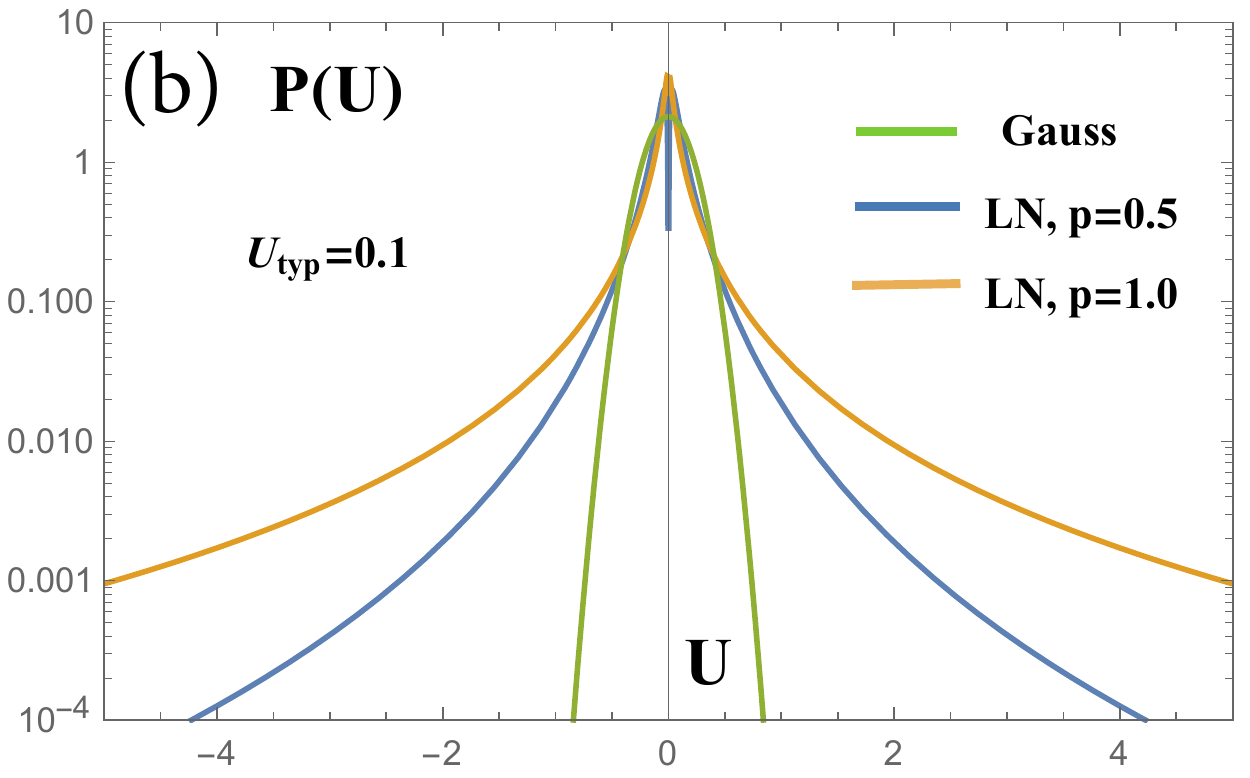}
\includegraphics[width=0.44\linewidth]{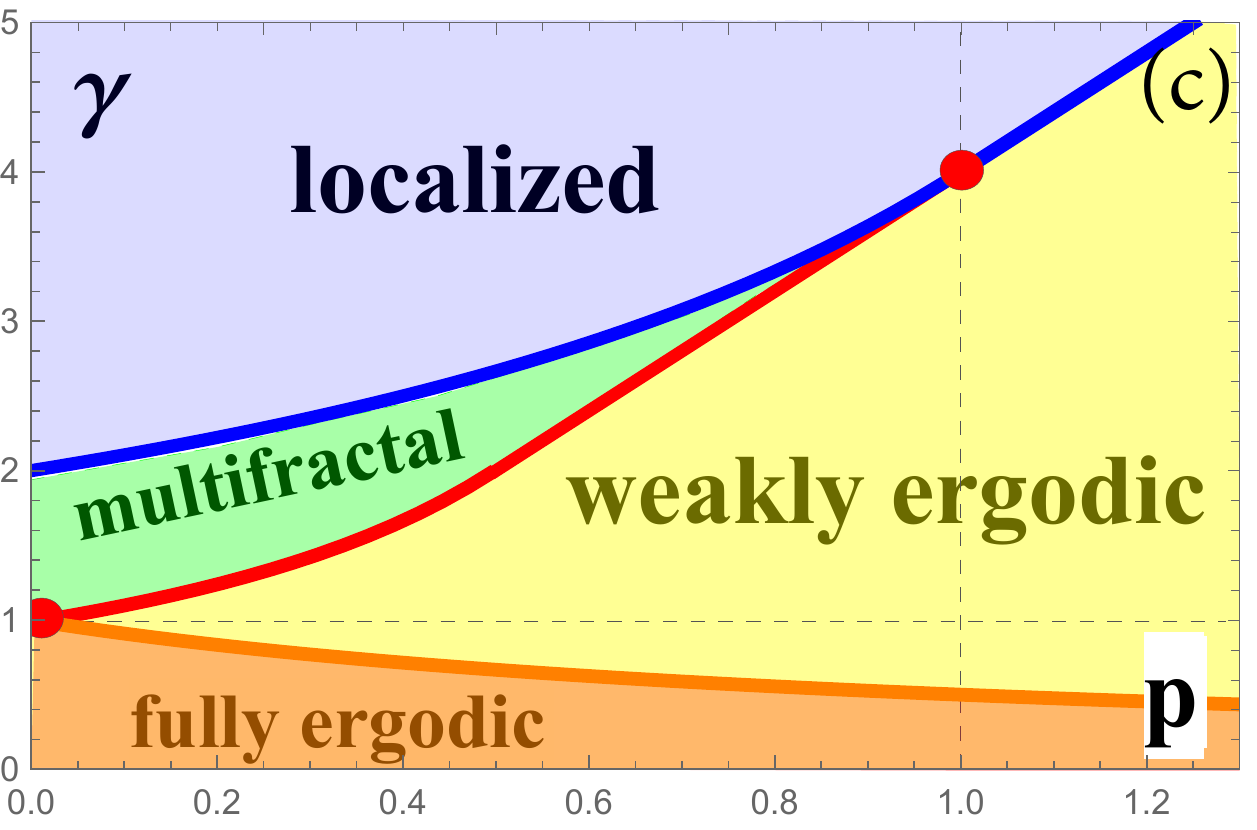} \\
\includegraphics[width=0.445\linewidth]{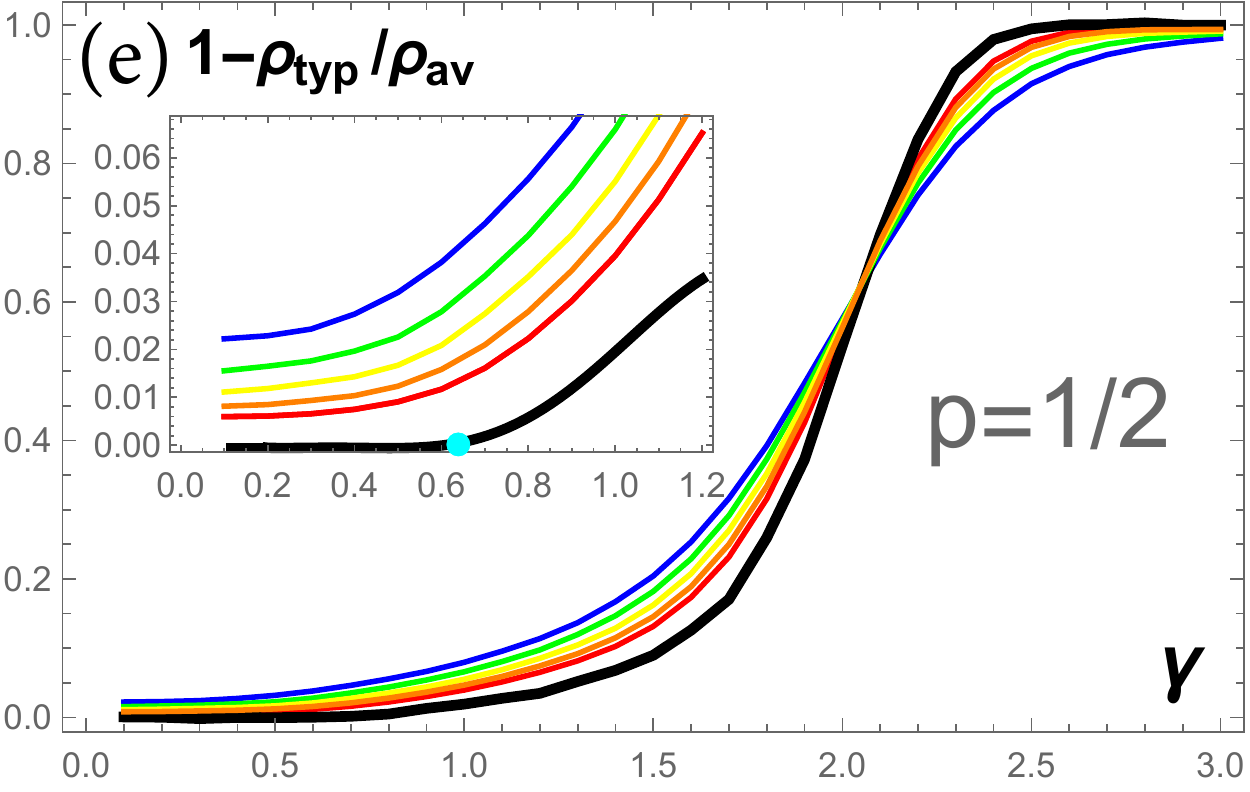}
\includegraphics[width=0.44\linewidth]{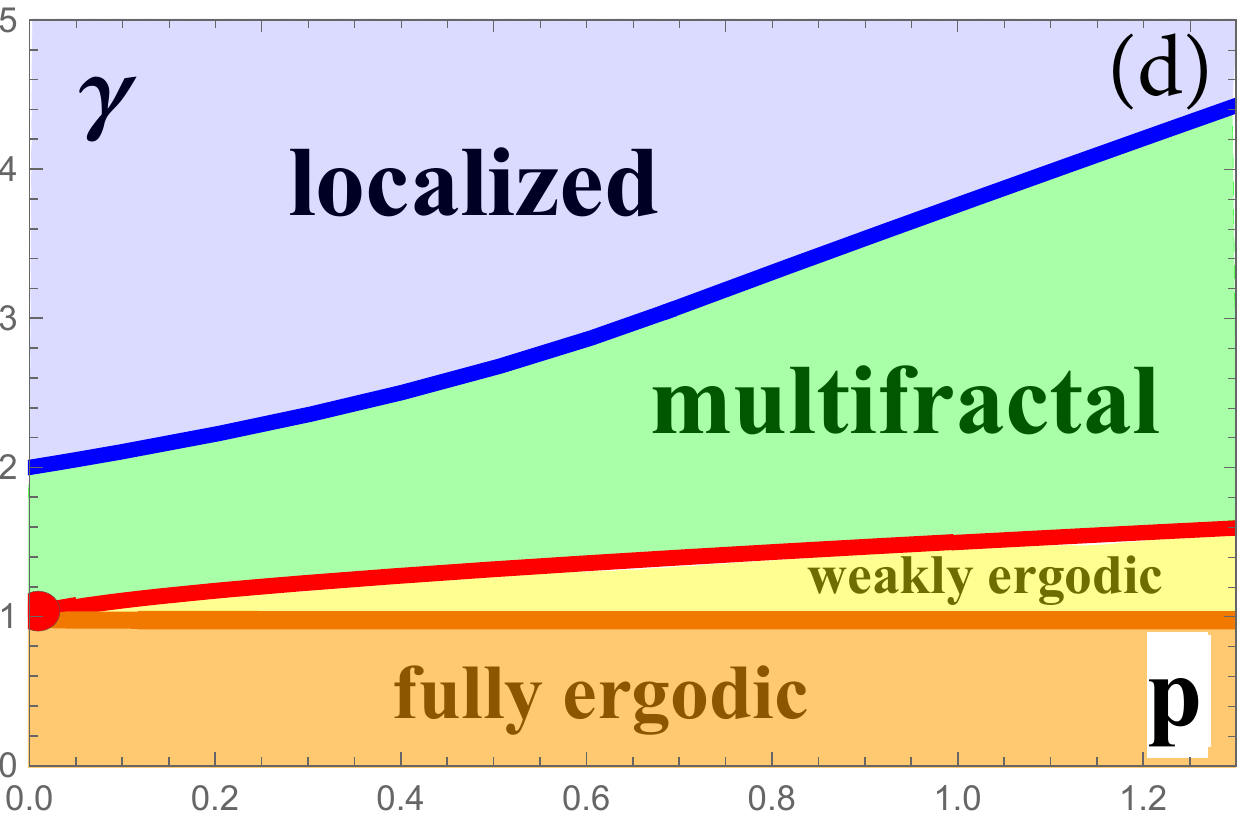}
\end{minipage}
\caption{(Color online){\bf Fragile extended phases and mini-bands in LN-RP model.}
 (a)~Cartoon of different extended states: fully-ergodic, ($D_{1}=1, f=1$),
weakly-ergodic ($D_{1}=1, f<1$) and multifractal ($D_{1}<1$).
 Sparse space structure of wave functions corresponds to sparse \revVK{fractal} structure of
mini-bands in the local spectrum.
 A compact mini-band in GRP (red) is compared with multiple  mini-bands in LN-RP (blue).
(b)~Gaussian and tailed log-normal (LN) distributions of $U=H_{nm}$. With increasing the parameter $p$
in~\eqref{LNdis} the weight of the tail at large $|U|$ increases.
Gaussian RP ensemble corresponds to $p\rightarrow 0$ and RRG is
associated with $p=1$.
(c)~Phase diagram of LN-RP $N\times N$ random matrix model~\eqref{LNdis}
in the middle of the spectrum.
The parameter $\gamma$ is an
effective disorder. The points $(0,1)$ and $(1,4)$ in $(p,\gamma)$ plane are
the tricritical points.
With increasing $p$ the weakly-ergodic (WE) phase proliferates and pushes out both the
multifractal (MF)
and the fully-ergodic (FE) phases. For $p>1$ the MF phase no longer exists.
(d)~Phase diagram of RP model in the middle of the spectrum with the LN distribution truncated so that
$|U|<N^{-\gamma_{{\rm tr}}}$, ($\gamma_{{\rm tr}}=0.95$). The WE phase
shrinks dramatically and gives the way to MF and FE phases.
(e)~Dependence on $\gamma$ of $1-\rho_{{\rm typ}}/\rho_{{\rm av}}$,
where $\rho_{{\rm typ}}$ and $\rho_{{\rm av}}$ are the typical and the mean
Local Density of States (LDoS), obtained by
exact diagonalization (blue to red curves) and extrapolated
to $N\rightarrow\infty$ (black curve). The intersection of curves signals of
the transition from MF to WE phase. In the inset: dependence of the order parameter
$\phi=1-\rho_{{\rm typ}}/\rho_{{\rm av}}$ on $\gamma$. Bright blue point is
the FWE transition
between FE ($\phi=0$) and WE ($\phi>0$) phases.}
\label{fig:phase_dia}
\end{figure}

The structure of many body wave function is important for a variety
of problems that range from many body localization (MBL) (see Ref.~\cite{BAA} and a recent
review~\cite{Abanin-rev})
to quantum computation. It was recently realized that in many of these
problems the wave function is neither localized nor completely \rvI{ergodic~\cite{Mace_Laflorencie2019_XXZ,Tarzia_2020,QIsing_2020}.}
Instead it is characterized by anomalous dimension,
$D_{1}<1$: $\sum_{i}\psi_{\mu}(i)^2\ln\psi_{\mu}(i)^2 =-D_{1} \ln \mathcal{N}$,
where $ \mathcal{N}$ is the full
dimension of the Hilbert space and $\psi_{\mu}(i)$ is
the wave function coefficient $\langle\mu |i\rangle$ of $\mu$-th state, reminiscent
of configurational entropy of glasses.
These fractal
wave functions (see Fig.~\ref{fig:phase_dia}(a)) were reported and intensively discussed
in the physical problems of localization
on random regular graphs ~\cite{DeLuca2014,AnnalRRG, PRL2016,
 AnnalRRG, Tikh-Mir1, Tikh-Mir2,
Scard-Par,RRG_R(t),
deTomasi2019subdiffusion,Lemarie2017,Lemarie2020_2loc_lengths}, the Josephson junction
chains~\cite{PinoIoffeAlt, Pino-Ioffe-VEK}, the random energy
model~\cite{smelyanskiy2018non,faoro2018non} and even in the Sachdev-Ye-Kitaev
model of quantum gravity~\cite{micklitz2019non,monteiro2020fock,feigelman2020}.
In quantum computation similar fractal wave functions appear in the
search algorithms based on the efficient population transfer and
it is believed that
the appearance of the fractal dimensions is linked with
quantum
 supremacy~\cite{kechedzhi2018efficient}.
Moreover, the wave function corresponding to a generic fault tolerant quantum
computation is fractal because it is confined to the
computational space that is much smaller than the full Hilbert space.
 However, despite the apparent
importance of this phenomena, its understanding and analytic description
is still in its infancy.

Generally, one expects that fractal wave
function might appear in the intermediate regime sandwitched between
fully ergodic and fully localized states. However, the only solvable
model that shows the appearance of such a regime in
a certain range of parameters, the Gaussian
Rosenzweig-Porter
(GRP) model~\cite{RP,gRP,Warzel,Biroli_RP,Ossipov_EPL2016_H+V,Amini2017,Monthus},
 is largely oversimplified. Firstly, such a phase in this model is {\it fractal} and not {\it multi}-fractal.
However, more importantly, few mini bands in the local spectrum of this model~\cite{gRP,return}
are compact and absolutely continuous in the energy space, and not multiple and fractal as in realistic many-body
systems~\cite{Pino-Ioffe-VEK} (see Fig.~\ref{fig:phase_dia}(a)). This behavior is intimately related
to the compactness of \revVK{distribution of the wave function coefficients on the support set}
and can be traced back to the property of the moments of the Gaussian distribution
$\langle |U|^{q}\rangle=\langle U^{2}\rangle^{q/2}$.

In this paper we introduce a natural generalization
of this model and show that it displays a much richer phase diagram
and a more realistic behavior.
In GRP model every site of the reference space
(represented by a matrix index) is connected to every
other site with the transition amplitude distributed according to
the Gaussian law.
Such model occurs
as the effective description of the systems without internal structure,
in which transition between resonance sites is due to a small number
of hops, such as random energy model~\cite{smelyanskiy2018non,faoro2018non}.
 In more realistic models delocalization of the wave function is
due to a long series of quantum transitions. Each transition has a
random amplitude, so their product is characterized by the log-normal  (LN)
distribution, rather than the Gaussian one as in GRP model. Inspired by
this argument in this paper we introduce and study the generalization
of RP model in which the transition amplitude between sites has a small typical
value, as in RP model, but with much
wider, log-normal distribution function that we define in Section
\ref{sec:LN-Rosenzweig-Porter} (see Fig.~\ref{fig:phase_dia}(b)).

It appears that the rare large hopping matrix elements from the tail of
this distribution alter the phase diagram of the system by
considerably shrinking the region of multifractal phase as the parameter $p$
that controls the weight in the tail, increases. For large enough $p$ the
multifractal phase is totally replaced by an ergodic one
(see Fig.~\ref{fig:phase_dia}(c)). However, this ergodic phase is fragile.
Because it is due to very rare hopping elements,
even a far cutoff of the LN distribution function restores the
multifractal phase and may even extend it in the phase diagram (Fig.~\ref{fig:phase_dia}(d)).

Generally, the mere statement that the eigenfunction fractal dimension
$D_{1}=1$
 is not sufficient for complete characterization of the
ergodic phase. As was shown in Ref.~\cite{Nosov2019correlations},
in certain {\it translational-invariant} RP models
$D_{1}=1$ in the reference basis, yet in the Fourier-transformed 'momentum' basis all
eigenvectors are localized. Consequently, the eigenvalue statistics is Poisson, despite
extended character of wave functions in the reference basis.
 On the other hand,
the ergodic
states in the GRP model remain ergodic
in any bas\rvI{is~\cite{WE_PT-footnote}},
like in the classic Wigner-Dyson (WD) random matrix ensemble.

This observation urged us to
distinguish between the {\it fully-ergodic} (FE) \revVK{ and the {\it weakly-ergodic} (WE) phases.
In FE phase} \rev{in the thermodynamic limit}:
${\bf (i)}$
\rev{the eigenfunction statistics is Porter-Thomas, i.e.
the fraction of {\it essentially}
populated sites in an eigenfunction is $f=1$},
${\bf (ii)}$
the eigenvalue statistics is WD \rvI{all the way to the bandwidth}
and
${\bf (iii)}$ eigenfunction statistics
is invariant
under basis rotation~\cite{WE_PT-footnote}. \revVK{In contrast, in WE phase }
 ~\cite{WE-footnote}
  this invariance is broken together with the \rvI{emergence of the energy scale $\Gamma$ smaller than the bandwidth, beyond which the WD eigenvalue statistics breaks down}
and $f<1$ ( Fig.~\ref{fig:phase_dia}(a)).
Furthermore,
since the so defined two ergodic phases differ by the symmetry with respect to basis rotation,
there should be a {\it  phase  transition} and not a crossover between them. We will refer to this
 transition between the fully- and weakly ergodic phases as the {\it FWE transition}.
Note that the basis-rotation invariance~\cite{WE_PT-footnote} is \rev{ not {\it manifest present} in the
formulation of any RP model with a special diagonal. It is  a non-trivial {\it emergent} symmetry
which may arise only in the thermodynamic limit $N\rightarrow\infty$. We argue that the existence
of FE phase is related to the absence of mobility edge: in systems with the mobility edge
only WE phase may exist~\cite{KhayKr_ME}. \revVK{Indeed, the localized states at the edge of the spectrum
should be orthogonal to extended states in the middle of the spectrum. This is only possible if
the states in the middle of the spectrum have {\it population holes}, $f<1$
(see Fig.\ref{fig:phase_dia}(a)),
exactly
where the wave function coefficients of localized states at the spectral edge are peaked. \rvI{The similar deviations from full ergodicity in generic many-body systems have been also explained it terms of the above orthogonality~\cite{Sent2020_Haque}}.}
An important example of a system where the mobility edge is
known to be absent at small enough disorder~\cite{Warzel2011EPL} and which may be considered
as a toy model for MBL~\cite{DeLuca2014}, is the Anderson model on the Bethe lattice.
We argue that the same is true for the models with long-range hopping, in particular for the LN-RP model
~\cite{KhayKr_ME}.

The weakly-ergodic phase is much more widespread: for instance the metallic phase in 3D Anderson model
is weakly-ergodic\rvI{~\cite{kravtsov1994level,Fyodorov1995}. Due to the presence of the mobility edge
the wave function coefficient distribution is not of the Porter-Thomas form~\cite{Fyodorov1995},
while the level statistics at small energies is still WD in the thermodynamic limit~\cite{kravtsov1994level}}.
A non-trivial feature of the LN-RP model is that {\it both ergodic phases}
are present in it separated by a {\it line} of a new {\it FWE}
 quantum phase transition} (see
Fig.~\ref{fig:phase_dia} (c)-(e)) \rvI{similarly to the Bethe lattice~\cite{Warzel2011PRL}}.

\rev{Note that a critical \rvI{point in the single}-particle Anderson model in $d$ dimensions \rvI{where the dimensionless conductance $g$ is size-independent, but no multifractality present,} which was anticipated in
~\cite{gang-of-four} and studied in~\cite{AKLA94, AKL95}, is surely
weakly-ergodic but {\it sub-diffusive}: $\langle r^{2}\rangle \propto t^{2/d}$.
It is probably similar kind of WE phase in the Hilbert space of interacting systems which is responsible for a so-called ``bad metal''
phase on the ergodic side of the localization transition.
In such a phase, both many-body systems~\cite{BarLev_Luitz2016anom_diff}
and hierarchical structures like RRG~\cite{RRG_R(t),deTomasi2019subdiffusion}
have been shown to
demonstrate the anomalous sub-diffusive transport.}


The analytical theory of the Ergodic (ET), Localization (AT), and FWE transitions
developed
in this paper is verified
by extensive numerics based on the Kullback-Leibler divergence~\cite{KLdiv, KLdiv_book}
of certain correlation functions ${\rm KL1}$ and ${\rm KL2}$ of wave function
coefficients~\cite{KLPino} and on
numerical investigation of the typical ($\rho_{{\rm typ}}$) and the mean
($\rho_{{\rm av}}$) Local Density of States (LDoS). The quantity
$\phi=1-\rho_{{\rm typ}}/\rho_{{\rm av}}$ is an order parameter for the FWE transition,
with $\phi=0$ in FE phase and $\phi>0$ in WE phase (see Fig.~\ref{fig:phase_dia} (e)), while the
onset of divergence (with the system size $N$) of
${\rm KL1}$ and ${\rm KL2}$ marks the AT and ET transitions, respectively (see
  Fig.~\ref{Fig:intersect}).

\section{Log-Normal Roseizweig-Porter model}
\label{sec:LN-Rosenzweig-Porter}
We introduce a modification of the RP random matrix ensemble~\cite{RP,gRP} in which the
Gaussian distribution of independent, identically distributed (i..i.d.) {\it off-diagonal} real entries $H_{nm}=U$ is replaced by
 the logarithmically-normal one:
\be \label{LNdis}
P(U)=\frac{A}{|U|}\,{\rm exp}\left[-\frac{\ln^{2}(|U|/U_{{\rm typ}})}
{2p\,\ln(U_{{\rm typ}}^{-1})}\right],
\;\; U_{{\rm typ}}\sim N^{-\gamma/2}.
\ee
It is characterized by two parameters: the disorder-parameter $\gamma$ which determines
the scaling of the typical off-diagonal matrix element with the matrix size $N$ and the
parameter $p$ that controls the weight of the tail.

The i.i.d. {\it diagonal} entries are supposed to remain Gaussian
distributed, as in the original RP model:
\be\label{Hnn}
\langle H_{nn}\rangle =0,\;\;\;\langle H_{nn}^{2}\rangle = W^{2}\sim N^{0}.
\ee
This LN-RP model is principally different from the L\'evy random matrix models
(see, e.g.,~\cite{Bouchaud_Levy_Mat,Biroli_Levy_Mat, Bouchaud} and references therein) exactly
because the Gaussian distribution~\eqref{Hnn} is {\it not tailed}. For numerical purposes
we will replace it by the box distribution which is plain in the interval $[-W/2,W/2]$.

The tailed distribution~\eqref{LNdis}
 gives rise to the moments $\langle |U|^{q}\rangle^{\frac{1}{q}}
\sim N^{-\gamma_{q}/2}$ that
scale differently with $N$ for different values of $q$:
\be\label{gamma_q}
 \gamma_{q}=\gamma\,(1-p q/2).
\ee
The limit $p\rightarrow 0$ in which $\gamma_{q}=
\gamma$, corresponds to the GRP model. It is shown in Ref.~\cite{arXiv2020} that
$p=1$ is associated with RRG due to the hidden $\beta$-symmetry on the local Cayley tree
(see Eqs.~(6.5)-(6.8) in Ref.~\cite{AbouChacra}, Eqs.~(D.2),~(D.17) in
Ref.~\cite{AnnalRRG} and Appendix C in Ref.~\cite{arXiv2020}). Finally, the limit $p\rightarrow\infty$ corresponds to
the L\'evy power-law distribution of $U$~\cite{Tarzia_Biroli_Levy_RP}.

For any physically meaningful quantity in the bulk of the spectrum
with a bandwidth  $E_{BW}$,
only the values $|U|<E_{BW}$  are relevant. For larger values of $|U|=|H_{nm}|$
the states are pushed to the Lifshits tails of the spectrum
which we are not interested
in this paper. As in the non-ergodic part of phase diagram  $E_{BW}\sim W$
is of the order of the spread of on-site energies, in these regimes
the distribution $P(U)$ is effectively cut off for $|U|>W$. However,
for ergodic states  $E_{BW}$ is determined by the off-diagonal matrix elements and
is divergent with $N$. In this case the effective cutoff   $E_{BW}$ in~\eqref{LNdis} should
be determined self-consistently ~\cite{delta_typ-footnote}.

\section{Criteria of Localization, Ergodic and FWE transitions for dense  random matrices}
\label{sec:Criteria}
In this section we consider simple
 criteria of the disorder-driven~\footnote{The problem of mobility edge and
energy-driven transitions in systems with broadly distributed hopping is non-trivial~\cite{Bouchaud}
and we leave it for future publications.} localization, ergodic and FWE transitions
for random $N\times N$ matrices with the random uncorrelated random hopping
$\langle H_{nm} \rangle=0$ and
diagonal disorder $\sim O(1)$.
More general
picture and examples of systems are presented in Refs.~\cite{BogomolnyPLRBM2018,Nosov2019correlations}.

The first criterion, which is referred to as the {\it Anderson localization criterion},
 states that if the sum :
\be\label{Anderson}
S_{1}=\sum_{m=1}^{N}\langle |H_{n,m}|\rangle_{0}=N\,\langle |U| \rangle_{0},
\ee
\rvI{goes to zero} in the limit $N\rightarrow\infty$ then the states are Anderson localized,
while \revVK{if the above sum diverges the states are surely delocalized.
The case   $S_{1}=O(1)$   which is relevant for the short-range,
  size-independent random Hamiltonians (e.g.for the 3D Anderson model), is system- specific.}

Here $\la ..\ra_{0}$ stands for the disorder averaging over the distribution,~\eqref{LNdis},
which is cut off at $|U|>W \sim N^{0}$. The reason for such a cutoff
is the following.
The physical meaning of~\eqref{Anderson} is that the number is sites
in resonance with a given site $n$ is finite. The probability that {\it two}
sites $n$
and $m$
are in resonance is:
\bea
P_{n\rightarrow m}= \int\limits_{-W/2}^{W/2} \frac{d\varepsilon_{n}}{W}\int\limits_{-W/2}^{W/2}
\frac{d\varepsilon_{m}}{W}
\int\limits_{|\omega|}^{\infty}\,P(H_{nm})\, d(H_{nm}),
\eea
where for simplicity we consider
 the
box-shaped distribution $F(\varepsilon)$ of on-site energies.
Then integration over $(\varepsilon_{n}+\varepsilon_{m})/2$ and integration by parts
over $\omega=\varepsilon_{n}-\varepsilon_{m}$ gives:
\be\label{Pres}
P_{n\rightarrow m}=\int\limits_{-W}^{W}dU\,P(U)\,\left( \frac{|U|}{W}-
\frac{U^{2}}{2W^{2}}\right)+
\int\limits_{W}^{\infty} P(U)\,dU.
\ee
One can easily see that at $U_{{\rm typ}}\sim N^{-\gamma/2}\ll O(1)$
the last integral in~\eqref{Pres} can always be neglected.
The values of $|U|$ involved in the first integral are bounded from above
$|U| <W$, which is equivalent to imposing a cutoff at $|U|>W$ on the distribution $P(U)$.
As the second term in this integral is
at most 1/2 of the first term, the number of sites in resonance with the given site,
 $\sum_{m}P_{n\rightarrow m}$, coincides with
~\eqref{Anderson} up to a pre-factor of order $O(1)$.

The second criterion
 referred to as the {\it the Mott's criterion}
is a sufficient criterion of ergodicity. It states that if the sum
\be\label{Mott}
S_{2}=\sum_{m=1}^{N}\langle |H_{nm}|^{2}\rangle_{E_{BW}}=N\,\langle U^{2}
\rangle_{E_{BW}} \rightarrow \infty
\ee
diverges in the limit $N\rightarrow\infty$ then the system is in the
 one of the ergodic phases~\cite{Nosov2019correlations}.

In Eq.(\ref{Mott})
the subscript $E_{BW}$ implies that the distribution, Eq.(\ref{LNdis}) should be
truncated at $U\sim E_{BW}$, where  $E_{BW}\sim W\sim N^{0}$
is the total spectral bandwidth in the non-ergodic phase.
The physical meaning of~\eqref{Mott} is that the {\it mean} Breit-Wigner
width $\Gamma\sim
E_{BW}^{-1}\,N\langle U^{2}\rangle_{E_{BW}}$~\cite{SM}
that quantifies
the escape rate of a particle created at a given site $n$, is much
larger than the
spread of energy levels $W\sim N^{0}$ due to disorder. Then
the fulfillment of the Mott's criterion implies that the width $\Gamma$ is
of the same order as the total spectral bandwidth  $E_{BW}\sim \sqrt{S_{2}}$  and thus there are no
{\it mini-bands} (which width is $\Gamma$) in the local spectrum (see Fig.~\ref{fig:phase_dia}(a)).
As the presence of such mini-bands is suggested~\cite{Pino-Ioffe-VEK, return, Nosov2019mixtures} as a
``smoking gun'' evidence of the {\it non-ergodic extended } (e.g. multifractal) phase,
the fulfillment of the Mott's criterion~\eqref{Mott} immediately
implies that the system is in the {\it ergodic extended} phase.

The {\it multifractal} phase realizes
provided that in the limit $N\rightarrow\infty$ {\it both}~\eqref{Anderson} and~\eqref{Mott}
are not fulfilled:
\be
S_{1}\rightarrow \infty,\;\;\;S_{2}<\infty.
\ee

Finally, the {\it fully ergodic} phase is realized when
$S_{1}, S_{2}\rightarrow \infty$ and also:
\be\label{FW}
S_{3}=
 \frac{\left( \sum _{m=1}^{N} \langle |H_{nm}|^{2}\rangle_{{\rm typ}} \right)^{2} }
{\sum_{m=1}^{N}\langle |H_{nm}|^{2} \rangle_{E_{BW}}}=\frac{N\,U_{{\rm typ}}^{4}}
{\langle U^{2} \rangle_{E_{BW}}} \rightarrow\infty,
\ee
is divergent in the $N\rightarrow\infty$ limit,
where
$\langle |H_{nm}|^{2} \rangle_{{\rm typ}}\equiv U_{{\rm typ}}^{2}=
{\rm exp}\langle \ln|H_{nm}|^{2} \rangle$~\cite{E_BW-footnote}.
If only $S_{1},S_{2}\rightarrow\infty$ but
$S_{3}$ is not, the {\it weakly ergodic} phase is realized.

The physics behind the condition Eq.(\ref{FW}) is that the {\it typical} escape rate
$\Gamma_{{\rm typ}}\sim E_{BW}^{-1}\,N\,U_{{\rm typ}}^{2}=\sqrt{S_{3}}$~\cite{SM} is much larger
than  the disorder strength $W\sim N^{0}$. The two conditions,
Eqs.(\ref{Mott}),(\ref{FW}), coincide for a Gaussian distribution of $U$ but are different
for the tailed ones, like LN distribution, Eq.(\ref{LNdis}).

\section{Phase diagram}
\label{section:phase_diagram}

For the log-normal distribution~\eqref{LNdis} one easily computes the
moments $\la |U|^{q}\ra_{0}$ truncated at $U_{{\rm max}}\sim N^{0}$:
\be\label{mom_q}
\langle U^{q}\rangle_{0}=\left\{\begin{array}{ll}
N^{-\frac{\gamma q}{2}\,\left(1-\frac{pq}{2} \right)},& {\rm if}\;\; pq<1\cr
N^{-\frac{\gamma}{4p}}, & {\rm if}\;\; pq\geq 1\end{array}\right.
\ee
and finds using~\eqref{gamma_q},~\eqref{Anderson},~\eqref{Mott},~\eqref{FW} and $U_{{\rm typ}}=N^{-\gamma/2}$
the following critical points of the localization ($\gamma_{AT}$), ergodic
($\gamma_{ET}$) and FWE ($\gamma_{FWE}$) transitions:
\be\label{AT}
\gamma_{AT}=\left\{\begin{array}{ll}\frac{4}{2-p}, & {\rm if}\;\; p<1\cr
4p,& {\rm if}\;\; p\geq 1 \end{array}\right.
\ee
\be\label{ET}
\gamma_{ET}=\left\{\begin{array}{ll} \frac{1}{1-p}, & {\rm if}\;\; p<1/2\cr
4p,& {\rm if}\;\;p\geq 1/2\end{array} \right.
\ee
\be\label{FWT}
\gamma_{FWE}=\frac{1}{1+p}.
\ee
The phase diagram at a fixed energy in the middle of spectrum
  resulting from
~\eqref{AT}-\eqref{FWT}, is presented in   Fig.~\ref{fig:phase_dia}(c).

The main conclusion we may draw from this phase diagram is the emergence and
proliferation of the weakly-ergodic phase that pushes away both the
multifractal (MF) phase
and the fully ergodic phase, as the strength of the tail $p$ in the distribution
~\eqref{LNdis} increases. For $p>1$ the MF phase is completely gone
replaced by the weakly ergodic one. However,  this WE phase is fragile.
Truncation of the tail
of this distribution, so that
$|U|<N^{-\gamma_{{\rm tr}}}$, $\gamma_{{\rm tr}}>0$, eliminates the WE phase and
restores the MF phase, as well as increases the range of
the fully-ergodic one (see  Fig.~\ref{fig:phase_dia}(d)
and Appendix~\ref{sec:Truncation} for details).

\section{Stability of non-ergodic states against hybridization}\label{sec:MF-hybridization}

\begin{figure}[t!]
\center{
\includegraphics[width=0.6 \linewidth]{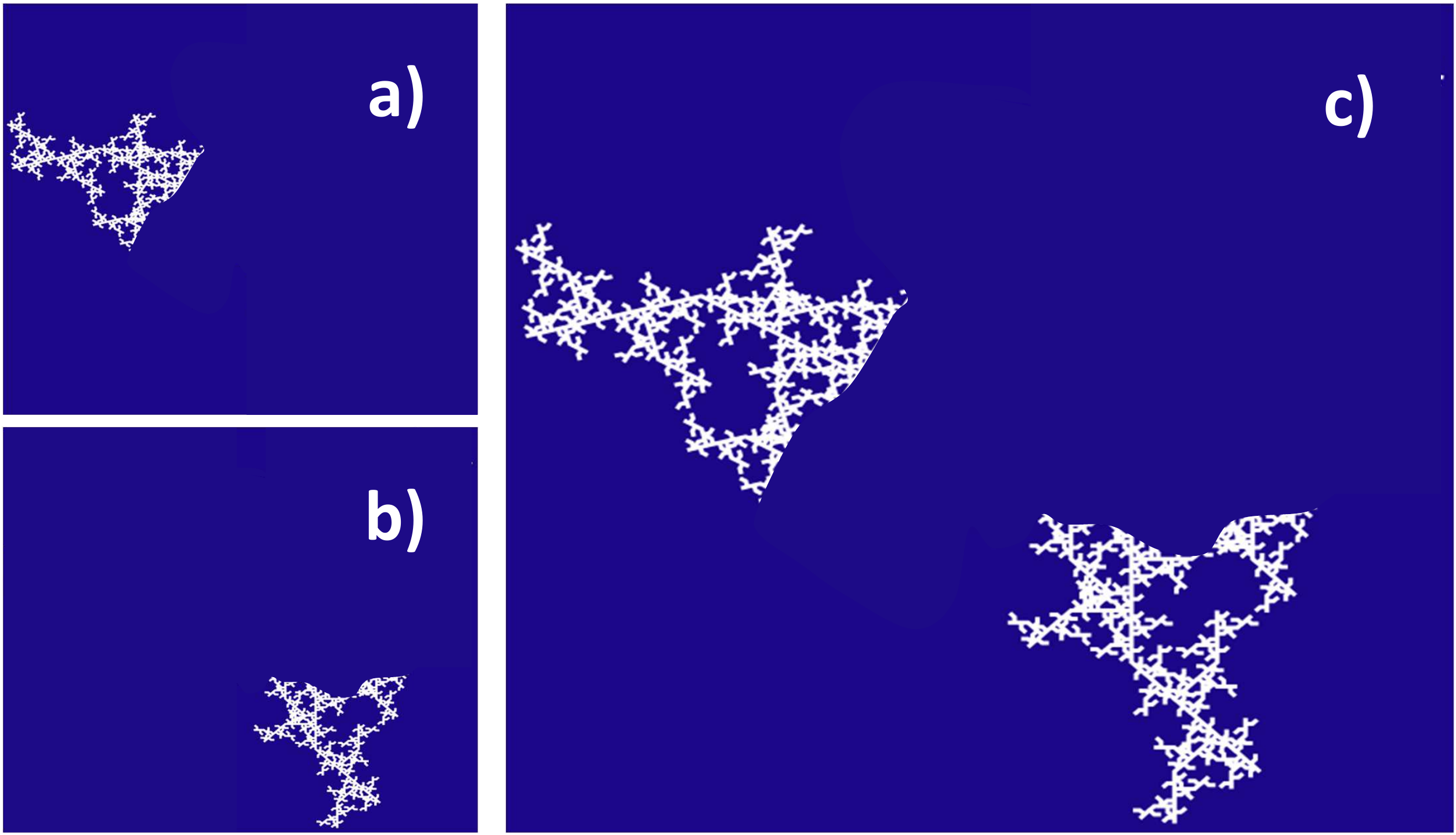}}
\caption{(Color online) \textbf{Hybridization of fractal support sets}
(a),~(b)~Two different fractal support sets,
(c)~The hybridized fractal support set.}
\label{Fig:hybrid}
\end{figure}
In this section we consider the stability of non-ergodic (multifractal and localized) states
 against hybridization. It allows us not only to derive expressions,~\eqref{Anderson} and~\eqref{Mott}, for the Anderson
localization and ergodic transitions in a different way
but also find the
fractal dimension $D_{1}(p,\gamma)$ of the
multifractal support set. \rev{The fractal dimension $D_{1}$ plays a special role, because it gives the scaling of the volume
$N^{D_{1}}$ of
the support set
of wave functions with the total system volume \rvI{$N$~\cite{Kravtsov2013support}}. The fundamental role of the support set is that
it gives the number of sites in the reference space and the number of states in the energy space that
is minimally sufficient for the normalization and completeness conditions. As a consequence, $D_{1}$
is directly related to the spectrum of fractal dimensions $f(\alpha)$ via
$D_{1}=\alpha_{1}=f(\alpha_{1})$ which significantly simplifies the analysis presented below. }
Furthermore, the new method
presented below is physically transparent and generic enough to be applied to analysis
of the multifractal
states in other systems.

\begin{figure*}[t]
\centering
\includegraphics[width=0.3\linewidth]{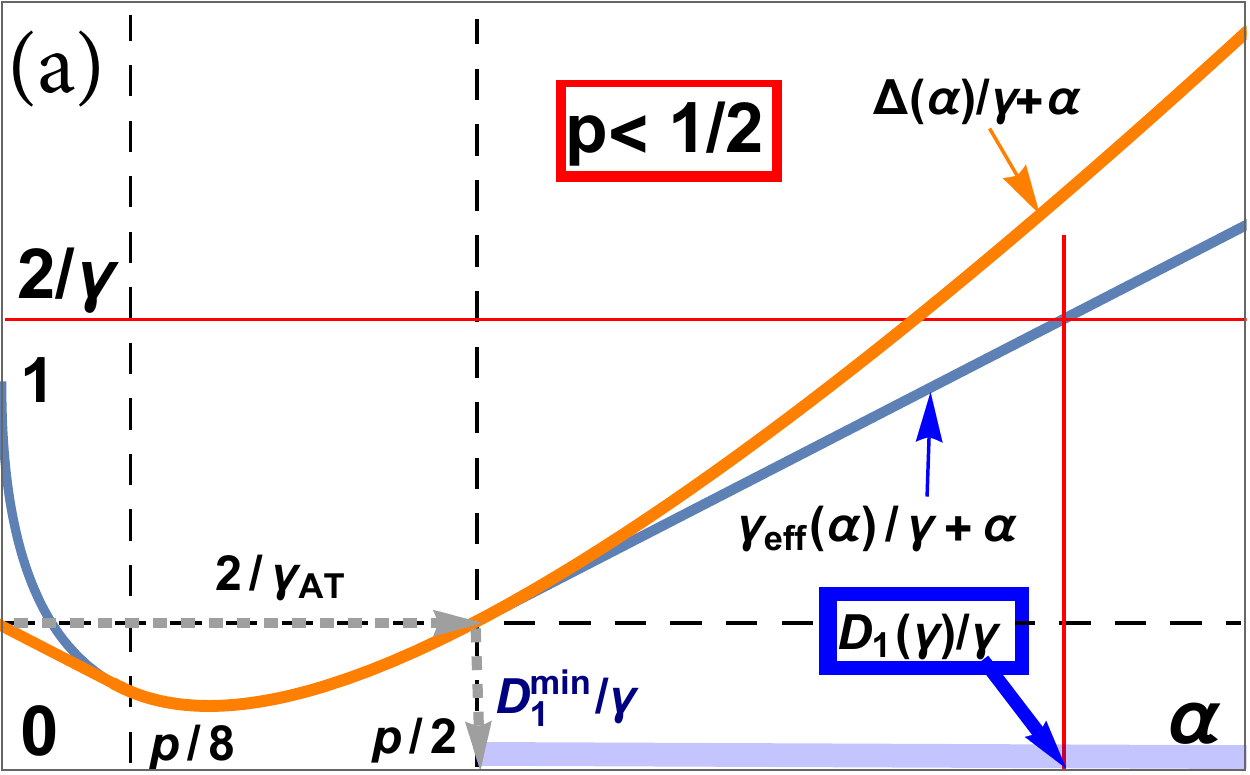}
\includegraphics[width=0.3\linewidth]{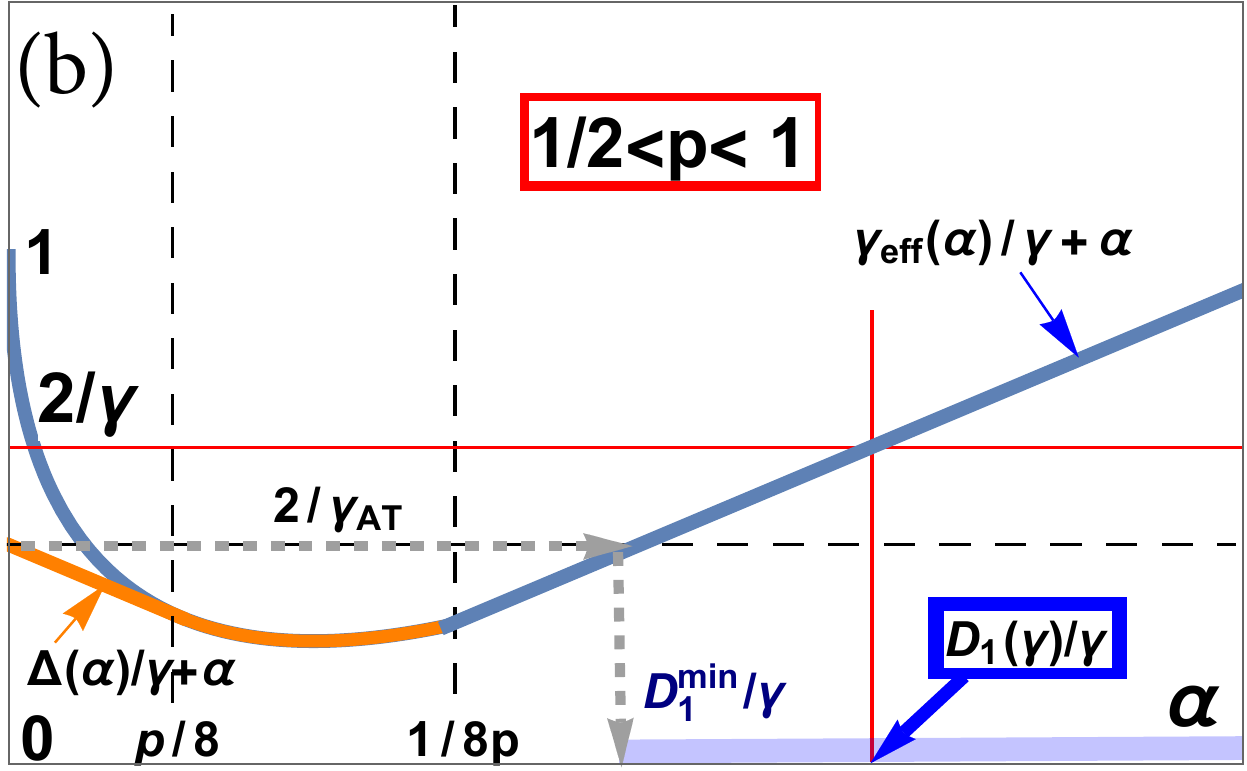}
\includegraphics[width=0.3\linewidth]{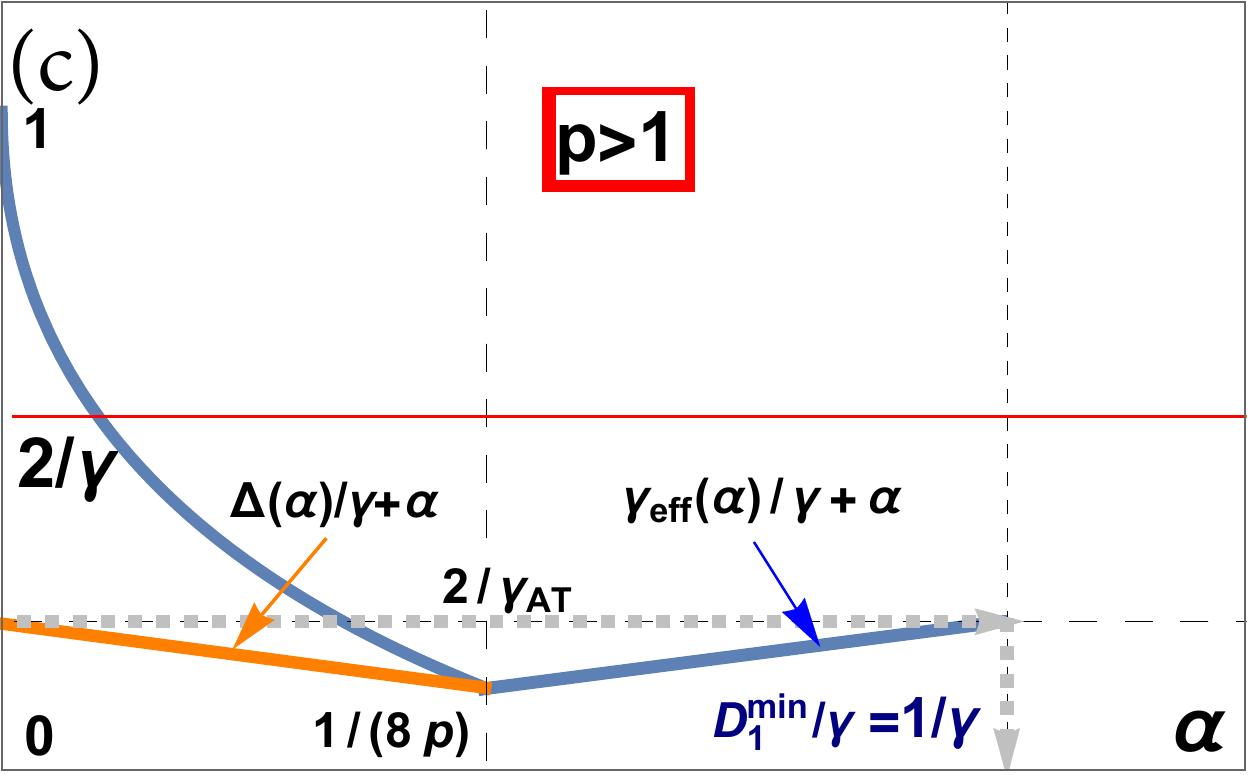}
\caption{(Color online) \textbf{The functions~\eqref{gamma_eff}
(blue curve) and~\eqref{Delta} 
(orange curve) entering inequalities
~\eqref{stab-MF-fin-Gauss},~\eqref{stab-MF-fin-LN}
in different regions of $p$:}
(a) $p<1/2$; (b) $1/2\leq p\leq 1$; (c) $p>1$.
Intervals of $\alpha = D_1/\gamma$ with different
functional dependence are shown by dashed vertical lines.
The Anderson localization transition corresponds to the lower of the blue and
orange curves equal to $2/\gamma$ at $\alpha=0$.
This transition is always determined by the orange curve representing the
log-normal part of the distribution $P(V)$.
On the contrary, the stable
fractal dimension $D_{1}(\gamma)=2-\gamma/\gamma_{ET}(p)$
for $\gamma\leq\gamma_{AT}$ is always determined by
the blue curve representing the Gaussian
part of the distribution $P(V)$.
The Anderson transition in all cases but $p=0$ is characterized by  the minimal
stable fractal dimension of the support set being
$D^{{\rm min}}_{1}=D_{1}(\gamma_{AT})=2-\gamma_{AT}/\gamma_{ET}(p)>0$
(shown by a gray dotted arrow).
The ergodic transition corresponds to $D_{1}(\gamma)=1$
and it is {\it continuous}.
For $p\geq 1$ there is no solution $D_{1}<1$ to the system of
inequalities~\eqref{stab-MF-fin-Gauss},~\eqref{stab-MF-fin-LN} in the region
of parameters where the localized
phase is unstable. In this case the multifractal phase is absent.
\label{Fig:stability} }
\end{figure*}

Let us consider two states $\psi_{\mu}$ and
$\psi_{\nu}$ on different fractal support
sets as it is shown in Fig.~\ref{Fig:hybrid}(a) and~(b).
We assume that both states are multifractal with $m \sim N^{D_{1}}$ sites
on a fractal support set where  the coefficients   $|\psi(i)|^{2}\sim N^{-D_{1}}$.

Here we apply a usual Mott's argument for hybridization of states,
Fig.~\ref{Fig:hybrid}(c), when
the disorder realization, in this case the off-diagonal matrix element,
changes from $H_{ij}$ to $H_{ij}' = H_{ij}+\delta\,H_{ij}$.
The key new element in the
theory we are introducing here is the hopping matrix element $V_{\mu,\nu}$
between the {\it states} and not between the {\it sites}
as is customary:
\be\label{U-mu-nu}
V_{\mu,\nu}=\sum_{i,j} \delta\,H_{ij}\,\psi_{\mu}(i)\,\psi_{\nu}(j).
\ee
Here $\psi_{\mu}(i)$ is the eigenfunction of the $\mu$-th state of $H_{ij}$, and
$\delta\,H_{ij}=H_{ij}'- H_{ij}$, where $H_{ij}'$
is drawn from the same log-normal distribution as $H_{ij}$.

Introducing $g_{ij}=-\ln \delta\,H_{ij}/\ln N $
and suppressing the indices $i,j$ for brevity
we conveniently rewrite~\eqref{LNdis} as follows~\footnote{Here we omit a
 small deviation  from the log-normal distribution
for $g_{ij}=-\ln |H_{ij}'-H_{ij}|/\ln N >\gamma/2$
which is not important in the current setting,
 see Appendix~\ref{App_sec:Stability} for details.}:
\be\label{P-g}
{\cal P}(g)={\rm const} \, N^{-\frac{1}{p\gamma}\,\left( g-\frac{\gamma}{2}
\right)^{2}},\;\;\;(g\geq 0).
\ee
By the constraint $g\geq 0$ we implemented the cutoff at $ |U|\sim O(N^0)$
discussed in Sec.~\ref{sec:Criteria}.

The typical number of terms in the sum~\eqref{U-mu-nu} in the interval
$dg$
is $N^{D_1}N^{D_1}{\cal P}(g)\sim N^{\sigma(g,D_{1})}\,d g$ where
\be
\sigma(g,D_{1})=2D_{1}-\frac{1}{p\gamma}\,
\left( g-\frac{\gamma}{2} \right)^{2}.
\ee
If $\sigma(g,D_{1})<0$,
the sum,~\eqref{U-mu-nu}, is dominated by a single term with the largest $|G_{ij}|$.
For positive
$\sigma(g,D_{1})>0$,
 many terms
contribute to this sum and the distribution $P(V\equiv |V_{\mu,\nu}|)$
becomes Gaussian. In general, there are both contributions
\be
\label{distr-V}
 P(V)= P_{{\rm LN}}(V) + P_{{\rm Gauss}}(V).
\ee
The condition of stability of the multifractal phase against hybridization
is derived similar to the Anderson criteria of stability,~\eqref{Anderson},
of the localized.
The difference is that now we have to replace the matrix element between the resonant
{\it sites} $U$ by the matrix element $V$ between the resonant {\it non-ergodic states}
and take into account that on each of $M=N^{1-D_{1}}$ different
support sets there are $m=N^{D_{1}}$ wave functions which belong to the same mini-band
and thus are {\it already in resonance} with each other.
Therefore the total number of {\it independent} states-candidates for hybridization
 with a given state should be smaller than the total number of states $M\,m=N$
and larger than the number of support sets $M$. This number is in fact equal to
their geometric mean $\sqrt{N M} = M\,\sqrt{m}=N^{1-\frac{D_{1}}{2}}$.

With this comment, the criterion of stability of the multifractal phase reads in the limit $N\rightarrow\infty$ as
\be\label{stab-MF}
N^{1-\frac{D_{1}}{2}}\int\limits_{0}^{W} dV\, V\, P(V)< \infty \ .
\ee
The contribution of the Gaussian part $P_{{\rm Gauss}}$ in~\eqref{distr-V}
to~\eqref{stab-MF} is:
\be\label{stab-MF-Gauss}
N^{1-\frac{D_{1}}{2}}\,\sqrt{\langle V^{2} \rangle}= N^{1-\frac{D_{1}}{2}-
\frac{1}{2}\gamma_{{\rm eff}}(D_{1})}<\infty,
\ee
where
\be\label{gamma_eff-V}
\langle V^{2} \rangle \equiv N^{-\gamma_{{\rm eff}}},
\ee
 and for stability it must be finite as
$N\rightarrow\infty$.
The contribution of $P_{{\rm LN}}$ in~\eqref{distr-V} to the stability
criterion~\eqref{stab-MF} is \\
$N^{1-\frac{D_{1}}{2}-\frac{\Delta(D_{1})}{2}}<\infty$, where
\be\label{stab-MF-LN}
\int\limits_{\sigma(g,D_{1})<0} dg\, N^{\sigma(g,D_{1})-g-D_{1}}
\equiv N^{-\frac{\Delta(D_{1})}{2}} \ .
\ee
Thus the multifractal phase is stable against hybridization
if the following inequalities are both fulfilled
\bea\label{stab-MF-fin-Gauss}
 D_{1}+ \gamma_{{\rm eff}}(D_{1}) &\geq& 2,\\
 D_{1}+ \Delta(D_{1}) &\geq & 2 \label{stab-MF-fin-LN}.
\eea
The functions $\gamma_{{\rm eff}}(D_{1})$ and $\Delta(D_{1})$ are computed in Appendix~\ref{App_sec:Stability} and discussed in the next
Section.

A particular case $D_{1}=0$ of~\eqref{stab-MF-fin-Gauss},~\eqref{stab-MF-fin-LN} describes the stability
criterion of the localized phase. If the localized phase is not stable, then
hybridization produces an avalanche of multifractal states living on fractal
support which
dimensionality grows until inequalities
~\eqref{stab-MF-fin-Gauss},~\eqref{stab-MF-fin-LN} are {\it both fulfilled}
for the first time
at some $0<D_{1}^{{\rm min}}<1$.
If this is possible in some parameter region then the multifractal state
is stable, otherwise the only stable extended
phase is ergodic.

\section{Fractal dimension of the NEE support set}\label{sec:D_1}
In this section we re-consider the phase diagram
Fig.~\ref{fig:phase_dia}(c)  from the viewpoint of
stability criteria given in the previous section by~\eqref{stab-MF-fin-Gauss},
~\eqref{stab-MF-fin-LN}
and derive the expression for the fractal dimension $D_{1}(\gamma)$.

To this end in Fig.~\ref{Fig:stability} we plot
\be\label{gamma_eff}
\frac{\gamma_{{\rm eff}}(\alpha)}{\gamma}+\alpha=
 \left\{\begin{array}{ll}1+3\alpha-2\sqrt{2\alpha p},&
 4\alpha< 2p, \frac{1}{2p} \\
 1/\gamma_{ET}(p)+\alpha,&
\text{otherwise}
\end{array}\right. \ ,
\ee
and
\be\label{Delta}
\frac{\Delta(\alpha)}{\gamma}+\alpha=
 \left\{\begin{array}{ll}
 1+3\alpha-2\sqrt{2 \alpha p},&
p<8\alpha<\frac1p\\
 {2}/{\gamma_{AT}(p)}-\alpha,&8\alpha<p,\frac1p\\
 1+3\alpha+2\sqrt{2 \alpha p},&
8\alpha>\frac1p
\end{array}\right. \ ,
\ee
 as functions of $\alpha=D_1/\gamma$.
Here $\gamma_{AT}(p)\geq 2$ and $\gamma_{ET}(p)\geq 1$ are
given by~\eqref{AT} and~\eqref{ET}, respectively (the details of derivation of
~\eqref{gamma_eff},~\eqref{Delta} from~\eqref{gamma_eff-V},~\eqref{stab-MF-LN}
are presented
in Appendix~\ref{App_sec:Stability}).
\begin{figure*}[t]
\centering
\begin{minipage}{0.79\textwidth}
\includegraphics[width=0.32\linewidth]{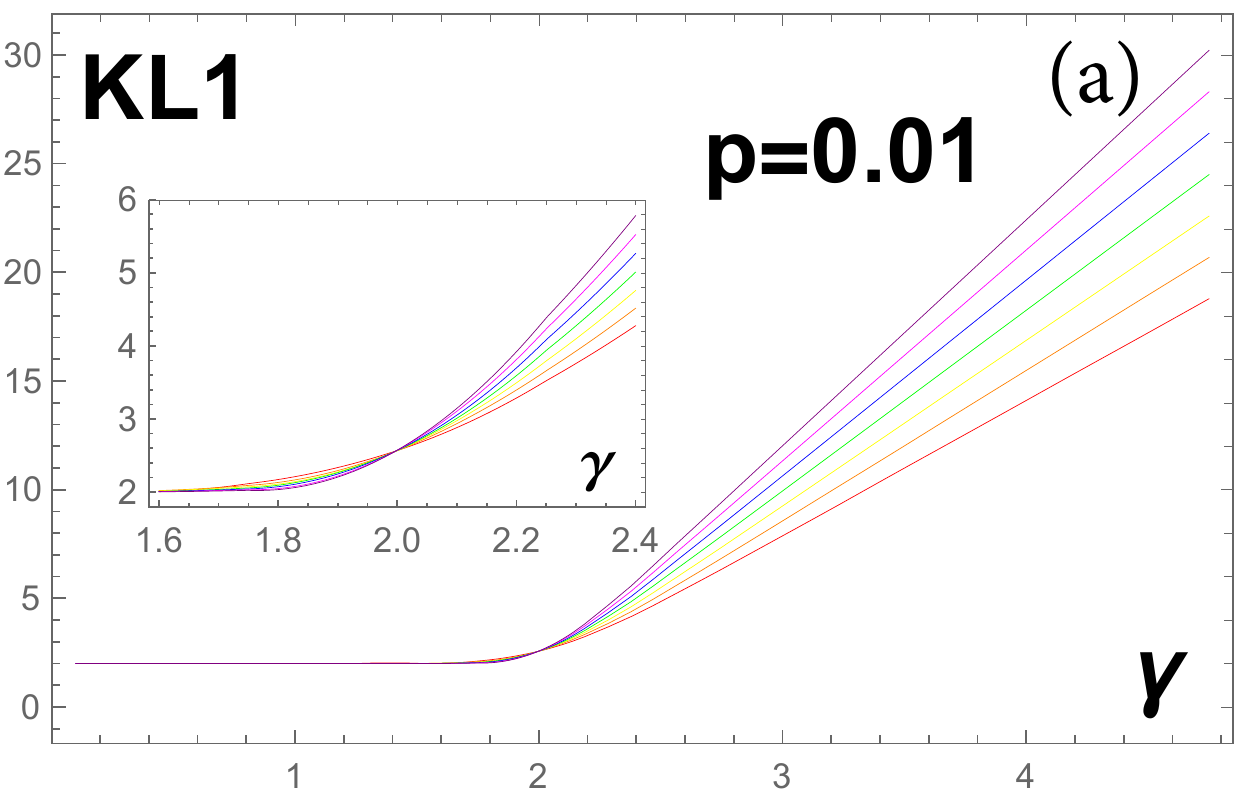}
\includegraphics[width=0.32\linewidth]{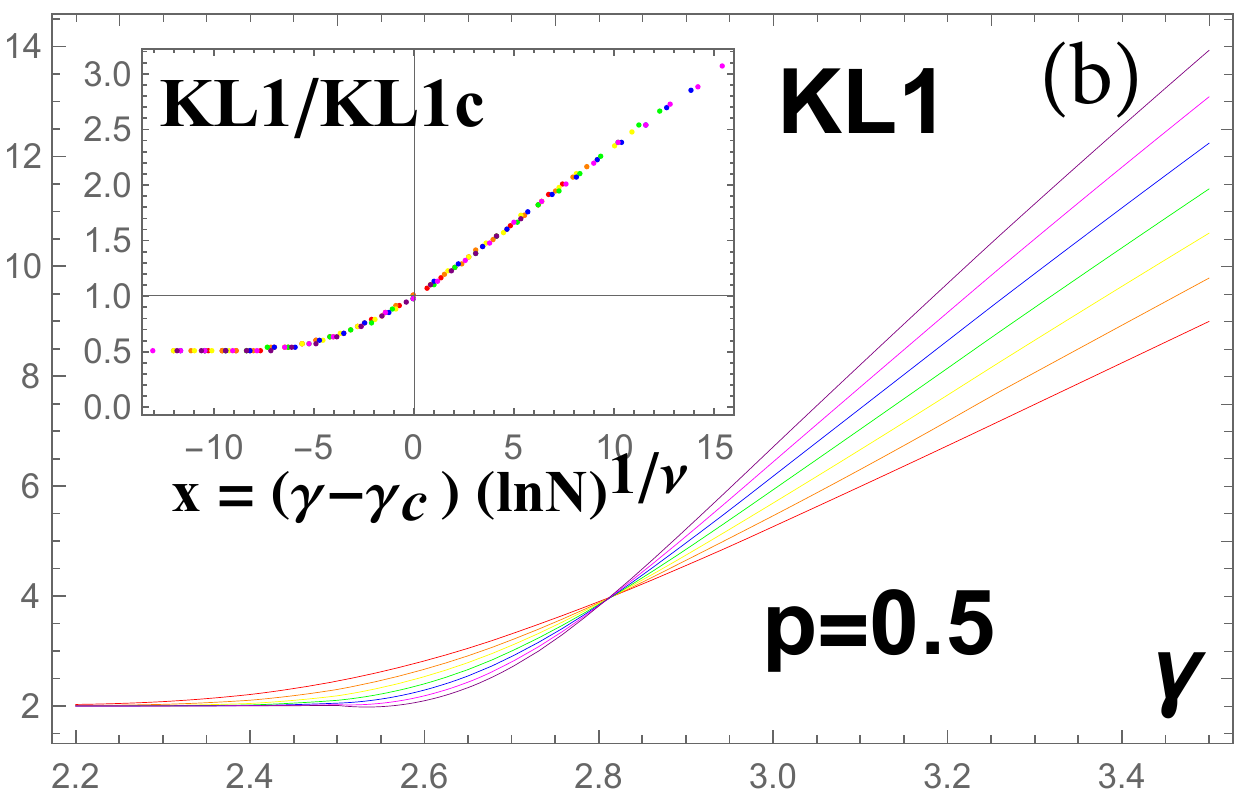}
\includegraphics[width=0.32\linewidth]{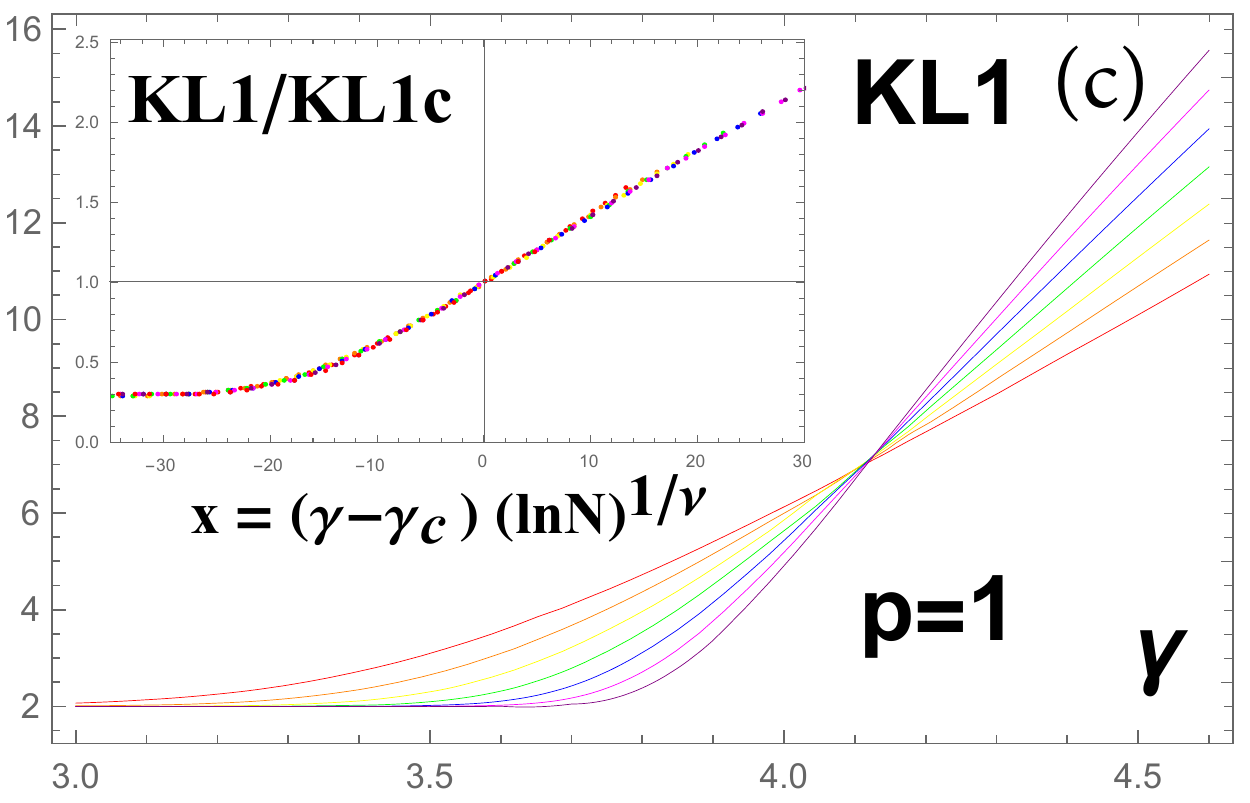}
\\
\includegraphics[width=0.32\linewidth]{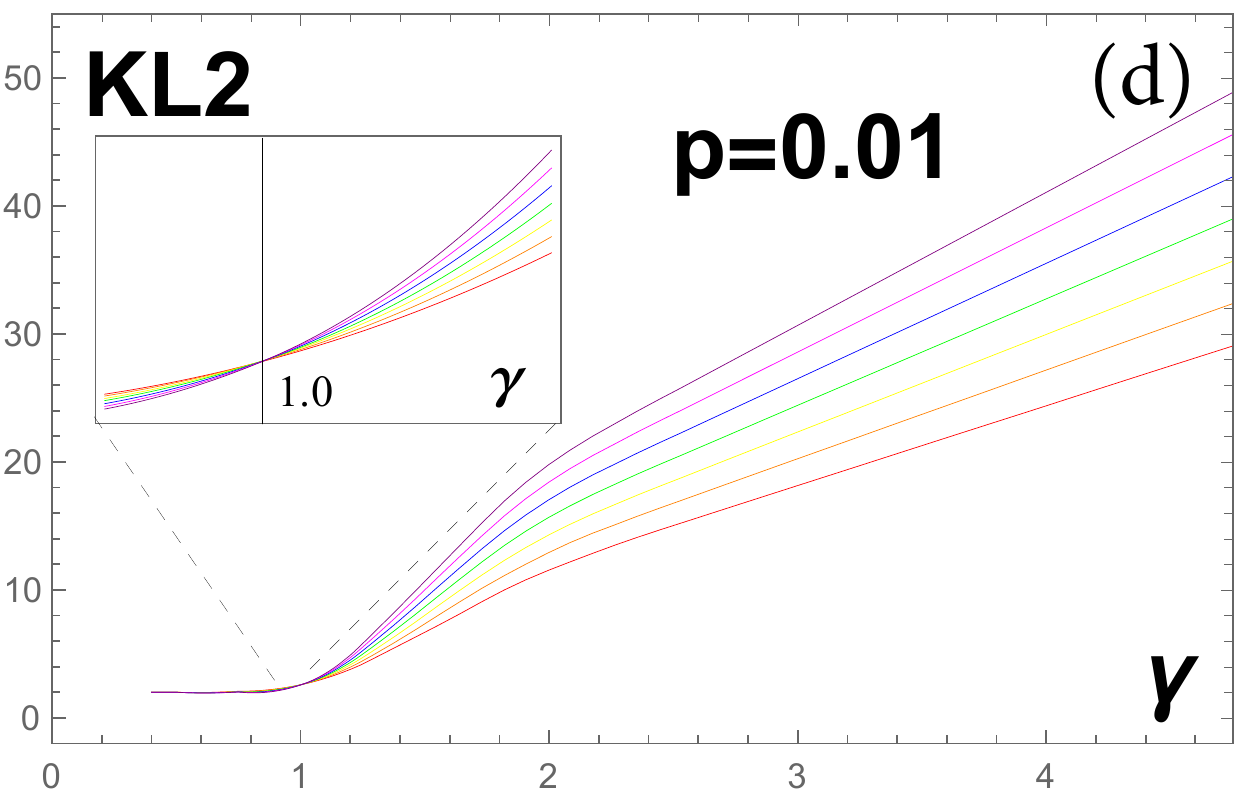}
\includegraphics[width=0.32\linewidth]{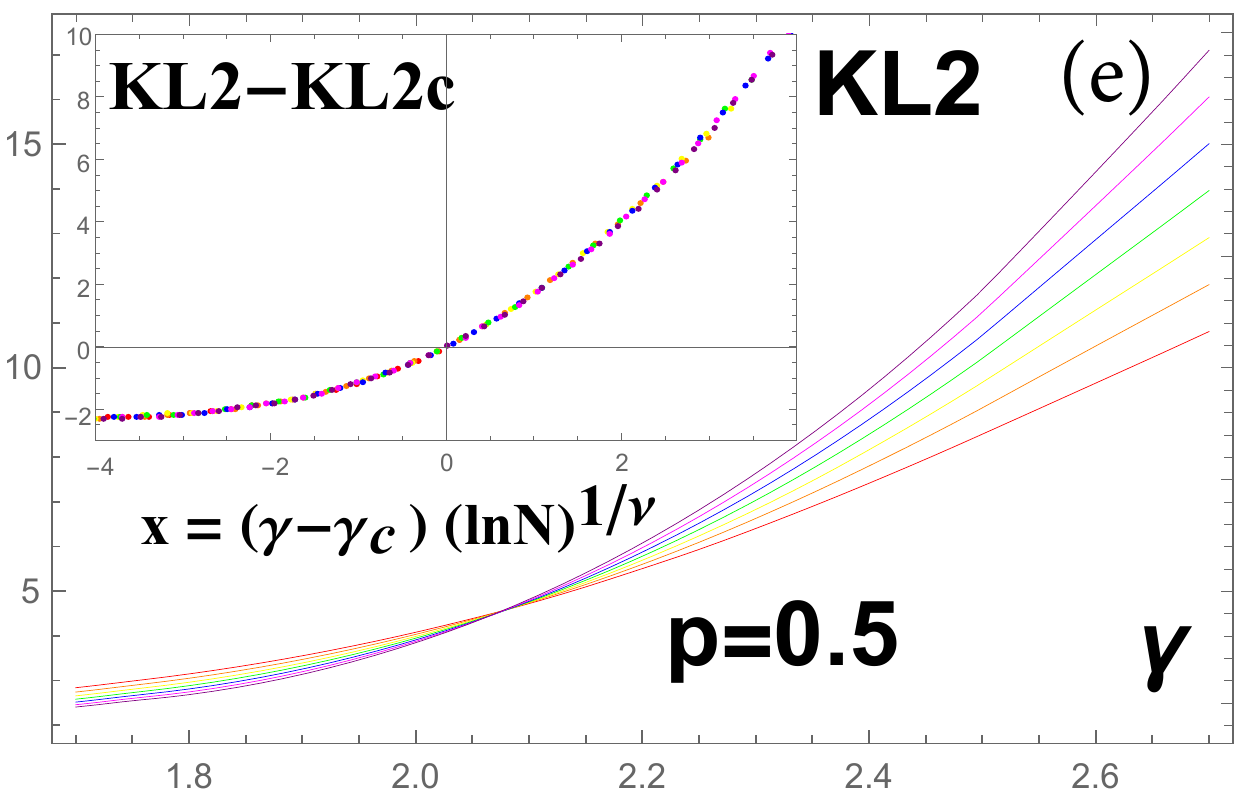}
\includegraphics[width=0.32\linewidth]{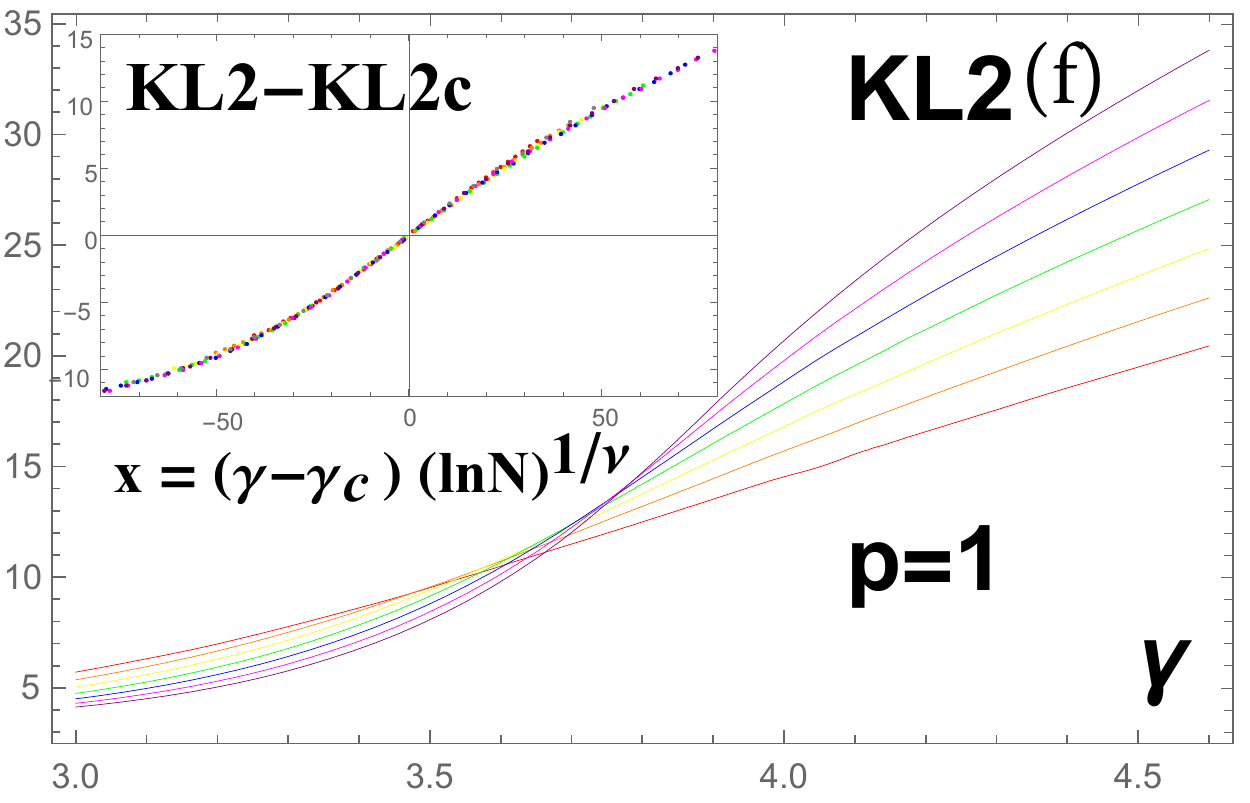}
\end{minipage}
\begin{minipage}{0.2\textwidth}
\includegraphics[width=\linewidth]{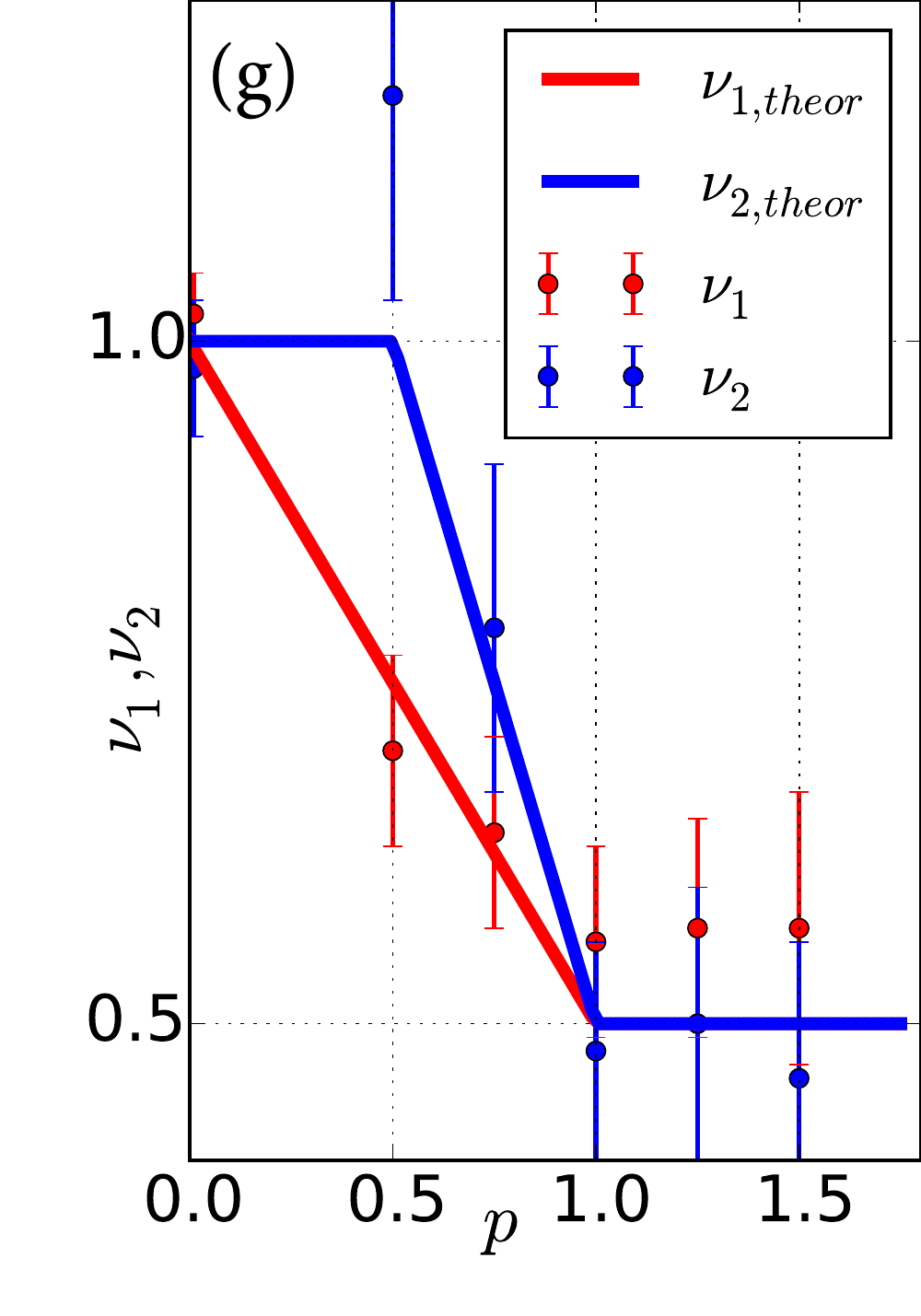}
\end{minipage}
 \caption{(Color online)
\textbf{Plots of ${\rm KL1}$ and ${\rm KL2}$ vs. $\gamma$ for LN-RP model}
at 
$N=2^L$, with $L$ from $9$ to $15$ with the step $1$ (from red to violet).
The logarithmic in $N$ divergence of
${\rm KL1}$ for $\gamma>\gamma_{AT}\approx 2$ and of ${\rm KL2}$ for $\gamma>\gamma_{ET}\approx 1$
is demonstrated in a wide interval of $\gamma$ for $p=0.01$, as well as
insensitivity of ${\rm KL1}$ to the ergodic transition. Intersection for
${\rm KL2}(\gamma$) curves is sharp at the
isolated continuous ergodic
transition at $\gamma_{ET}\approx 1$ for $p=0.01$ and at $\gamma_{ET}\approx 2.1$
for $p=0.5$, it is smeared out for $p=1.0$ when the ergodic transition
merges with the localization transition.
Intersection of curves for ${\rm KL1}$ at the Anderson localization transition
($\gamma_{AT}
\approx 2.0$ for $p=0.01$, $\gamma_{AT}\approx 2.8$ for $p=0.5$,
$\gamma_{AT}\approx 4.1$ for $p=1$ )
is sharp in all the cases. The insets show the collapse of the curves at the proper
choice of $\gamma_{c}$ and the critical exponents $\nu_{1}$ and $\nu_{2}$ for ${\rm KL1}$ and
${\rm KL2}$ at the AT and ET, respectively. The plot  Fig.\ref{Fig:intersect}(g) \rev{presents
the simplest conjecture for
$\nu_{1}$ and $\nu_{2}$ vs. $p$  consistent with the results of
finite-size scaling
presented in Table 1 and shown on the plot.} In the limit $p\to 0$ the critical exponents approach
their values $\nu_{1}=\nu_{2}=1$ for the Gaussian RP model~\cite{KLPino}.
For $p\geq 1$
we conjecture the mean-field values $\nu_{1}=\nu_{2}=1/2$.
\label{Fig:intersect} }
 \end{figure*}

According to the stability criteria~\eqref{stab-MF-fin-Gauss},~\eqref{stab-MF-fin-LN}
the functions~\eqref{gamma_eff},~\eqref{Delta}
should be compared to $2/\gamma$, see Fig.~\ref{Fig:stability}.
First, we note that the localized phase which formally corresponds to $D_{1}=0$,
is stable if the lowest of the blue and orange curves in Fig.~\ref{Fig:stability} is
higher than $2/\gamma$ at $\alpha=0$ and it is unstable otherwise.
One can see that at $\alpha=0$ for all values of $p$
the log-normal contribution to~\eqref{distr-V} (orange curve)
is lower than the Gaussian one (blue curve).
This means that the stability of the localized phase is always
determined by the log-normal part of $P(V)$.
Moreover, since at $\alpha=0$
~\eqref{gamma_eff},~\eqref{Delta} reduce to
$\alpha+{\gamma_{{\rm eff}}(\alpha)}/{\gamma} = 1$ and
$\alpha+\Delta(\alpha)/\gamma = 2/\gamma_{AT}$, respectively,
the stability of the localized phase implies that
$\gamma>\gamma_{AT}(p)\geq 2$ in agreement with~\eqref{AT}.


If the localized phase is unstable then
different localized states hybridize and form a multifractal state with $D_{1}>0$.
Those states are, however, unstable until their support set reaches the fractal
dimension $D_{1}^{min}>0$ where~\eqref{stab-MF-fin-Gauss},
~\eqref{stab-MF-fin-LN} are both fulfilled for the first time.

As the parameter $\gamma$ decreases below the critical value $\gamma_{AT}$,
the stable fractal dimension $D_{1}(\gamma)$ increases from $D^{{\rm min}}_{1}$
being always determined by the intersection of the horizontal line $y=2/\gamma>2/\gamma_{AT}(p)$ (red line in
Fig.~\ref{Fig:stability}) with the blue line. Thus the stable fractal dimension
$D_{1}(\gamma)$ is always determined by the Gaussian part of $P(V)$ and according to
the second line of~\eqref{gamma_eff} and Fig.~\ref{Fig:stability} is equal to:
\be\label{eq:D1}
D_{1}(\gamma)=2-\gamma_{{\rm eff}}=2-\frac{\gamma}{\gamma_{ET}(p)} \ , \quad p\leq 1
 \ .
\ee
At $\gamma=\gamma_{ET}$
the fractal dimension $D_{1}(\gamma)$ reaches unity, and at this point a {\it continuous}
ergodic transition happens. Thus the critical point of ergodic transition
coincides with that determined by~\eqref{ET}.


\rev{Note that, unlike   the ergodic transition,
 the localization transition is characterized by a {\it jump} in  the fractal dimension $D_{1}$
between the multifractal and the localized phase (where $D_{1}=0$).
 The stable fractal dimension $D_{1}(\gamma)$ is
{\it non-zero} just \rvI{below} the transition and is equal to:}
\be\label{eq:Dq_jump}
D_{1}^{min}=
\left\{\begin{array}{ll}
2-\frac{\gamma_{AT}(p)}{\gamma_{ET}(p)},&0<p<1\cr
1, & p\geq 1
\end{array} \right.
\ee
\begin{figure}[t!]
\center{
\includegraphics[width=0.44 \linewidth]{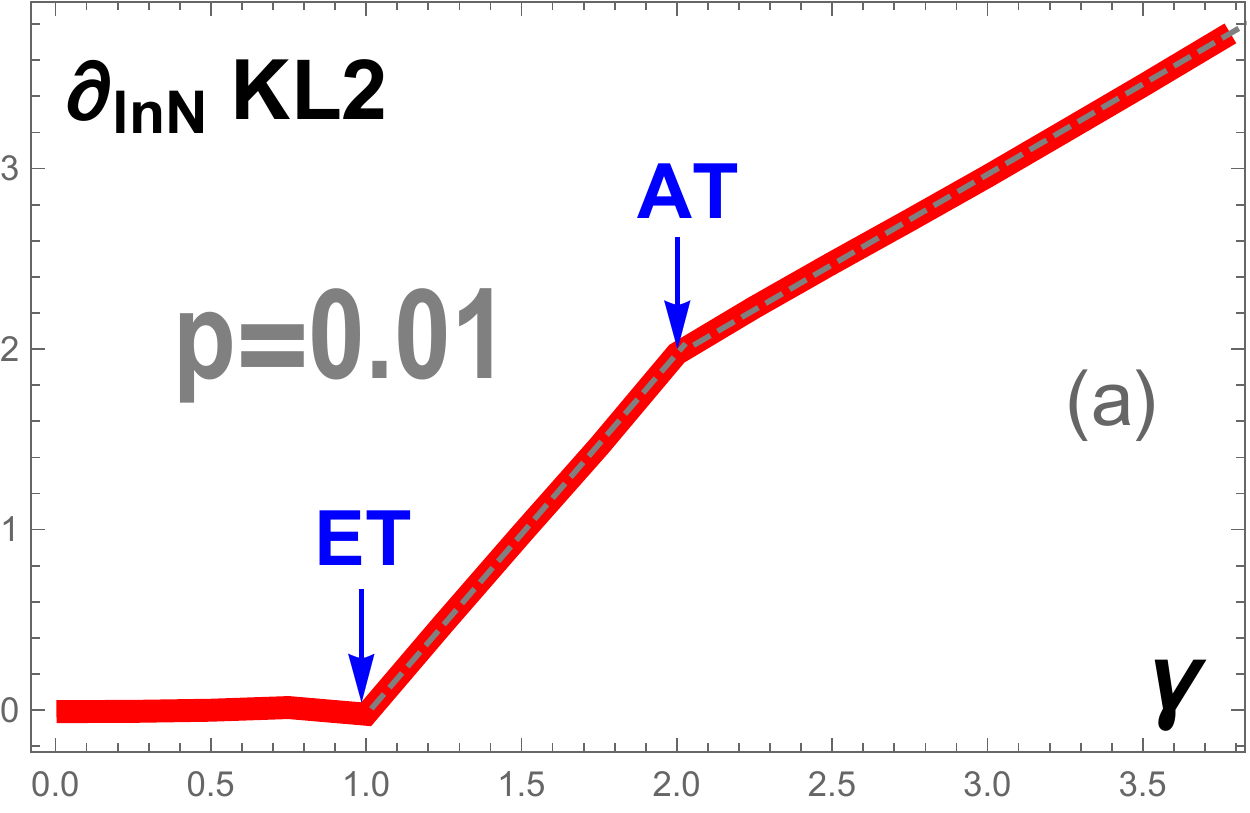}
\includegraphics[width=0.45 \linewidth]{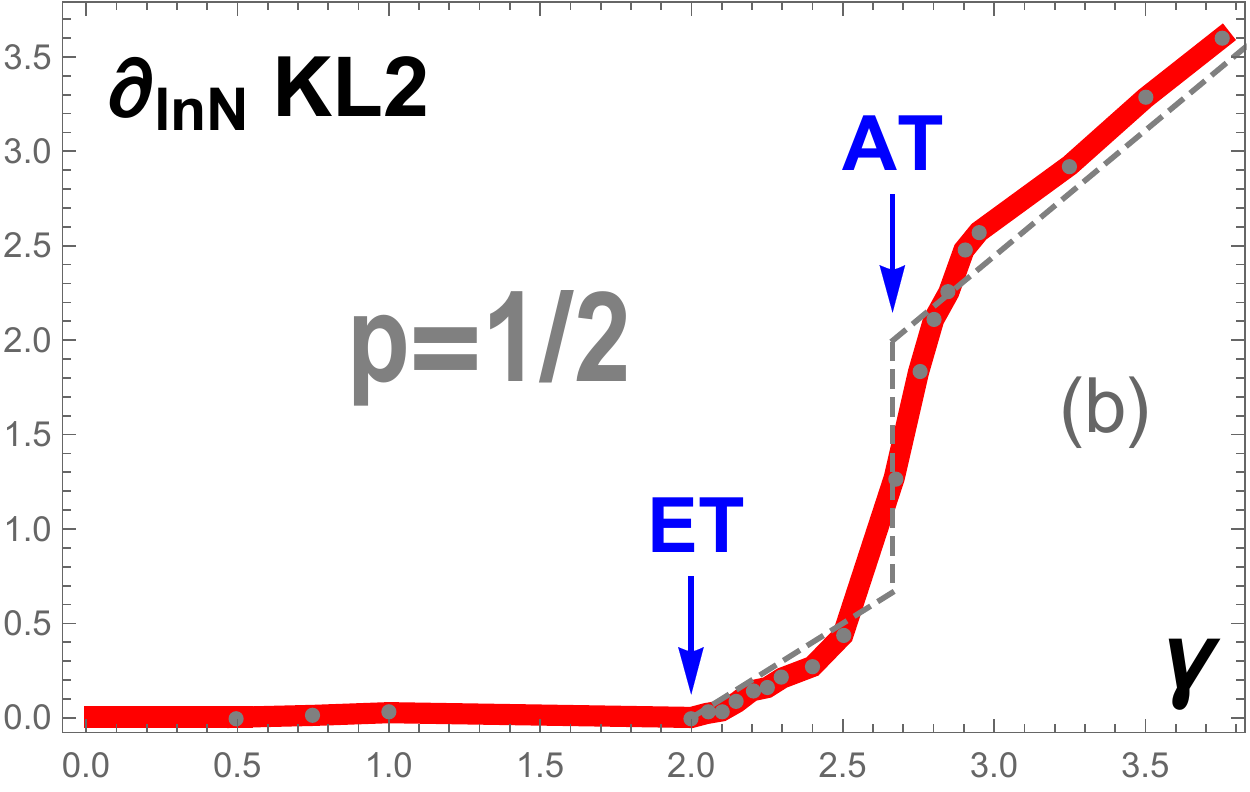}
}
\caption{(Color online) \textbf{Derivative of ${\rm KL2}$ w.r.t.
$\ln N$ vs. $\gamma$ for LN-RP model}
extrapolated from pairs of sizes $N=512~-~16384$ (red solid lines) for
(a)~$p=0.01$ and (b)~$p=0.5$
with the theoretical predictions~\eqref{KL-L} and~\eqref{KL-MF} (grey dashed lines).
The jump is related \rev{to} the jump in $D_1$,~\eqref{eq:Dq_jump}, for all $p>0$.
At $p\rightarrow 0$ \rev{the minimal fractal dimension $D_{min}\rightarrow 0$} and instead of
the jump in the function $d{\rm KL2}/d\ln N$ vs. $\gamma$ there is only a
jump in its $\gamma$-derivative at the AT. At $p=1/2$ \rev{and a at finite $N$}, instead, the jump
  manifests itself in the dramatic increase of slope near $\gamma=\gamma_{AT}$.
}
\label{Fig:KL_jump}
\end{figure}

This \revVK{{\it minimal fractal dimension} of the support set} is shown by the gray dotted arrow in Fig.~\ref{Fig:stability}.
As we show in the next section \revVK{it} reveals itself in the slope of the
Kullback-Leibler divergence, see Fig.~\ref{Fig:KL_jump} and~\eqref{KL-L}.
Fig.~\ref{Fig:stability}(c) demonstrates that for $p\geq 1$ the minimal
fractal dimension $D_{1}^{min}=1$, so that the multifractal phase is no
longer possible in LN-RP model~\eqref{LNdis}. However, it is restored if
the LN distribution is truncated at
$|U|\sim N^{-\gamma_{{\rm tr}}}$ with $\gamma_{{\rm tr}}>0$ (see
Appendix~\ref{sec:Truncation} for details).

\section{Kullback-Leibler (KL) measure}\label{sec:KL}

The numerical verification of~\eqref{AT},~\eqref{ET} and determination of the
critical exponents
at the Anderson localization and ergodic transitions is done in this paper using the
Kullback-Leibler divergence (KL)~\cite{KLdiv, KLdiv_book, Luitz2014, KLPino}~\footnote{For more detailed multifractal analysis of this
model see~\cite{KrKhay}}.

The Kullback-Leibler correlation functions ${\rm KL1}$ and ${\rm KL2}$
are defined as follows
~\cite{Luitz2014, KLPino}.
The first one is defined in terms of wave functions of two
{\it neighboring in energy} states $\psi_{\mu}(i)$ and $\psi_{\mu+1}(i)$
at the same disorder realization:
\be\label{KL1}
{\rm KL1}=\left\langle \sum_{i}|\psi_{\mu}(i)|^{2}\,\ln\left(\frac{|\psi_{\mu}(i)|^{2}}
{|\psi_{\mu+1}(i)|^{2}}\right)\right\rangle.
\ee
The second one is similar but the states $\psi$ and $\tilde{\psi}$ correspond to
different (and totally uncorrelated) disorder realizations:
\be\label{KL2}
{\rm KL2}=\left\langle\sum_{i}|\psi (i)|^{2}\,\ln\left(\frac{|\psi (i)|^{2}}
{|\tilde{\psi}(i)|^{2}}\right)\right\rangle.
\ee
The idea to define such two measures is the following. In the ergodic phases each of
the states has an amplitude $|\psi(i)|^{2}\sim N^{-1}$ of the same order of magnitude.
Then the logarithm of their ratio is of order $O(1)$, and for the normalized states
\be
{\rm KL1}\sim {\rm KL2} \sim O(1).
\ee
For fully-ergodic states the eigenfunction coefficients are
fully uncorrelated, even for the neighboring in energy states. Thus there is no
difference between ${\rm KL1}$ and ${\rm KL2}$. Using the Porter-Thomas distribution one finds:
\be\label{KL-EE}
{\rm KL1}={\rm KL2}=2.
\ee
For weakly-ergodic states ${\rm KL2}$ is still $O(1)$ but is larger than
the Porter-Thomas
value due to the fact that there are 'population holes' where $N|\psi(i)|^{2}$ is
$N$-independent but small, Fig.~\ref{fig:phase_dia}(a).

Deeply in the localized phase
$\ln |\psi_{\mu}(i)|^{2}\sim -|i-i_{\mu}|/\xi $, where $i_{\mu}$
is the position of the
localization center.
Since the positions of localization centers $i_{\mu}$ are not correlated
even for the states
neighboring in the energy, the logarithm of the ratio of the two wave function
coefficients in~\eqref{KL1},~\eqref{KL2} is
divergent in the thermodynamic limit.
For Anderson localized states on
finite-dimensional {\it lattices} this divergence is linear in the system size $L$.
However, localization on graphs such as RRG and RP models is not a conventional
localization~\cite{DeLuca2014, gRP}. In this case there is a power-law in $1/N$
background with the most probable (typical) value of
$|\psi (i) |^{2}_{{\rm typ}}\sim N^{-\alpha_0}$ far from the localization center and therefore:
\be\label{KL-L}
{\rm KL1}\sim {\rm KL2} = \alpha_0\,\ln N \rightarrow\infty
\ee
with $\alpha_0=(\gamma_{AT}/2)(\gamma-\gamma_{AT})+2$ for LN-RP model.

A qualitative difference between ${\rm KL1}$ and ${\rm KL2}$ is in the
multifractal phase.
In this phase the neighboring in energy states $|\psi_{\mu}(i)|^{2}$ and
$|\psi_{\mu+1}(i)|^{2}$ are
most probably belonging
to the same support set~\cite{MBL-footnote}
and hence they
are strongly overlapping: $|\psi_{\mu}(i)|^{2}\sim |\psi_{\mu+1}(i)|^{2}$.
Furthermore, eigenfunctions on the same fractal support set can be represented as:
$\psi_{\mu}(i)=\Psi(i)\,\phi_{\mu}(i)$, where $\Psi(i)$ is the multifractal
envelope on the support set and $\phi_{\mu}(i)$ is the fast oscillating function
with the
Porter-Thomas statistics~\cite{DeLuca2014}. Thus the ratio
$|\psi_{\mu}(i)|/|\psi_{\mu+1}(i)|$ and hence ${\rm KL1}$ in MF phase
has the same statistics
as in the ergodic one. We conclude that ${\rm KL1}$ is {\it not sensitive}
to the {\it ergodic} transition but is {\it very sensitive} to the {\it localization}
one, Fig.~\ref{Fig:intersect}.

In contrast, the eigenfunctions $\psi(i)$ and
$\tilde{\psi}(i)$ in ${\rm KL2}$ corresponding to {\it different realizations}
of a random
Hamiltonian,
overlap very poorly in MF phase. This is because the fractal support sets which
contain a vanishing fraction of all the sites,
do not typically overlap when taken at random. Therefore
\be\label{KL-MF}
{\rm KL2}=(\alpha_0 - D_1)\,\ln N = 2(1-D_1)\,\ln N
\ee
is divergent in the thermodynamic limit in the {\it multifractal} phase of RP models, with $(\alpha_0 - D_1) = 2(\gamma/\gamma_{ET}-1)>0$,~\eqref{eq:D1}, very much like in
the localized one.
This makes ${\rm KL2}$ {\it very sensitive} to the {\it ergodic} transition.
The properties of ${\rm KL1}$ and ${\rm KL2}$,
~\eqref{KL-EE},~\eqref{KL-L}, are fully confirmed by
numerics presented in Fig.~\ref{Fig:intersect}.
The jump in the slope $\alpha_0(\gamma_{AT}+0)-\alpha_0(\gamma_{AT}-0)+D_1^{min} = 2 D_1^{min}$ at the Anderson transition, $\gamma = \gamma_{AT}$, originates from the jump in $D_1$,~\eqref{eq:Dq_jump}.
Numerically it is clearly seen in the derivative of {\rm KL2} over $\ln N$ versus $\gamma$ shown in Fig.~\ref{Fig:KL_jump}.
We also show in Fig.~\ref{Fig:ratio} that ${\rm KL2}$ is sensitive to
the FWE transition and can be operative in identifying it.

A more detailed theory of ${\rm KL1}$ and ${\rm KL2}$ in the multifractal phase
is given in Appendix~\ref{App_sec:theory_KL}. The main conclusion of this analysis
 is that the curves for ${\rm KL1}(\gamma,N$) for different $N$ have an
intersection point at
 the critical
point $\gamma=\gamma_{AT}$ of the Anderson localization transition.
 At the same time, the intersection point for curves for
${\rm KL2}(\gamma,N$) coincides with the ergodic transition~\cite{KLPino},
provided that it is continuous
and well separated from the Anderson localization transition.
If the localization and ergodic
transition merge together and the multifractal state exists only at
the transition point, then intersection of
${\rm KL2}$ curves is
smeared out and may disappear whatsoever (as in 3D Anderson model).
However, the intersection of
${\rm KL1}$ curves remains sharp
in this case too (see Fig.~\ref{Fig:intersect}).

\begin{table}[t]
\centering
{
 \begin{tabular}{V{2.5} c V{2.5} l r V{2.5} l r V{2.5} c V{2.5} c V{2.5}}
\Xhline{3\arrayrulewidth}
\rowcolor{blue!30!white!50}
 p
 & $\gamma_{AT}$ &\extr{ext}, \theo{th}
 & $\gamma_{ET}$ &\extr{ext}, \theo{th}
 & $\nu_1$
 & $\nu_2$
 \\
\Xhline{3\arrayrulewidth}
 $0.01$ &
 $2.00$ & \theo{2}&
 $1.00$ & \theo{1}&
 $1.02\pm 0.03$ &
 $0.98 \pm 0.05$ \\
 \hline
\rowcolor{blue!30!white!50}
 $1/2$ &
 $2.80$ &\theo{2.67}&
 $2.06$ & \theo{2}&
 $0.70\pm 0.07$ &
 $1.18 \pm 0.15$ \\
 \hline
 $3/4$ &
 $3.43$ & \theo{3.2}&
 $2.96$ & \theo{3}&
 $0.64\pm 0.07$ &
 $0.79 \pm 0.12$ \\
 \hline
\rowcolor{blue!30!white!50}

 $1$ &
 $4.10$ &\extr{4.04}, \theo{4}&
 $3.90$ &\extr{4.10}, \theo{4}&
 $0.56\pm 0.07$ &
 $0.48 \pm 0.08$ \\
 \hline

 $5/4$ &
 $4.81$ &\extr{5.02}, \theo{5}&
 $4.45$ &\extr{5.19}, \theo{5}&
 $0.57\pm 0.08$ &
 $0.50 \pm 0.10$ \\
 \hline
\rowcolor{blue!30!white!50}

 $3/2$ &
 $5.51$ &\extr{5.84}, \theo{6}&
 $5.18$ &\extr{5.83}, \theo{6}&
 $0.57\pm 0.10$ &
 $0.46 \pm 0.10$\\
\Xhline{3\arrayrulewidth}
 \end{tabular}
 }
\caption{ Comparison of analytical predictions (blue),~\eqref{AT},~\eqref{ET},
and numerical data for the transition points $\gamma_{AT}$ and
$\gamma_{ET}$ and the corresponding critical exponents $\nu_{1}$ and $\nu_{2}$
for LN-RP model.
Numerical data (black) is obtained by
exact diagonalization of LN-RP random matrices with $N=512-32768$
from the intersection points in ${\rm KL1}$ and ${\rm KL2} $
and from finite-size scaling by the best collapse of the curves, Fig.~\ref{Fig:intersect}.
For $p>1$ a linear in $1/\ln N$ extrapolation to $N\rightarrow \infty$ of
the position of the
intersection point for two consecutive $N$ is shown in red.
}
\label{Fig:table1}
\end{table}

 The intersection of finite-size curves for ${\rm KL1}$ and ${\rm KL2}$ helps to
locate numerically
the critical points $\gamma_{AT}$ and $\gamma_{ET}$. More precise determination of
the critical points and the corresponding critical exponents $\nu_{1}$ and $\nu_{2}$
is done by the finite-size scaling (FSS) data collapse
(see insets in Fig.~\ref{Fig:intersect} and Appendix~\ref{App_sec:KL_collapse}).
The results are shown in the Table ~\ref{Fig:table1}.
On the basis of these
numerical results we conclude that our expressions~\eqref{AT},~\eqref{ET}
for the Anderson and ergodic
transition points
 are accurate and conjecture on the $p$-dependence of the critical exponents
$\nu_{1}$ and $\nu_{2}$ of AT and ET obtained from ${\rm KL1}$ and ${\rm KL2}$.
 (see  Fig.~\ref{Fig:intersect}(g)).
\section{Numerical location of the FWE transition}
\begin{figure}[t]
\center{
\includegraphics[width=0.455\linewidth]{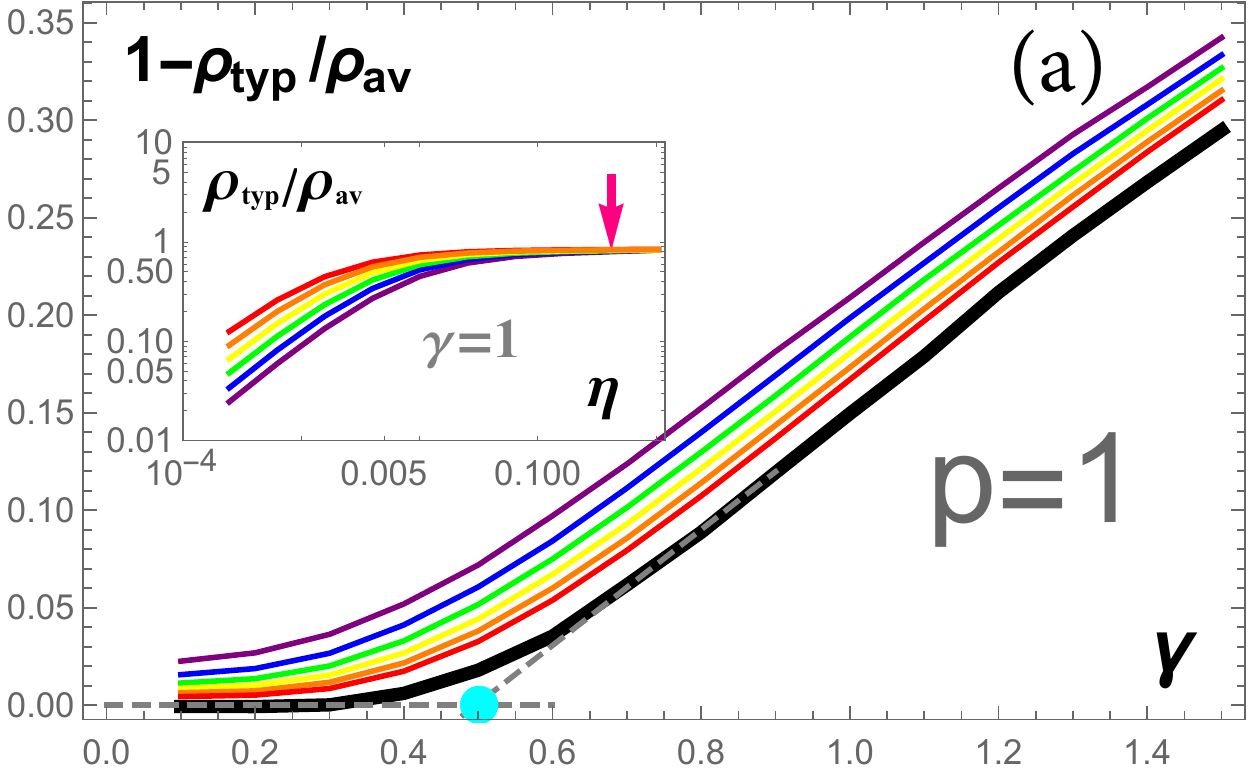}
\includegraphics[width=0.445\linewidth]{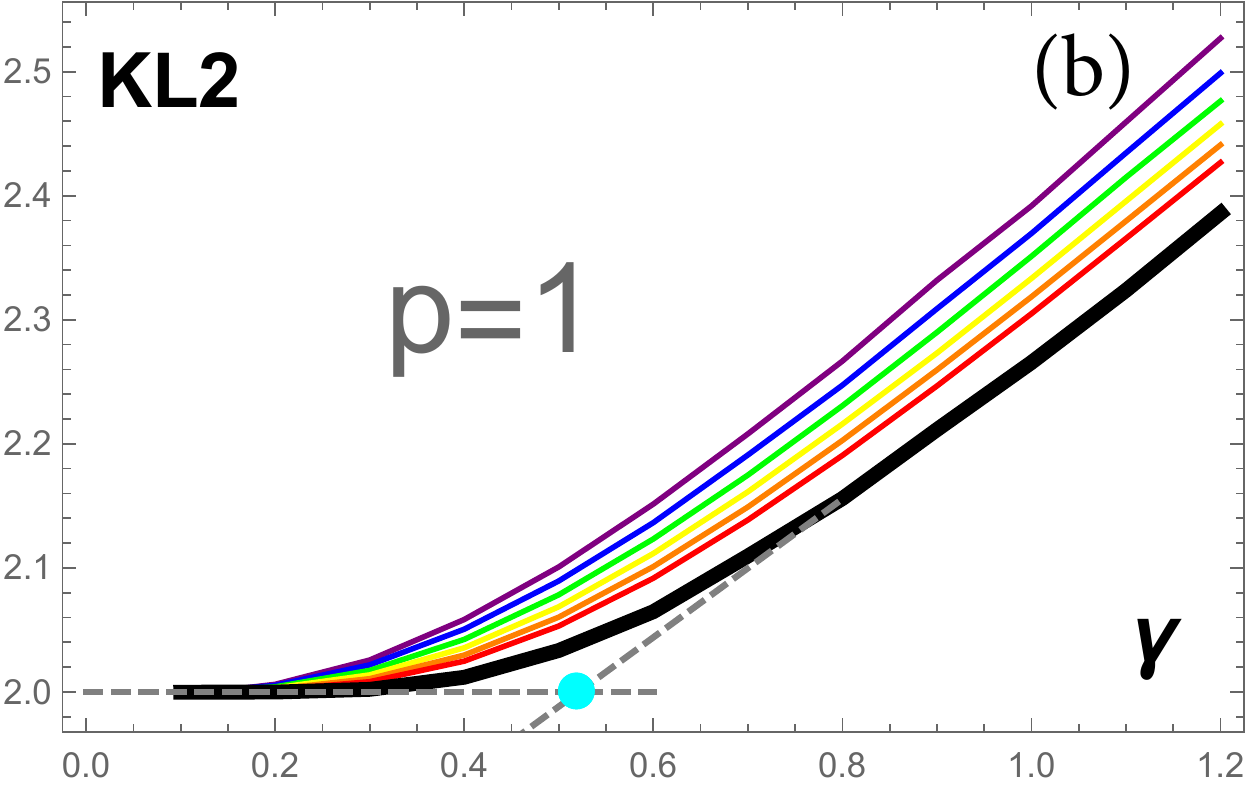}
 }
\caption{(Color online) \textbf{(a): The ratio of the typical and average LDoS}
 as a function of $\gamma$ for $p=1$ at different values of $N= 512-32768$
(purple through red)
and extrapolated to $N=\infty$ (black). Intersection of
dashed lines gives the position of FE-WE transition point
$\gamma_{FWE}\approx 0.5$ (shown by a bright blue point)
as predicted by~\eqref{FWT}. Inset:
dependence on the level width $\eta$. The main plot is done for $\eta$ shown by an
arrow at the plateau of $\eta$-dependence. \textbf{(b):
The Zoom of   Fig.~\ref{Fig:intersect}(f) 
{\rm KL2} vs.
$\gamma$ for p=1} for the same values of $N$ and their extrapolation
to $N\rightarrow\infty$. Intersection of dashed lines gives
the same position of FWE transition $\gamma_{FWE}\approx 0.5$ as on
Fig.\ref{Fig:ratio}(a). }
 \label{Fig:ratio}
 \end{figure}

For numerical verification of~\eqref{FWT} for FWE transition point we
make use of the
ratio of the typical $\rho_{{\rm typ}}$ and mean $\rho_{{\rm av}}$ average
\be
\ln\rho_{{\rm typ}}=\la \ln \rho(x,E+i\eta)\ra \ , \quad \rho_{{\rm av}}=\la \rho(x,E+i\eta)\ra \ ,
\ee
of {\it local} density of states (LDOS)
\be
\rho(x,E+i\eta) = \Im\sum_{\mu} |\psi_\mu(x)|^2/(E+i\eta-E_\mu) \ .
\ee
As is shown in Ref.~\cite{AnnalRRG}, at small
bare level width
$\eta \ll E_{BW}/N$, where $E_{BW} = \max(\Gamma,W)$ is the total spectrum bandwidth, this ratio $\rho_{{\rm typ}}/\rho_{{\rm av}}\sim \eta\,N^{D_{1}}/E_{BW}$
grows linearly with $\eta$ but then saturates at $\rho_{{\rm typ}}/\rho_{{\rm av}}
\sim N^{-1+D_{1}}$. In the ergodic phase $D_{1}=1$ and the plateau in
$\rho_{{\rm typ}}/\rho_{{\rm av}}$ tends to a finite limit as $N\rightarrow\infty$.
This behavior is well seen in the inset of Fig.~\ref{Fig:ratio}. We used
the properly defined \footnote{at the maximum of the second derivative of this
ratio vs. $\eta$, see Appendix~\ref{sec:LDOS_ratio} for details} plateau value of $\phi =1-\rho_{{\rm typ}}/\rho_{{\rm av}}$ as the
 {\it order parameter} for the FWE transition. For $\gamma<\gamma_{FWE}$ this parameter
$\phi =0$, signaling of the {\it fully-ergodic} phase. For $\gamma>\gamma_{FWE}$
the order parameter is non-zero. This behavior is shown in Fig.~\ref{Fig:ratio}
(see also an inset in   Fig.~\ref{fig:phase_dia}(e) and figures in
Appendix~\ref{sec:FWE_trans}), where
the black curve represents $\phi=\phi_{\infty}(\gamma)$ extrapolated to $N=\infty$ from
the finite $N$
values $\phi_{N}(\gamma)$ obtained by exact diagonalization.
In spite of imperfect extrapolation that does
not allow to get a true  non-analyticity  at $\gamma=\gamma_{FWE}$, the
 dashed gray lines of continuation of the black curve intersect exactly at
$\gamma=1/2$
which is the predicted value of $\gamma_{FWE}$ at $p=1$.
A similar intersection at $\gamma\approx 1/2$ is shown in the
${\rm KL2}$ vs. $\gamma$ plot in Fig.~\ref{Fig:ratio}.
They all suggest that the FWE transition does exist and is described
by~\eqref{FWT}.

\begin{figure*}[t]
\center{
\includegraphics[width=0.32\linewidth]{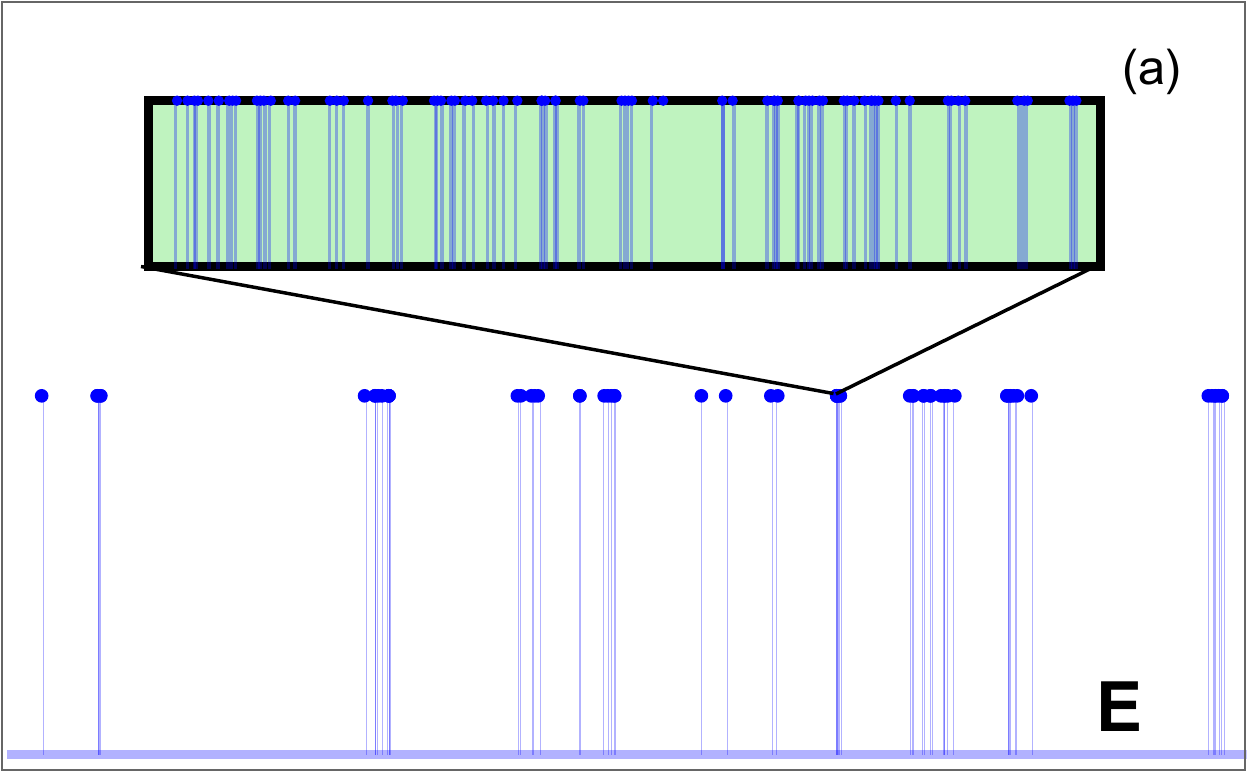}
\includegraphics[width=0.33\linewidth]{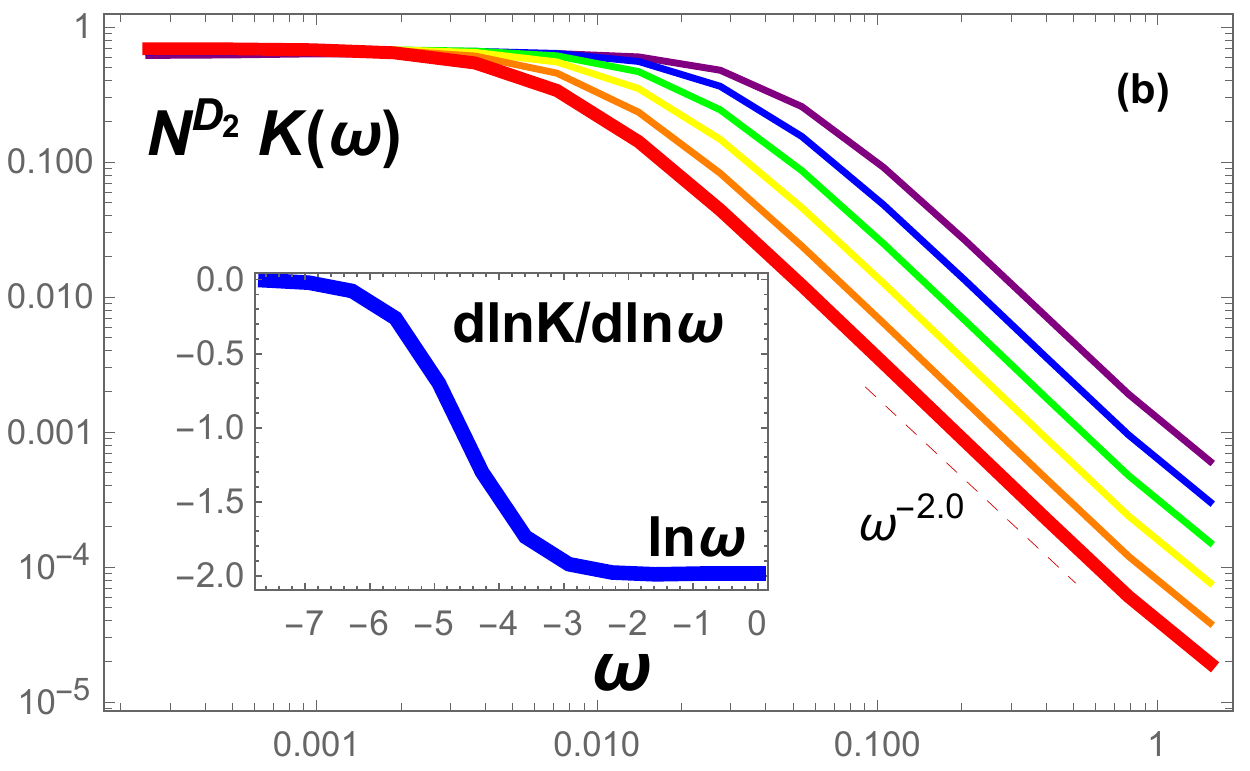}
\includegraphics[width=0.33\linewidth]{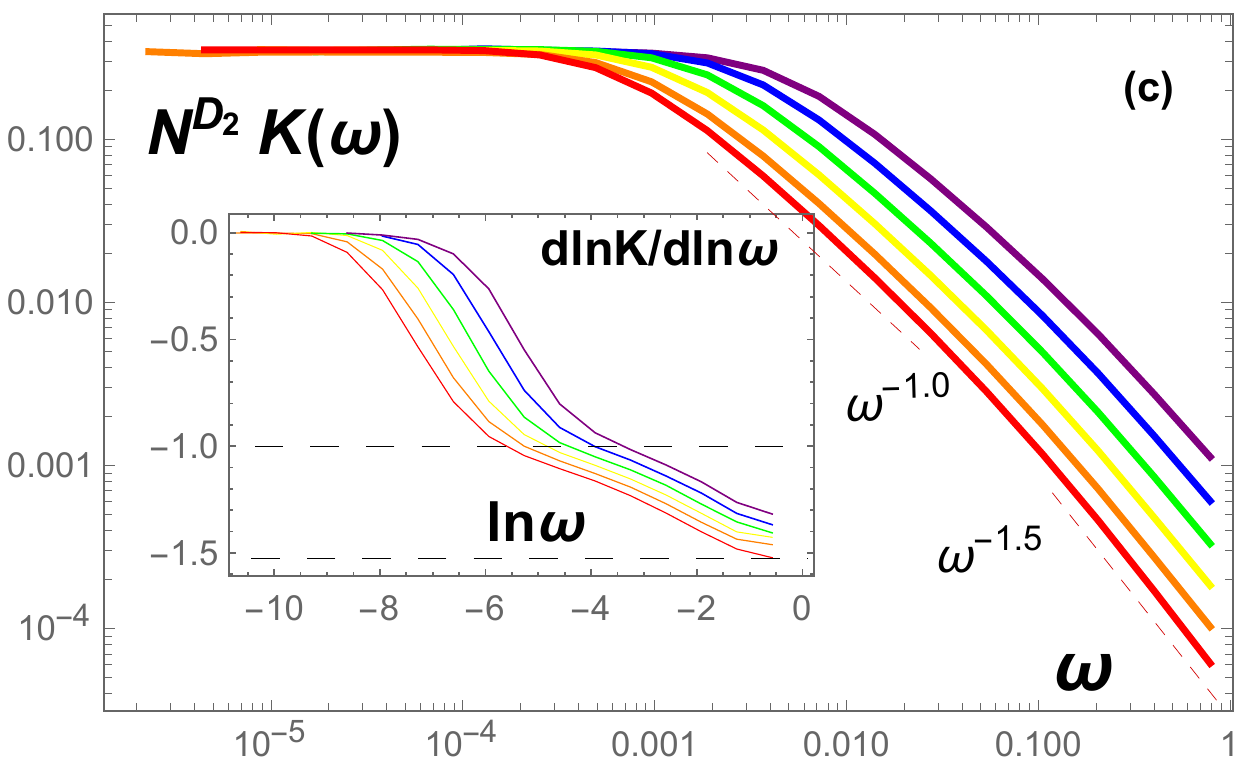}}
\caption{(Color online){\bf Fractal miniband structure.}
(a)~The \rvI{sketch of fractal set of fractal mini-bands in the local spectrum in the} case when fractal dimension $D_{s,in}$ of the set of
 levels {\it within} a mini-band is larger than the fractal dimension $D_{s,out}$ of the set of mini-bands.
 The width of mini-bands \rvI{$\Gamma$} tends to zero in the
limit $N\rightarrow\infty$ while the number of levels \rvI{$\Gamma/\delta$} in a mini-band tends to infinity;
(b)~The correlation function $K(\omega)$ for LN-RP model with $p=0.01$, $\gamma=1.5$, 
which corresponds to the eigenfunction fractal dimension $D_{2}\approx 2-\gamma=0.5$.
$K(\omega)$ in this \rvI{limit} is almost
indistinguishable from the one for GRP. It corresponds to a single Lorenzian mini-band with
$D_{s,in}\rightarrow 1$ and is characterized by  a single (trivial) power-law
$K(\omega)\propto \omega^{-2}$.
(c)~$K(\omega)$ in the multifractal phase of LN-RP model with $p=0.5, \gamma=2.5$.
It has a low-energy plateau which onset scales as $\omega \sim N^{-0.56}\gg N^{-1}$.
It can be interpreted as the width a  mini-band. At $\omega\rightarrow 0$ all plots collapse in one horizontal line after rescaling
$N^{D_{2}}\,K(\omega)$, where $D_{2}\approx 0.5$ for (b) and $D_{2}\approx 0.30$ for (c).
The falling part of $K(\omega)$ at larger $\omega$ cannot be described by
a single power-law with a trivial exponent \rvI{$\mu_{out}=2$}.  This is compatible with existence of extensive
number of mini-bands with the fractal structure as  shown in (a).
In panels (b) and (c) the color code corresponds to $N=512$, $1024$, $2048$, $4096$, $8192$, $16384$ from purple to red.
In the insets: the derivative
$\mu(\omega)=-d\ln K(\omega)/d\ln\omega$ which gives a running with $\omega$ exponent of
a local power-law.
It demonstrates a
formation of a plateau at $\mu\approx 1$ with
increasing $N$. At large $\omega\sim E_{BW}$ a {\it high}-$\omega$ plateau emerges
which level goes down \revVK{towards $-2$} as $N$
increases.
The correlation function $K(\omega)$ is a proxy for
the correlation function of local operators $\int dt\,e^{i\omega t}\,\langle O(r,t)\,O(r,0)\rangle$
in the problem of many-body localization, e.g. the
spin-spin correlation function~\cite{Abanin-Papic-Serbyn2017,Polkovnikov2020}.
\label{Fig:K_omega}}
\end{figure*}

\section{Fractal structure of mini-bands in the local spectrum. }
\rev{In non-ergodic phases the spectral statistics of local operators differs drastically from its
global counterpart.
This is because in a given observation point many states have negligible amplitude
in the limit $N\rightarrow\infty$ and thus cannot be seen. So emerges the {\it pure-point} spectrum in the
localized phase. In the multifractal phase the hierarchical structure \revVK{of distribution of
wave function  coefficients in the reference space} imposes, due to completeness, the fractal
structure of the local spectrum with power-law distribution of large inter-level spacing.
\revVK{Generically, both the distribution of levels inside a mini-band and
the distribution of mini-bands in the local spectrum may have a fractal structure}.

The simplest model of the local spectrum is shown in Fig.\ref{Fig:K_omega}(a). It consists of the set of
 mini-bands with the
width  $\Gamma$ that vanishes in the limit $N\rightarrow\infty$. Yet, in any of such mini-band there is an
\rvI{extensive number of levels, $\Gamma/\delta$, going to infinity} in this limit\rvI{,
due to $\delta \ll\Gamma\ll E_{BW}$}.
This allows one to define the fractal dimension $D_{s,in}$ of the local
spectrum {\it inside} a mini-band, as well as the fractal dimension $D_{s,out}$ of the set
of mini-bands. The global spectrum is a union of such sets which pave
densely all the spectral band.
However, in each given observation point one can see a sparse set of  mini-bands
  or even
one single mini band, as in the GRP model. The same set of mini-bands can be seen
in observation points which constitute a {\it fractal support set } in the reference space. There
are many support sets which the reference space is divided into, each of them corresponding to a
certain set of mini-bands. The {\it stratification} of the reference space first
suggested in~\cite{CueKrav2007} is a typical feature of
the multifractal phase.

 This qualitative picture can be tested by the correlation function $K(\omega)$~\cite{CueKrav2007}:
\be\label{Komega}
K(\omega)=\frac{ \sum_{n,m}\langle |\psi_{n}(r)|^{2}\,|\psi_{m}(r)|^{2}\delta(E-E_{n})
\delta(E+\omega-E_{m})\rangle}
{ N^{-1}\sum_{n,m} \langle \delta(E-E_{n})\delta(E+\omega-E_{m})\rangle},
\ee
where $\psi_{n}(r)$ and $E_{n}$ is the eigenfunction coefficient and the
eigenenergy of the $n$-th state.

One can show~\cite{Pino-Ioffe-VEK, KrKhay} that the fractal spectrum
like in Fig.\ref{Fig:K_omega}(a)   leads to
$K(\omega)$ which in the simplest approximation could be represented by two different
power-laws in $\omega$. For $N^{-1}<\omega<\Gamma$ smaller
than the width $\Gamma$ of a mini-band, the exponent $\mu_{in}$ of the power-law
$K(\omega)\sim \omega^{-\mu_{in}}$ that reflects the fractal structure of spectrum
inside a mini-band, is equal to
$\mu_{in}=1-D_{s,in}$. For GRP where $D_{s,in}=1$, one finds a trivial value $\mu_{in}=0$ which just extends the low-$\omega$
plateau beyond its natural limit $\omega=N^{-1}$.

At larger $\omega\gg \Gamma $
 the exponent $\mu_{out}$, reflecting the fractal structure of the set of mini-bands,
 \revVK{can take any values $0\leq \mu_{out}\leq 2$.  For the Gaussian RP, $K(\omega)$ is just a Lorenzian
~\cite{gRP, return, Biroli_RP}, and $\mu_{out}$ reaches its maximal value $\mu_{out}=2$}.
A similar behavior arises for LN-RP in the limit $p\rightarrow 0$
(see Fig.\ref{Fig:K_omega}(b)).

\revVK{In the MF phase of LN-RP model with $0<p<1$  the exponent $\mu_{out}$ appears
 to be  non-trivial ~\cite{KrKhay}, while $D_{s,in}$ is still equal to 1.}
A typical $K(\omega)$  is shown in Fig.\ref{Fig:K_omega}(c)
obtained \rvI{for $p=0.5$ and $\gamma=2.5$, $\gamma_{ET}<\gamma<\gamma_{AT}$} by exact diagonalization.
The low-$\omega$ plateau corresponding to $\mu_{in}=0$
is terminated at $\omega\sim N^{-0.56}$. \revVK{It
is followed by the under-developed power law with $\mu_{out}\approx 1.2$. At yet larger
$\omega\sim E_{BW}$ of the order of the total band-width, the slope of the
log-log plot of $K(\omega)$ decreases towards $-2$. }
Note that the fact that the low-energy plateau is extended till $\omega\sim N^{-0.56}$ much greater than the mean
level spacing \revVK{$\sim N^{-1}$}, tells us that the spectrum inside a mini-band
\revVK{of the width $\Gamma\sim N^{-0.56}$} has a fractal
dimension $D_{s,in}=1$, as in the case of GRP. \revVK{On the contrary,
the complex behavior of \revVK{$\mu(\omega)=-d\ln K(\omega)/d\ln \omega$} for $\omega>\Gamma$ which shows
 a shoulder at a
non-trivial $\mu_{out}\approx 1.2$,  signals of the fractal structure of
the set of mini-bands. }

Concluding this section, we would like to note that $K(\omega)$ can be considered as a proxy
for the correlation function of a local observable in many-body systems,
e.g. the Fourier-transform of the spin auto-correlation function\rvI{~\cite{Biroli_RRG_R(t)}}.
This quantity is very popular in the MBL literature~\cite{Abanin-Papic-Serbyn2017}.
In particular, it has been very recently studied in connection to the
discussion of the true nature of the MBL phase~\cite{Polkovnikov2020}.
}
\section{Conclusion and Discussion }\label{sec:Conclusion}
In this paper we introduce
a log-normal Rosenzweig-Porter (LN-RP) random matrix ensemble characterized by
a long-tailed distribution of off-diagonal matrix elements.
We obtain analytically the phase diagram of LN-RP using the
Anderson localization and Mott ergodicity criteria for random matrices complemented
by the new criterion for the transition between the fully- and weakly-ergodic phases.
This phase diagram is confirmed it by extensive numerics.

An alternative approach to localization and ergodic transitions based on the
analysis of stability with respect to
hybridization of
multifractal wave functions developed
in this paper
gives results identical to those obtained from the above criteria.
Using this approach we computed analytically
the dimension $D_{1}$ of the eigenfunction fractal support set and showed that
the Anderson localization
transition in our model \rev{is characterized by a jump in the fractal dimension $D_{1}$}
with the minimal fractal dimension
$D_{1}^{{\rm min}}>0$.

Our results show how the rare off-diagonal matrix elements which are much larger than
the typical ones,  give  rise to
a  \rev{ phase diagram with the fully-ergodic as well as a fragile weakly-ergodic and multifractal
extended phases and a
new FWE phase transition between the two ergodic phases}. \rev{These results shed light on the nature of the
extended states in the Anderson model on Random Regular Graph, as well as in the Hilbert space of
interacting systems in the problem of Many-Body Localization.}

\revVK{
Here we would like to mention the correspondence of the LN-RP model to a generic many-body
localization problem.  Similar to~\cite{Tarzia_2020} starting from the {\it short-range} Hamiltonian
of interacting particles in a
many-body  problem one can compute the parameters of the effective {\it long-ranged} LN-RP model.
 As the MBL phase breaks down the ergodicity and given the emerging evidence
\cite{Mace_Laflorencie2019_XXZ} that
the wave function in this phase has a multifractal structure
in the Hilbert space,
the MBL transition should be associated with the ergodic transition, $\gamma_{ET}$, of our model.
The localization in the Hilbert space of a generic many-body system can be achieved
only at the disorder strength $W$ scaling with the system size $L$~\cite{QIsing_2020}, thus,
$\gamma_{AT}$ corresponds to $W\sim L$.
It is also tempting to associate the FWE transition  at $\gamma=\gamma_{FWE}$
between the fully and the weakly ergodic phases
discovered in this paper, with the diffusion - sub-diffusion transition in
the many-body setting~\cite{BarLev_Luitz2016anom_diff}. However,
this correspondence has limitations related with correlated nature of off-diagonal matrix elements
in the Hilbert space due to the locality of Hamiltonian in the real space. Such correlations, as well as
the notion of locality is absent in our model.}


\begin{acknowledgments}
V.E.K. and I.M.K are grateful for support and hospitality to GGI of INFN and University of Florence (Italy) where this work was initiated.
V.E.K and B.L.A. acknowledge the support and hospitality of Russian Quantum Center during the work on this paper and G. V. Shlyapnikov for illuminating discussions there.
V.E.K gratefully acknowledges support from the Simons Center for Geometry and Physics, Stony Brook University at which part of the research for this paper was performed.
L.B.I. acknowledges a support of Basic Research Program of HSE.
This research was supported
by the DFG project KH~425/1-1 (I.~M.~K.),
by the Russian Foundation for Basic Research Grant No. 17-52-12044 (I. M. K.),
and by Google Quantum Research Award ``Ergodicity breaking in Quantum Many-Body Systems'' (V. E. K.).
\end{acknowledgments}

\bibliography{LN-RP}

\begin{thebibliography}{76}
\bibitem [{E_B()}]{E_BW-footnote}%
  \BibitemOpen
  \href@noop {} {}\bibinfo {note} {In the ergodic phase the bandwidth $E_{BW}$
  is growing with $N$ faster than $U$ that makes the main contribution to
  $\langle U^{2}\rangle$ and the average in Eq.(\ref{FW}) can be done using the
  full log-normal distribution~\cite{SM}.}\BibitemShut {Stop}%
\bibitem [{foo()}]{footnote:trunc}%
  \BibitemOpen
  \href@noop {} {}\bibinfo {note} {Note that the truncation at $U_{{\rm
  max}}\gtrsim O(1)$, $\gamma_{tr}\leq 0$, does not alter ergodic and
  localization transitions in the phase diagram in Fig.~1.}\BibitemShut {Stop}%
\end{thebibliography}
\appendix


\section{Truncated LN-RP and fragility of ergodic phase.}\label{sec:Truncation}

The phase diagram shown in Fig.~\ref{fig:phase_dia} of the main text confirmed numerically by calculations of the KL-divergence and by the ratio of typical and mean local density of states (LDOS)
demonstrates the collapse of the multifractal phase at $p\geq 1$ and
existence of the tricritical points in LN-RP model at $p=0$ and $p=1$.

In this section we show that the weakly-ergodic (WE) phase that emerges at $p>0$
and replaces fully the multifractal (MF) phase and partly the fully-ergodic (FE) one at $p\geq 1$
is unstable with respect to a deformation of LN-RP model such that $P(U)$ is cut from
above at:
\be
U_{\rm max}\sim N^{-\gamma_{tr}/2}\ll O(1), \;\;\;\;(\gamma_{tr}>0).
\ee
As the result of this truncation the multifractal phase re-appears by substituting
a part of the ergodic phase in a non-truncated LN-RP model
(see   Fig.~\ref{fig:phase_dia}(d)) ~\cite{footnote:trunc}.
To this end we use the expression that generalizes~\eqref{mom_q}:
\begin{eqnarray}\label{us-id}
&&\int\limits_{0}^{\min(N^{-\gamma_{\rm tr}/2},W)}dU\,U^{q}\,P(U)\sim \nonumber \\ &&\sim \left\{
\begin{array}{ll} N^{-\frac{q\gamma}{2}\left(1-\frac{pq}{2} \right)},
& \gamma(1-pq)>\gamma_{\rm tr},0 \\
N^{-\frac{1}{p\gamma} \left[\frac{(\gamma-\gamma_{\rm tr})^{2}}{4}+\frac{1}{2}\,pq\,\gamma\gamma_{\rm tr}\right]},
& \gamma_{\rm tr}>\gamma(1-pq),0
\\
N^{-\frac{\gamma}{4p}},
& \gamma_{\rm tr},\gamma(1-pq)<0
\end{array} \right.
\end{eqnarray}
and apply the same criteria~\eqref{Anderson},~\eqref{Mott},~\eqref{FW} to find the critical points
of the localization and both ergodic transitions.

Then we obtain that the critical point $\gamma_{AT}$ of
the Anderson localization transition is affected as follows
\be\label{AT-trunc}
 \gamma_{AT} = 2p-(p-1)\gamma_{tr}+\sqrt{(2p-(p-1)\gamma_{tr})^{2}-\gamma_{tr}^{2}},
\ee
only if $\gamma_{tr}>\gamma_{AT}\,(1-p),0$. In the opposite case truncation does not affect
$\gamma_{AT}$.

For the critical point $\gamma_{ET}$ of the ergodic transition in the same way we find
the effect only for $\gamma_{tr}>\gamma_{ET}\,(1-2p),0$ given by
\be\label{ET-trunc}
 \gamma_{ET}=2p-(2p-1)\gamma_{tr}+\sqrt{(2p-(2p-1)\gamma_{tr})^{2}-\gamma_{tr}^{2}}.
\ee

The criterion for the fully-weakly ergodic (FWE) transition does not have any truncation of $\la U^2 \ra$ at $U\sim W$, thus
it is affected by the truncation at all $\gamma_{tr}>\gamma_{FWE}\,(1-2p)$ (even negative ones if $p>1/2$).
As a result FWE transition occurs for $\gamma_{tr}>\gamma_{FWE}\,(1-2p)$ at
\be\label{FWT-trunc}
 \gamma_{FWE}=\frac{2p+(2p-1)\gamma_{tr}+\sqrt{(2p+(2p-1)\gamma_{tr})^{2}+(8p-1)\gamma_{tr}^{2}}}{8p-1}.
\ee

Note that~\eqref{AT-trunc} and~\eqref{ET-trunc} give real solutions for $\gamma_{tr}<\gamma_{AT}(0) = 2$ and $\gamma_{tr}<\gamma_{ET}(0) = 1$, respectively,
and both these solutions increase with the tail weight $p$.
At the same time FWE transition replaces ET one for all $\gamma_{tr}>1$ as
$\gamma_{FWE}(\gamma_{tr} = 1) = \gamma_{ET}(\gamma_{tr} = 1) = 1$ for all $p$.
Similar thing happens for $\gamma_{tr}>2$, when FWE transition replaces ALT as well, with $\gamma_{AT}(\gamma_{tr} = 2) = 2$, but in this case $\gamma_{FWE}(\gamma_{tr} = 2) = 2$ only for $p\to 0$.

The results of~\eqref{AT-trunc},~\eqref{ET-trunc}, and~\eqref{FWT-trunc} are plotted in
  Fig.~\ref{fig:phase_dia}(d).

One can see that at {\it any} positive non-zero $\gamma_{tr}$
the multifractal NEE phase emerges at $p\geq 1$ in between of the localized and
ergodic ones.
Indeed, at small $\gamma_{tr}\ll 1$ the line of localization transition is almost
insensitive to truncation close to $p=1$ ($p>1-\gamma_{tr}/(4p)$)
\be
\gamma_{AT} \simeq 4p - 2(p-1)\gamma_{tr} - \frac{\gamma_{tr}^2}{4p}+O\lp\gamma_{tr}^3\rp \ ,
\ee
while the line of ergodic transition is pushed to
smaller values of $\gamma$ at $2p>1-\gamma_{tr}/(4p)$
\be
\gamma_{AT} \simeq 4p - 2(2p-1)\gamma_{tr} - \frac{\gamma_{tr}^2}{4p}+O\lp\gamma_{tr}^3\rp \ ,
\ee
corresponding to larger typical transition matrix elements $U$ (smaller effective disorder).
Thus, the width of the MF phase increases linearly with $\gamma_{tr}\ll 1$
\be
\gamma_{AT} - \gamma_{ET} = 2p\gamma_{tr}+O\lp\gamma_{tr}^3\rp \ .
\ee
This proves the fact that the weakly ergodic phase in LN-RP with $p\geq 1$ is very fragile and exists only due to atypically large
transition matrix elements. It is substituted by the multifractal NEE phase as soon
as such matrix elements are made improbable by truncation.

In the limit $\gamma_{tr}\ll 1$ the width of the WE phase can be approximated at $2p>1-\gamma_{tr}/(4p)$ as
\be
\gamma_{ET} - \gamma_{FWE} = \frac{8p}{8p-1}\lb (4p-1)-2(2p-1)\gamma_{tr}\rb + O\lp\gamma_{tr}^3\rp \ ,
\ee
showing linear decrease with $\gamma_{tr}$ and
giving a reasonable approximation of the value of $\gamma_{tr} \simeq 1$ where this phase disappears.
Here we use
\be
\gamma_{FWE} = \frac{4p +2(2p-1)\gamma_{tr}}{8p-1}+\frac{\gamma_{tr}^2}{4p} + O\lp\gamma_{tr}^3\rp \ .
\ee

\section{Analysis of stability}\label{App_sec:Stability}
\begin{figure}[b!]
\center{
\includegraphics[width=0.4\linewidth,angle=0]{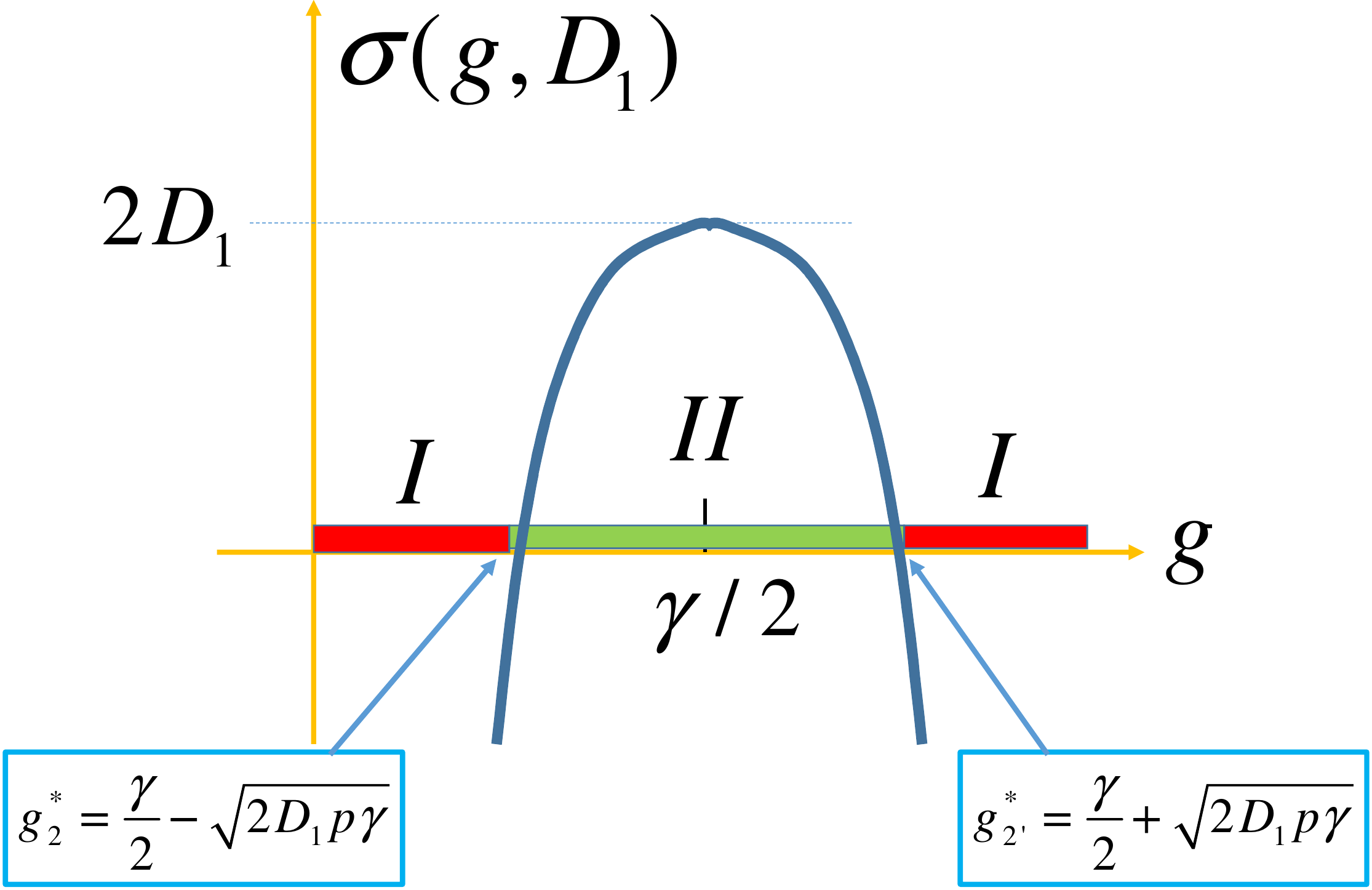}}
\caption{(Color online) Regions of $g$ contributing to the log-normal (I) and
Gaussian (II) parts of the distribution function $P(U_{\mu,\nu})$. }
\label{Fig:regions}
\end{figure}

In this section we calculate the contributions to $P(V)$ from the
log-normal $P_{{\rm LN}}(V)$ and Gaussian $P_{{\rm Gauss}}(V)$ parts to~\eqref{distr-V}.

\begin{figure*}[t]
\center{
\includegraphics[width=0.49 \linewidth,angle=0]{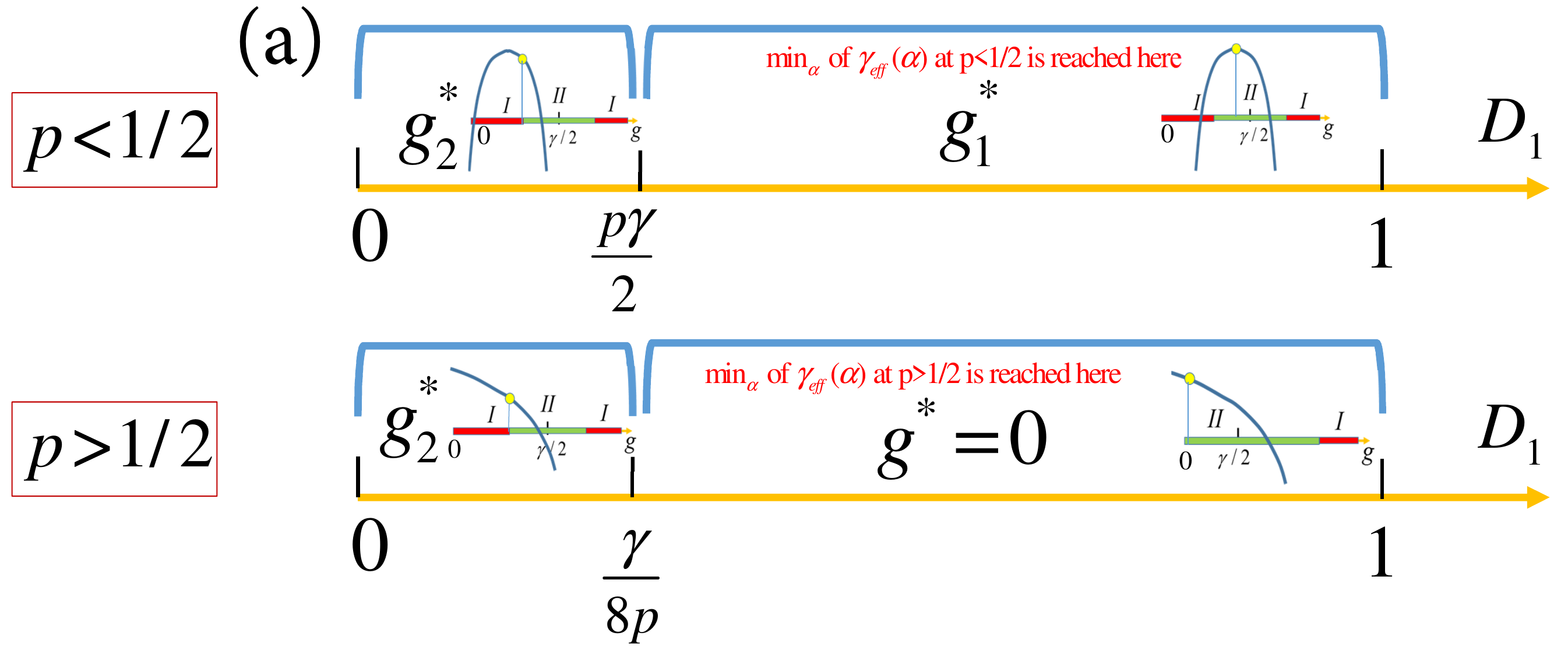}
\includegraphics[width=0.49 \linewidth,angle=0]{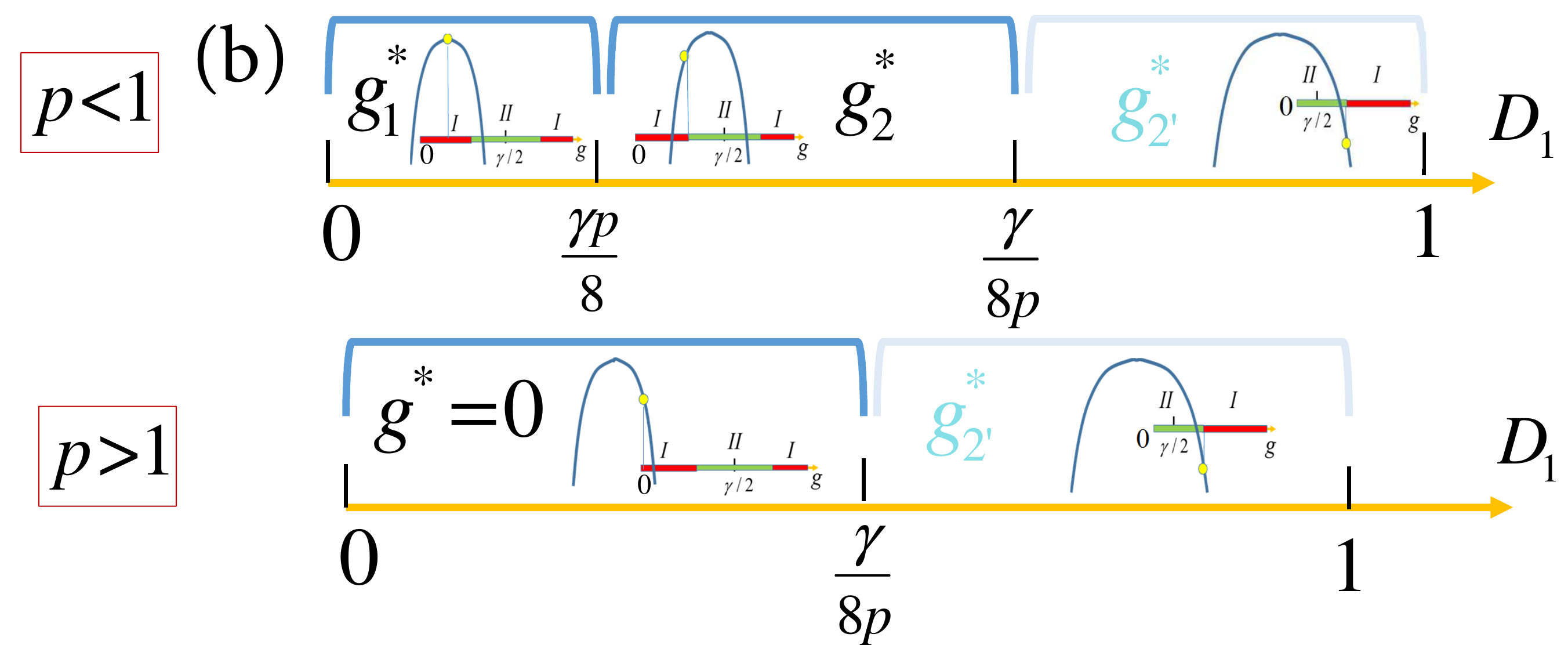}}
\caption{(Color online) (a) Different possible positions $g^{*}_{1}$, $g^{*}_{2}$
or $g^{*}=0$ that maximize
~\eqref{max} in region II depending on $p$, $\gamma$ and $D_{1}$. The reference
of maximum realized in each sector of parameters is shown by an ikon
in the corresponding sector.
(b) Different possible positions $g^{*}_{1}$, $g^{*}_{2}$
or $g^{*}=0$ that maximize
~\eqref{max-LN} in region I. The reference
of maximum realized in each sector of parameters is shown by an ikon
in the corresponding sector. For $D_{1}>\gamma/8p$ the maximum in
~\eqref{max-LN} is reached at the
edge of the {\it right} segment of region I, $g=g_{2'}^{*}$ (not to be
confused with the edge of the {\it left} segment $g=g_{2}^{*}$,
see Fig.~\ref{Fig:regions} ). It leads to a higher branch of
the orange curve $\Delta(\alpha)/\gamma+\alpha$ in Fig.~\ref{Fig:stability}
(not shown in Fig.~\ref{Fig:stability})
which is separated by a gap from the blue curve in Fig.~\ref{Fig:stability}
and thus is irrelevant for our analysis.
\label{Fig:maximize_Gauss_LN}}
\end{figure*}

One can easily compute the variance of the Gaussian part of $P_{{\rm Gauss}}(V)$
leaving in it only the
bi-diagonal terms with $i=i'$ and $j=j'$:
\bea\label{max}
&&\langle V^{2} \rangle =\int\limits_{g\in II} dg\, N^{-\frac{1}{p\gamma}
\left(g-\frac{\gamma}{2} \right)^{2} -2g}\\ \nonumber
&\sim & {\rm max}_{g\in II}\left\{ N^{-\frac{1}{p\gamma}
\left(g-\frac{\gamma}{2} \right)^{2} -2g}\right\}\equiv
N^{-\gamma_{{\rm eff}}}.
\eea
The maximum in~\eqref{max} at $g$ belonging to region II in Fig.~\ref{Fig:regions}
can be reached
(i) inside the region II at $g=g^{*}_{1}$, (ii) at the border of this region at
$g=g^{*}_{2}$,
and (iii) at the
cut-off of $P(g)$ at $g^{*}=0$ (see Fig.~\ref{Fig:regions} and
Fig.~\ref{Fig:maximize_Gauss_LN}(left)).

The expression for $\gamma_{{\rm eff}}(D_{1)}$ takes the form:
\be\label{gamma_eff_App}
 \gamma_{{\rm eff}}(D_{1})=\left\{\begin{array}{ll}
 \gamma(1-p),& \frac{p\gamma}{2} <D_{1}<1,\; p<\frac{1}{2}\cr
 2D_{1}+\gamma-2\sqrt{2D_{1}\gamma p},& D_{1}<\min\lp\frac{p\gamma}{2},\frac{\gamma}{8p}\rp\cr
 \frac{\gamma}{4p},&\frac{\gamma}{8p}<D_{1}<1,\;
p\geq\frac{1}{2}\end{array}\right..
\ee

Next we compute the function
\be\label{max-LN}
\Delta(D_{1})=-2{\rm max}_{g\in I}\left\{ \sigma(g,D_{1})-g-D_{1}\right\}.
\ee
in~\eqref{stab-MF-LN}.

The details of the calculation which is similar to calculation of
$\gamma_{{\rm eff}}(D_{1})$ in~\eqref{max} are illustrated in
Fig.~\ref{Fig:maximize_Gauss_LN}(right). The resulting expression for $\Delta(D_{i})$
is:
\be\label{Delta_App}
 \Delta(D_{1})=\left\{\begin{array}{ll}
 \frac{\gamma}{2}\left(1-\frac{p}{2}\right)-D_{1},&0<D_{1}<\frac{\gamma p}{8},\;p<1\cr
 D_{1}+\frac{\gamma}{2}-\sqrt{2D_{1}\gamma p},&\frac{\gamma p}{8}<D_{1}<\frac{\gamma}{8p},\; p<1\cr
 \frac{\gamma}{4p}-D_{1},& 0<D_{1}<\frac{\gamma}{8p},\;
p\geq 1\end{array}\right..
\ee

In the end of this section we consider the question of the distribution of $g_{ij} = -\ln |H_{ij} - H'_{ij}|/\ln N$
with log-normal distributed $H = N^{-g_1}$ and $H' = N^{-g_2}$.
 Applying the usual logarithmic approximation $\ln|H-H'|\approx\ln {\rm max}\{|H|,|H'| \}$,
we approximate
\be
g_{ij} = \min(g_1, g_2)
\ee
and, thus, the distribution $P(g)$ is given by
\be
P(g) = P(g_1 = g) \int\limits_{g}^\infty P(g_2) d g_2 \simeq
\left\{
\begin{array}{lr}
P(g_1 = g) & g<\gamma/2 \\
P^2(g_1 = g) & g>\gamma/2 \\
\end{array}
\right.
\ee
As one can see from Fig.~\ref{Fig:maximize_Gauss_LN} the latter region
$g>\gamma/2$ is actual only for the upper branch $g_{2'}^*$ of $\Delta(D_1) + D_1$ for
$D_1>\gamma/(8 p)$ which never contributes to the phase diagram.

\section{Kullback-Leibler measures in the multifractal phase}\label{App_sec:theory_KL}

In this section we give a more detailed quantitative description of ${\rm KL1}$ and
${\rm KL2}$ measures.

We begin by considering the simpler correlation function, ${\rm KL2}$.
For that we employ the ansatz for the wavefunction moments:
\be\label{moment-ansatz}
M_{q}=\la \sum_{i}|\psi_{\mu}(i)|^{2q}\ra= N^{-D_{q}(q-1)}\,f_{q}
 (L/\xi_{q}),
\ee
where $D_{q}$ is the fractal dimension in the corresponding phase and $f_{q}(x)$ is the crossover scaling function:
\be\label{asympt}
f_{q}(L/\xi_{q}\rightarrow \infty)\rightarrow
\left\{\begin{array}{ll}{\rm const}. &\;\;{\rm multifractal\;\; phase}\cr
{\rm const }.\, N^{(q-1)(D_{q}-1)},& \;\; {\rm ergodic\;\; phase}\cr
{\rm const }.\, N^{(q-1)D_{q}} &\;\;{\rm localized\;\; phase}
\end{array}\right.
\ee
that tends to a constant as $L\rightarrow\infty$.

Note that graphs with the local tree structure
and for LN-RP matrices the length scale $L\propto \ln N$,
so that
the scaling function is in general a function of {\it two arguments} $\ln N/\xi_{q}$
and $N/e^{\xi_{q}}$ representing the {\it length}- and {\it volume} scaling~\cite{Lemarie2017,Lemarie2020_2loc_lengths}. On
the finite-dimensional lattices $N\propto L^{d}$, and the volume scaling can be represented
as the length scaling in the modified scaling function. In this case a single argument
$L/\xi_{q}$ is sufficient.

When $L\propto \ln N$ the volume scaling is the
leading one for $L\gg \xi_{q}$, and
it is this scaling that provides
the asymptotic behavior~\eqref{asympt}.
The length scaling is
important in the crossover region $L\lesssim \xi_{q}$. Below for brevity we will use the short-hand notation
$L/\xi_{q}$ in all the cases.

There are two trivial cases:
$M_{0}=N$ and $M_{1}=1$ (which follows from the normalization of wave function).
As a consequence we have $D_{0}=1$ and
\be
f_{0}(x)=f_{1}(x)\equiv 1.
\ee

Next using the statistical independence of $\psi$ and $\tilde{\psi}$ in~\eqref{KL2} and normalization
of wave functions we represent
\be\label{expr-KL2}
{\rm KL2} =\la \sum_{i}|\psi(i)|^{2}\,\ln|\psi(i)|^{2} \ra
-N^{-1}\la \sum_{i} \ln|\psi(i)|^{2}\ra.
\ee
Now we express both terms in~\eqref{expr-KL2} in terms of $M_{q}$ using the identity:
\be\label{log-expr}
\ln |\psi_{\alpha}(i)|^{2} = \lim_{\epsilon\rightarrow 0}
\epsilon^{-1}\,(|\psi_{\alpha}(i)|^{\epsilon}-1)
\ee
The first term
is equal to:
\be
\la\sum_{i}\lim_{\epsilon\rightarrow\infty}\frac{|\psi(i)|^{2(1+\epsilon)}-
|\psi(i)|^{2}}
{\epsilon}\ra=\lim_{\epsilon\rightarrow\infty}\left[\frac{1}{\epsilon}\left(
M_{1+\epsilon}-1\right)\right].
\ee
The second term can be expressed as:
\be
-\frac{1}{N}\la\sum_{i}\lim_{\epsilon\rightarrow\infty}\frac{
|\psi(i)|^{2\epsilon}-1}{\epsilon}\ra=-\lim_{\epsilon\rightarrow 0}\left[\frac{1}
{\epsilon}\left( N^{-1}M_{\epsilon}-1\right)\right].
\ee
Now expanding $M_{1+\epsilon}$ and $M_{\epsilon}$ in the vicinity of $q=0,1$ and defining
\bea
f_{1+\epsilon}(x)&=&1+\epsilon\,\varphi_{1}(x)+O(\epsilon^2);\\
 f_{\epsilon}(x)&=&1-\epsilon\,\varphi_{0}(x)+O(\epsilon^2),
\eea
we obtain:
\be
{\rm KL2}={\rm KL2}_{c}(N) +\varphi_{0}(L/\xi_{0})+\varphi_{1}(L/\xi_{1}),
\ee
where ${\rm KL2}_{c}$ is logarithmically divergent, as in~\eqref{KL-MF}:
\bea\label{KL2_crit}
{\rm KL2}_{c}&=&\ln N \,(1-\partial_{\epsilon} D_{\epsilon}|_{\epsilon=0} -
D_{1})+{\rm const}.\nonumber \\ &=&\ln N\,(\alpha_{0}-D_{1})+{\rm const}.
\eea
Here we used the identity
\bea
\alpha_{0}=\frac{d\tau_{\epsilon}}{d\epsilon}\left|_{\epsilon=0}\right.
=\partial_{\epsilon}[D_{\epsilon}(\epsilon-1)]\left|_{\epsilon=0}\right..
\eea
for $\alpha_{0}$ describing the typical value of the
wave function amplitude:
\be
|\psi(i)|_{typ}^{2}=N^{-\alpha_{0}}
\ee

Note that, generally speaking, the characteristic lengths
$\xi_{0}\sim |\gamma-\gamma_{c}|^{-\nu^{(0)}}$ and $\xi_{1}
\sim|\gamma-\gamma_{c}|^{-\nu^{(1)}}$ in $\phi_{0}$ and $\phi_{1}$ may have different critical exponents
$\nu^{(0)}$ and $\nu^{(1)}$. If this is the case, the smallest one will dominate
the finite-size corrections near the critical point:
\be\label{KL2_fin}
{\rm KL2} - {\rm KL2}_c (N) =
\Phi_{2}(L |\gamma-\gamma_c|^{\nu_{2}}),\;\;\;
 \nu_{2}={\rm min}\{\nu^{(0)},\nu^{(1)} \}.
\ee

~\eqref{KL2_fin} is employed in this paper for the numerical characterization of
the phases by finite size scaling (FSS). One can see from~\eqref{KL2_crit} that
 ${\rm KL2}$ is
logarithmically divergent in the multifractal phase, as
$\alpha_{0}>1$ and $D_{1}<1$ and the scaling functions $\varphi_{0}(x)$
and $\varphi_{1}(x)$ tend to a finite $N$-independent limit. It is also
logarithmically divergent in the localized phase, as in~\eqref{KL-L}, where one can formally set
$D_{1}=0$ in~\eqref{KL2_crit}:
\be
{\rm KL2}_{c}=\alpha_{0}\,\ln N.
\ee

 However, in the ergodic phase the logarithmic divergence of ${\rm KL2}$ is gone,
since
 in
this case $\alpha_{0}=D_{1}=1$ in~\eqref{KL2_crit}.
One can easily show using the Porter-Thomas distribution:
\be
P_{PT}(x=N|\psi(i)|^{2})= \frac{e^{-x/2}}{\sqrt{2\pi \,x}}
\ee
 that ${\rm KL2}=2$ in the fully-ergodic phase.

At the {\it continuous}
ergodic transition, where the correlation length $\xi=\infty$ and
$\alpha_{0}=D_{1}=1$,
the critical value ${\rm KL2}_c(N)$ of ${\rm KL2}$ is independent of $N$. This results in
{\it crossing} at $\gamma=\gamma_{ET}$ of all the curves for ${\rm KL2}$ at different values of
$N$ which helps to identify the {\it ergodic} transition~\cite{KLPino}.

\begin{figure*}[t]
\includegraphics[width=0.32\linewidth]{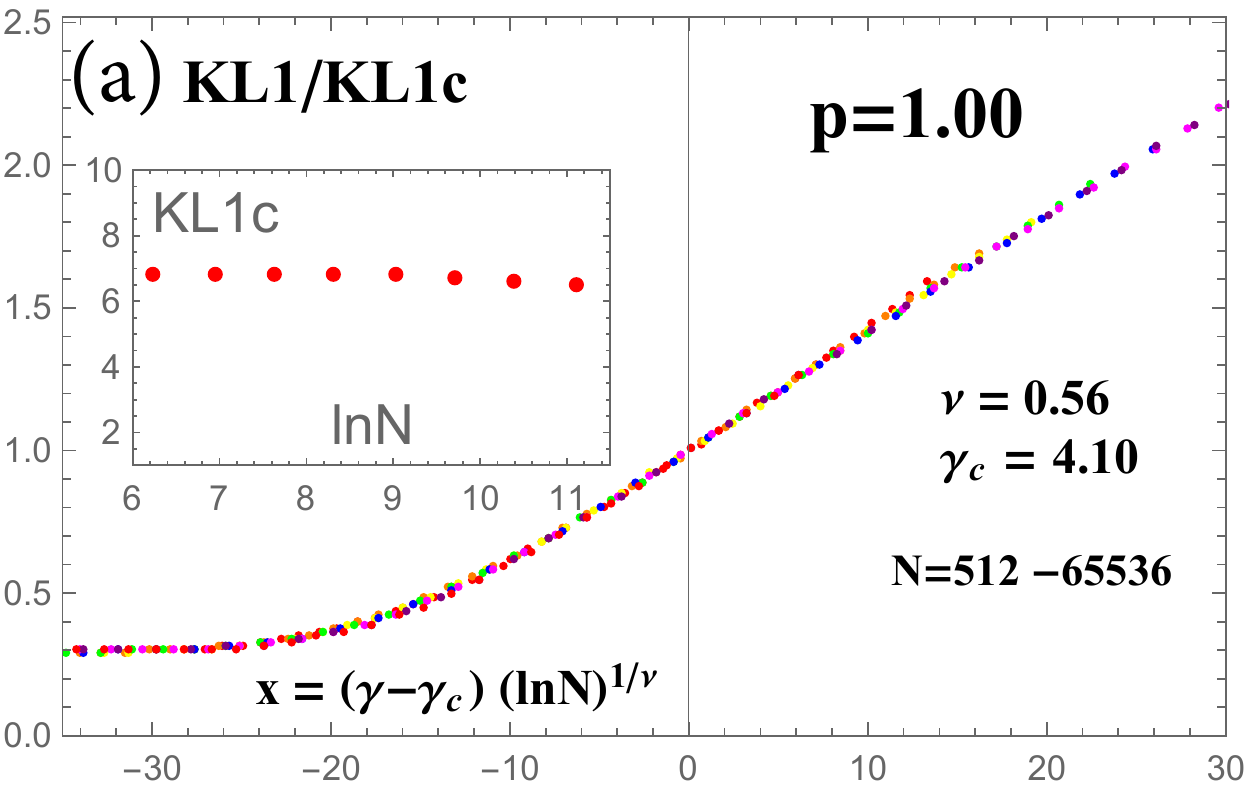}
\includegraphics[width=0.32\linewidth]{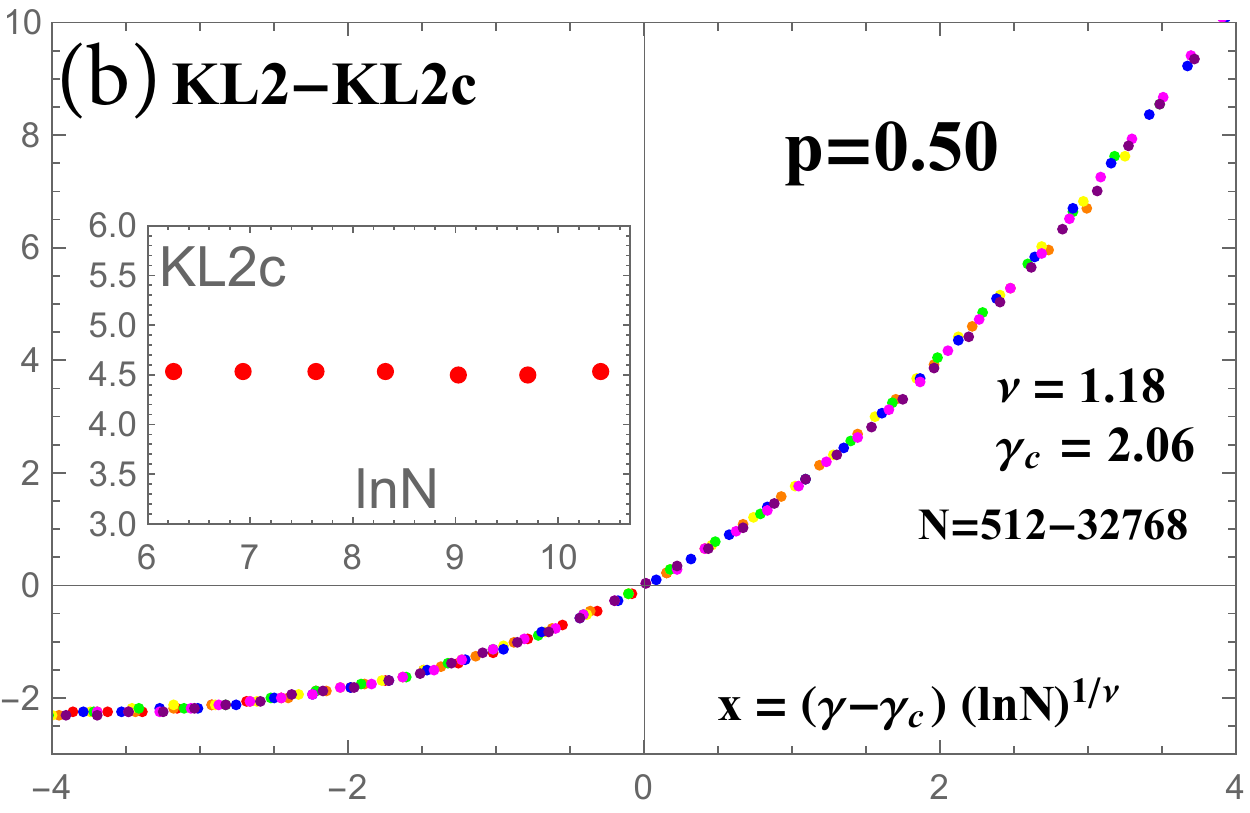}
\includegraphics[width=0.32\linewidth]{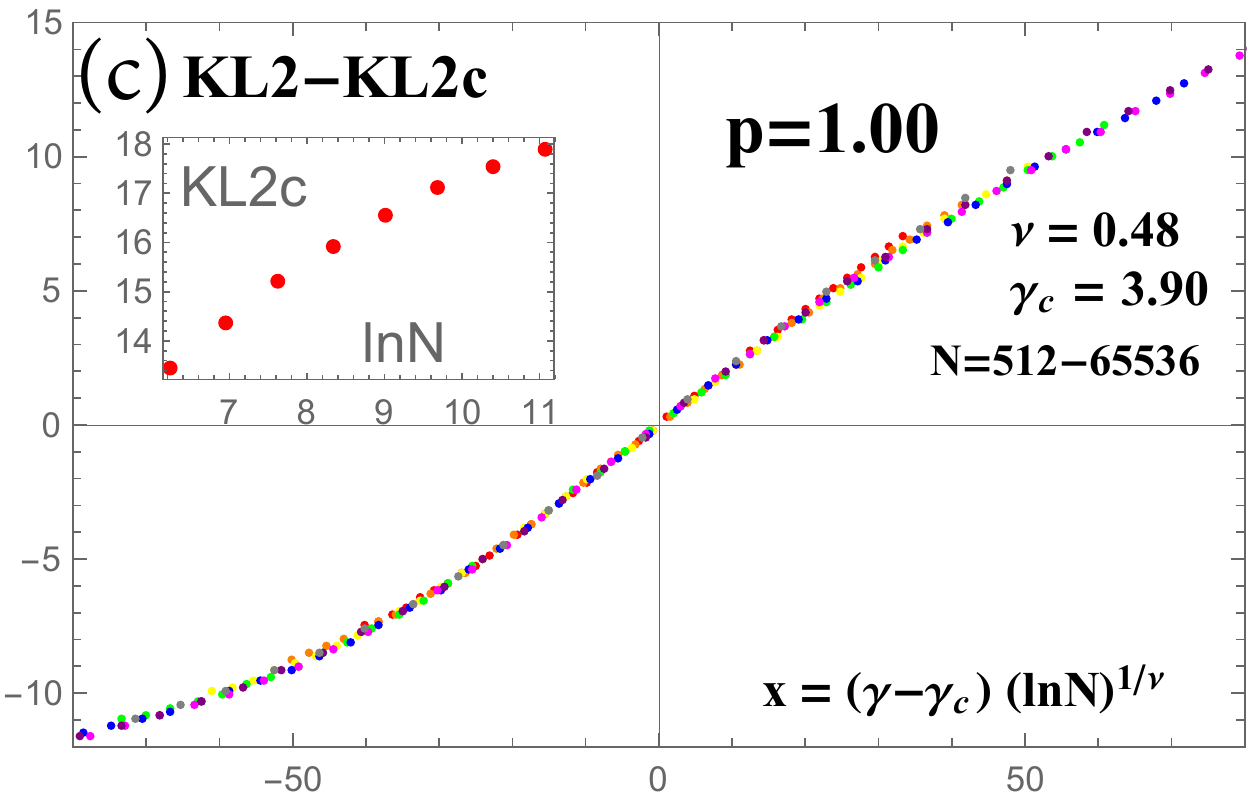}
\caption{(Color online) \textbf{The best collapse of the $KL1$ and $KL2$ data for LN-RP} with
$p=1$ and $p=0.5$. The collapse for $KL1$ and $KL2$ is done in the vicinity of
the localization (for $KL1$) and ergodic (for $KL2$)
transitions by recursive procedure that finds $\gamma_{c}$ and $\nu$ by minimizing
the mean square deviation of data from a smooth scaling function which is updated at any
step of the procedure.
(insets) The critical value of $KL1$ and $KL2$ as a function of $\ln N$.
It stays almost a constant for $KL1$ and for $KL2$ at $p=0.5$ when the ergodic transition is continuous
and well separated from the localized one
but it grows logarithmically in $N$ at $p=1$ when the ergodic and localization
transitions merge together.
This growth is the reason of smearing of the intersection of
$KL2$ curves in Fig.~\ref{Fig:intersect}.
The exponent $\nu$ significantly depends on $p$ and is
consistent with $\nu_{1}\approx \nu_{2}=0.5$ at $p=1$ and $\nu_{2}=1$ at $p=0.5$.
\label{Fig:collapse_LN-RP} }
\end{figure*}

However, if the ergodic transition coincides with the Anderson localization transition
and is \rev{characterized by a jump in fractal dimension}, (i.e. $\alpha_{0}$ and $D_{1}$
are not equal to 1 at the transition), the critical value ${\rm KL2}_c (N)$
is no longer $N$-independent. In this case the crossing is smeared out and can
disappear whatsoever. Nonetheless, by subtracting ${\rm KL2}_c$
from ${\rm KL2}$ one can still
locate
the transition point from the best collapse of ${\rm KL2}$
vs. $\gamma$ curves by choosing an optimal
$\gamma_{c}$ and $\nu_{2}$
in~\eqref{KL2_fin}. However, it is safer to use ${\rm KL1}$ in this case.

The derivation of finite size scaling (FSS) for ${\rm KL1}$ proceeds in the same way by plugging the identity
~\eqref{log-expr}
into:
\be \label{expr-KL1}
{\rm KL}1 =\la \sum_{i}|\psi_{\alpha}(i)|^{2}\,\ln|\psi_{\alpha}(i)|^{2}
- \sum_{i} |\psi_{\alpha}|^{2}\,\ln|\psi_{\alpha+1}(i)|^{2}\ra.
\ee
and employing the ansatz:
\bea
&&\la\sum_{i}|\psi_{E}(i)|^{2q_{1}}\,|\psi_{E+\omega}(i)|^{2q_{2}}
\ra \sim N^{1+\beta}\,N_{\omega}^{\alpha}\nonumber \\ &&\times \,F_{q_{1},q_{2}}
(L/\xi_{q_{1}},L/\xi_{q_{2}}),
\eea
where $N_{\omega}=1/(\rho\omega)$ and $\rho$ is the mean DoS.

Applying for large $\omega\sim \rho^{-1}$ ($N_\omega\simeq 1$) the ``decoupling rule'':
\bea
&&\la\sum_{i}|\psi_{E}(i)|^{2q_{1}}\,|\psi_{E+\omega}(i)|^{2q_{2}}
\ra \sim \nonumber \\ &&\sum_{i}\la |\psi_{E}(i)|^{2q_{1}}\ra\,
\la|\psi_{E+\omega}(i)|^{2q_{2}}
\ra,
\eea
and for small $\omega\sim \delta$ ($N_\omega \simeq N$) the ``fusion rule'':
\be
\la\sum_{i}|\psi_{E}(i)|^{2q_{1}}\,|\psi_{E+\omega}(i)|^{2q_{2}}
\ra \sim \\\la\sum_{i} |\psi_{E}(i)|^{2q_{1}+2q_{2}}\ra,
\ee
one easily finds:
\bea
\beta &=& -2 +D_{q_{1}}(1-q_{1})+D_{q_{2}}(1-q_{2}),\\ \nonumber
\alpha+\beta &=& -1+D_{q_{1}+q_{2}} (1-q_{1}-q_{2}).
\eea
Due to the ``fusion rule'' for $\psi_{\alpha}$ and $\psi_{\alpha+1}$ we obtain from
~\eqref{moment-ansatz}:
\bea\label{fuss}
 &&\la\sum_{i}|\psi_{\alpha}(i)|^{2q_{1}}\,|\psi_{\alpha+1}(i)|^{2q_{2}}
\ra \sim F_{q_{1},q_{2}}
(L/\xi_{q_{1}+q_{2}} )\nonumber \\ &&\times
 N^{-D_{q_{1}+q_{2}}(q_{1}+q_{2}-1)}.
\eea

Substituting~\eqref{fuss} in~\eqref{log-expr},~\eqref{expr-KL1}
we observe cancelation of the leading logarithmic in $N$ terms in ${\rm KL1}$ in
the multifractal phase:
\be
{\rm KL1}_c ={\rm const}.
\ee
We obtain:
\be\label{KL1_fin}
{\rm KL1} = \Phi_{1}(L|\gamma-\gamma_c|^{\nu_{1}}).
\ee
where $\nu_{1} = \nu^{(1)} \geq \nu_{2}$ and the crossover scaling function $\Phi_{1}(x)$ is:
\be
\Phi_{1}(x)=\partial_{\epsilon}f_{1+\epsilon}(x)-\partial_{\epsilon}f_{1,\epsilon}(x)|
_{\epsilon=0}.
\ee
As it is seen from~\eqref{KL1_fin}, ${\rm KL1}$ is independent of $N$ at the Anderson
transition point
$\gamma=\gamma_{AT}$. Thus all curves for ${\rm KL1}$ at different values of $N$
intersect at
$\gamma=\gamma_{AT}$. This gives us a powerful instrument to identify the Anderson
localization transition point.

Note that the coefficient in front of $\ln N $ in
${\rm KL2}$ may help to detect discontinuity of the Anderson transition. Indeed, one can use the
Mirlin-Fyodorov symmetry of fractal dimensions to establish the relation, see~\eqref{KL-MF}:
\be
\alpha_{0}=2-D_{1},\;\;\;\Rightarrow \alpha_{0}-D_{1}=2(1-D_{1}).
\ee
This tells us immediately that for {\it continuous} Anderson transition
which is characterized by vanishing
$D_{1}$ both just below and just above the transition, the coefficient
in front of $\ln N$ in ${\rm KL2}$ is equal to $2$.
In particular, we conclude that $\alpha_{0}$ on the localized side of the
transition is equal to $2$. It appears that in LN-RP this value
\be
\alpha_{0}=2,\;\;\;\;(\gamma=\gamma_{AT}+0).
\ee
 in the localized phase just above the transition remains equal to $2$ \rev{in all the cases}. This is
in contrast to the corresponding coefficient $2(1-D_{1})$ in front of $\ln N$ in ${\rm KL2}$
just below the transition
which is smaller than $2$ \rev{if there is a jump in the fractal dimension at the transition}. Such a jump in the coefficient
in front of $\ln N$
in ${\rm KL2}$ is a signature of the discontinuity of the transition which
is the most easily detectable numerically, see Fig.~\ref{Fig:KL_jump}.
\section{Finite-size scaling collapse for $KL1$ and $KL2$.}
\label{App_sec:KL_collapse}

The next step is to analyze the {\it finite-size scaling} (FSS) by a collapse of the
data for ${\rm KL1}$ and ${\rm KL2}$ at different $N$ in the
vicinity of the localization and ergodic transition, respectively.
To this end we use the form of FSS derived in IS~\ref{App_sec:theory_KL}.
\bes\label{collapse-form}
\begin{align}
{\rm KL1} &=
\Phi_{1}(\ln N |\gamma-\gamma_{AT}|^{\nu_{1}}),\\
{\rm KL2} - {\rm KL2_c (N)} &=
\Phi_{2}(\ln N |\gamma-\gamma_{ET}|^{\nu_{2}}).
\end{align}
\ees
The input data for the collapse is ${\rm KL1}$ and ${\rm KL2}$
versus $\gamma$ for $7$
values of $N$ is shown in Fig.~\ref{Fig:intersect}.
The fitting parameters extracted from the best collapse
are $\nu_{1}$ ($\nu_{2}$) and the critical points $\gamma_{AT}$ ($\gamma_{ET}$).
The critical value of ${\rm KL2}_c(N)={\rm KL2}(\gamma_{ET},N)$
is determined by the best fitting for $\gamma_{ET}$.
For the localization transition
where the critical point $\gamma_{AT}$ is well defined by
the intersection in ${\rm KL1}$, one may look for the best collapse by fitting only $\nu_{1}$.

The plots of Fig.~\ref{Fig:collapse_LN-RP} demonstrate the quality of the collapse
for several representative cases. In the insets of the figures we show the $\ln N$-
dependence of the critical values of ${\rm KL1}$, ${\rm KL2}$
 which were obtained numerically from
${\rm KL1}(\gamma=\gamma_{AT},N)$ and ${\rm KL2}(\gamma=\gamma_{ET},N)$, respectively,
with $\gamma_{AT}$ and $\gamma_{ET}$ found from the best collapse. It is demonstrated
that the critical value of ${\rm KL1}$ is almost $N$-independent, as well as the
critical value of ${\rm KL2}$ at $p=1/2$ when the continuous ergodic transition
is well separated from the Anderson localization one. However, at $p=1$ when
ET and AT merge together the
critical value of ${\rm KL2}$ increases linearly with $\ln N$, signaling of the critical
multifractal state at the Anderson transition point, very similar to the case of
3D Anderson transition. This $\ln N$-dependence of ${\rm KL2}_{c}$ is the reason of
smearing out of the intersection point in ${\rm KL2}$ shown in Fig.~\ref{Fig:intersect}.
\section{Ratio of typical and mean LDOS}\label{sec:LDOS_ratio}
\begin{figure*}[t]
\centering
\includegraphics[width=0.24\linewidth]{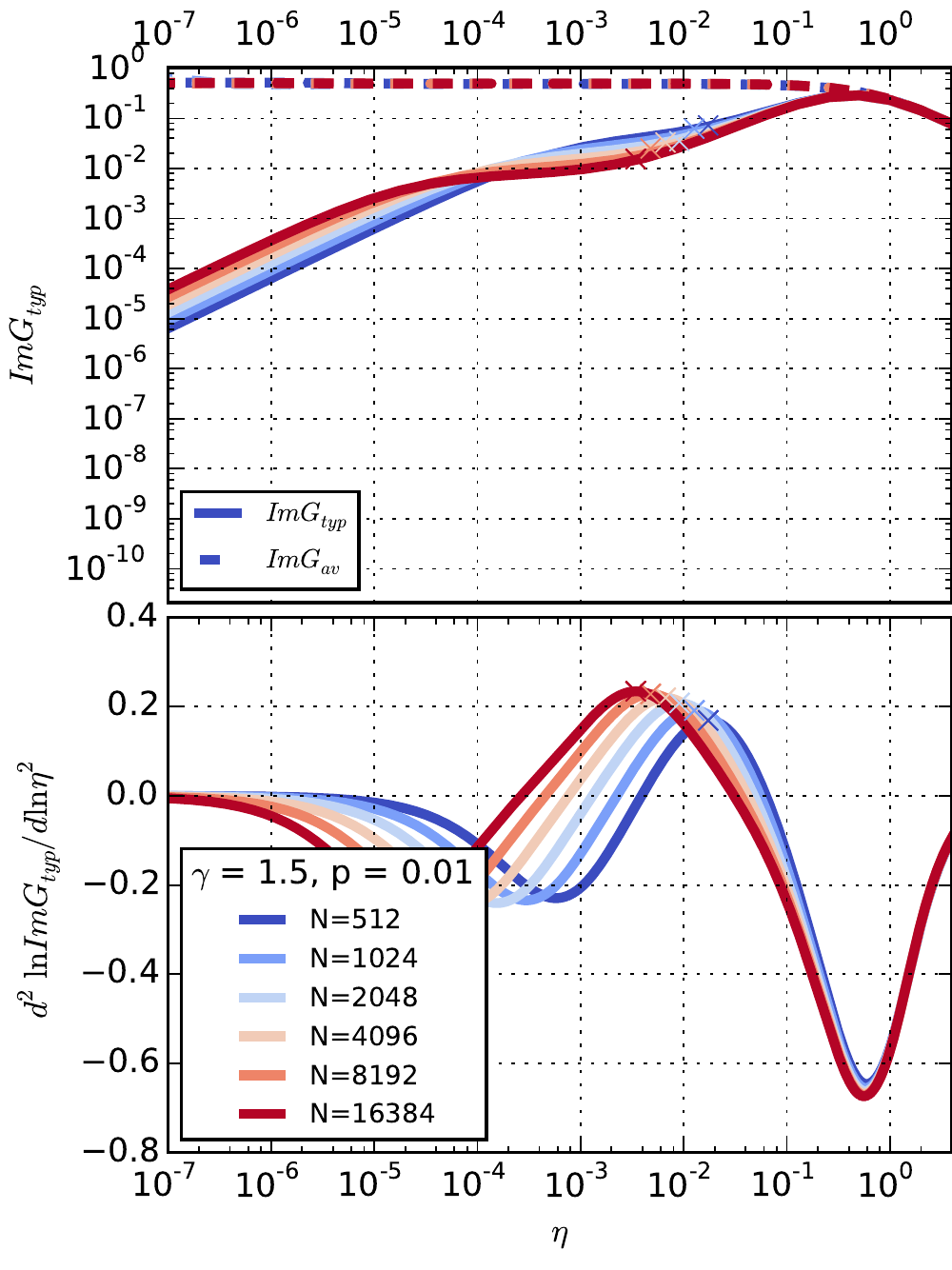}
\includegraphics[width=0.24\linewidth]{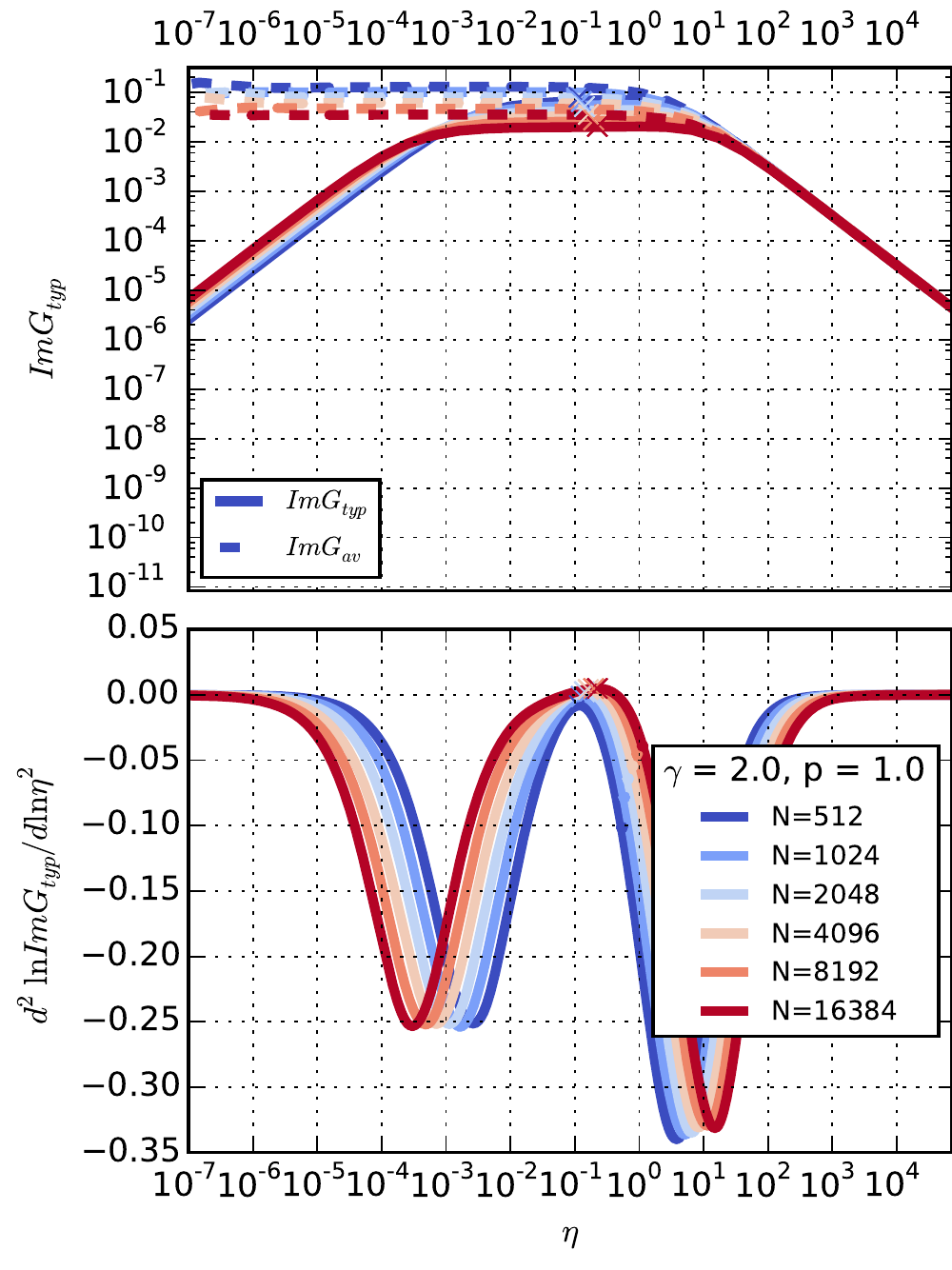}
\includegraphics[width=0.24\linewidth]{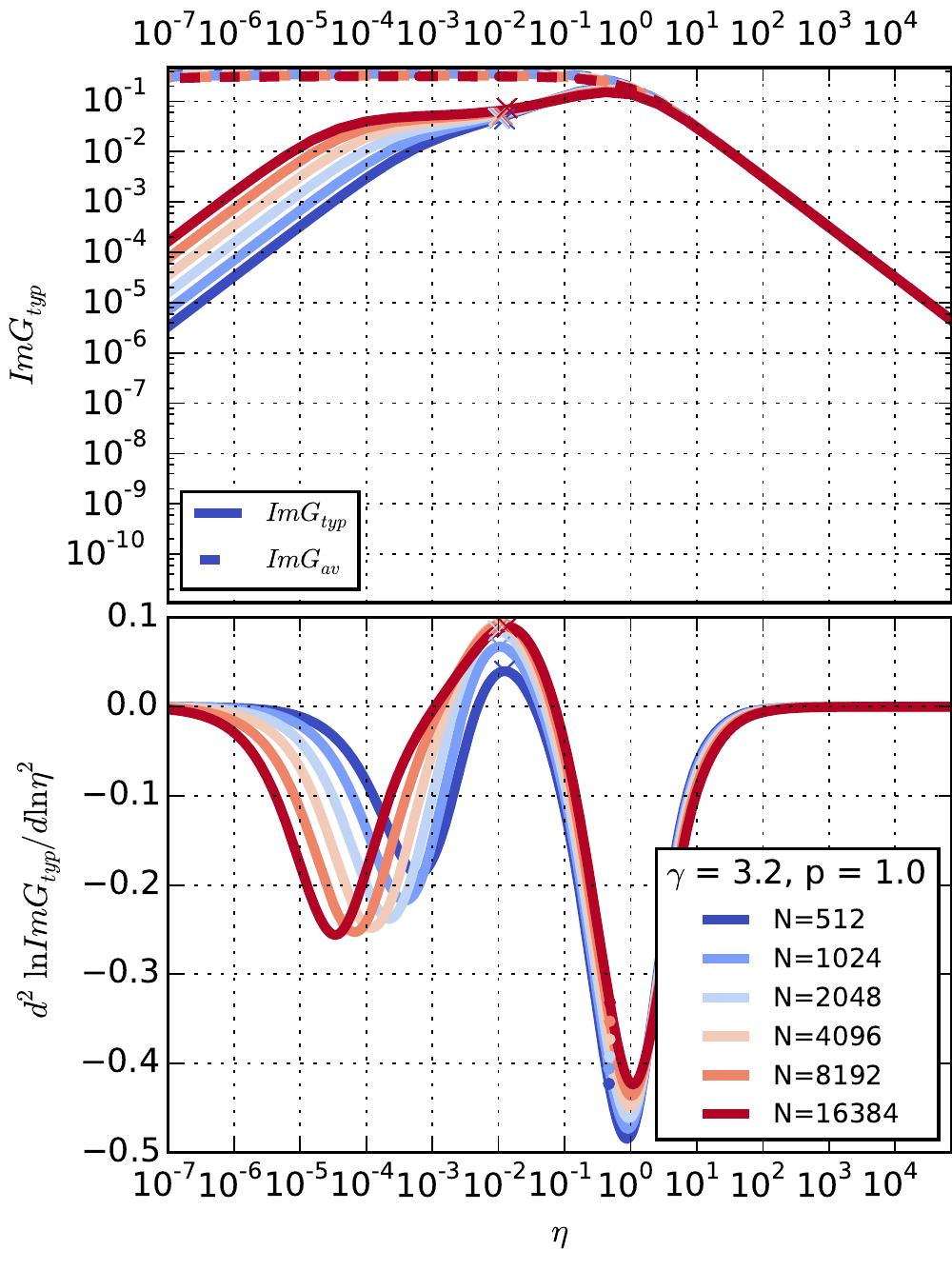}
\includegraphics[width=0.24\linewidth]{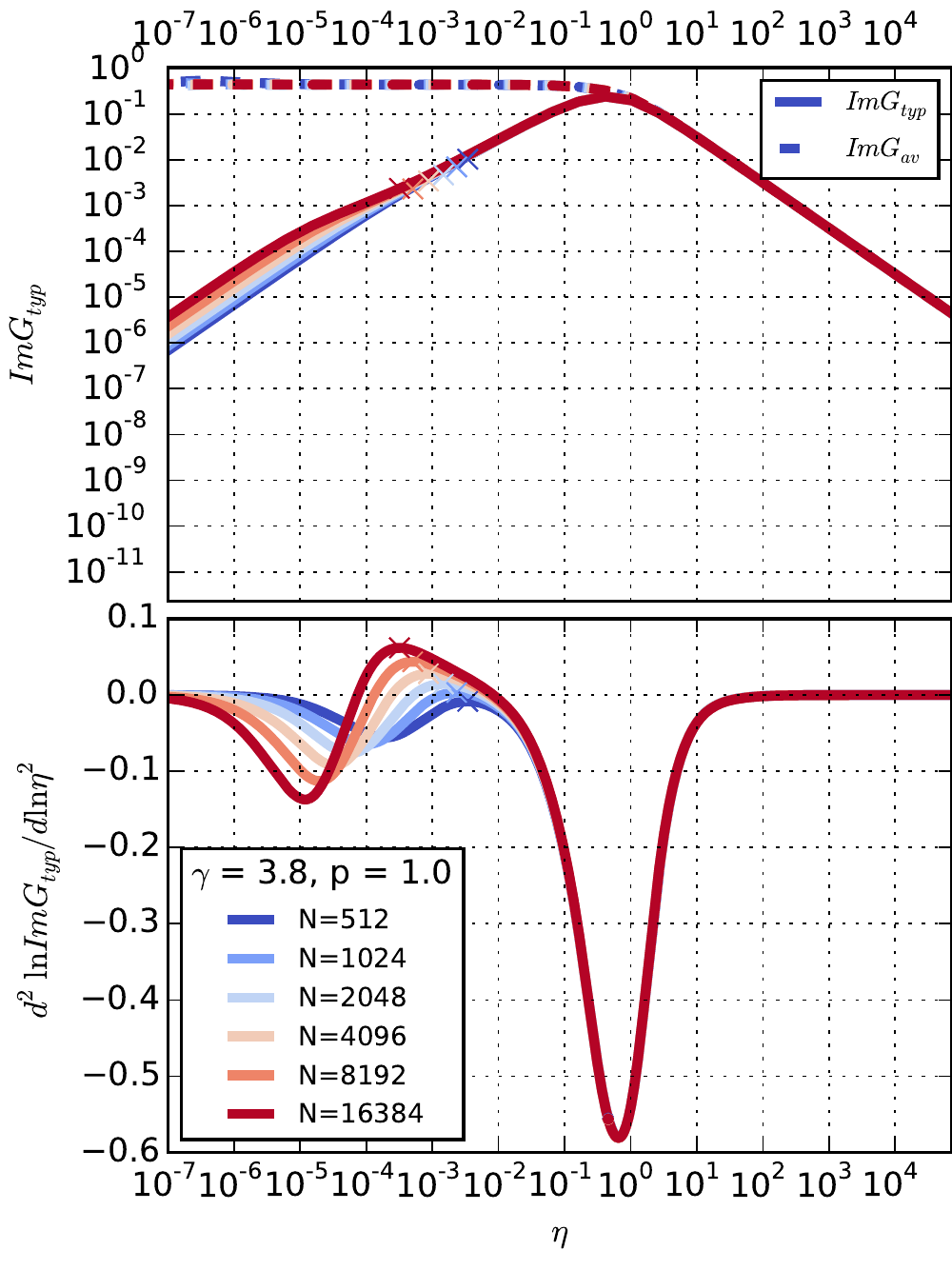}
 \caption{
\textbf{  [upper panels] Plots of the typical $\rho_{{\rm typ}}$ (solid lines) and
mean $\rho_{{\rm av}}$ (dashed lines)
LDOS  and  [lower panels] of the second derivative of
$\rho_{{\rm typ}}/\rho_{{\rm av}}$ w.r.t.
$\eta$}
for the log-normal RP model 
 at
(left panel) $p=0.01$ in the multifractal phase
$\gamma_{ET}<\gamma = 1.5<\gamma_{AT}$, (other panels) $p=1.0$
 in the weakly ergodic  phase $\gamma_{FWE}<\gamma = 2.0$, $3.2$,
$3.8<\gamma_{ET} = \gamma_{AT}$.
The positions of the maxima of the second derivative are shown by
crosses of the corresponding color for all system sizes in the range
from $N=512$ (dark blue) to $N=16384$ (red). Notice a plateau developing in
$\rho_{{\rm typ}}$ for intermediate values of $\eta$ with increasing the system size.
The maximum of the second derivative is always inside the plateau region or on its
right end.  The plateau gradually shrinks with increasing $\gamma$ and
disappears in the localized phase.
\label{Fig:rho_2der} }
 \end{figure*}

In this section we consider in more details the technical issue with the
determination of the order parameter for the FWE transition
\be\label{OP}
\phi(\eta) = 1 - \frac{\rho_{{\rm typ}}}{\rho_{{\rm av}}} \ ,
\ee
being the ratio of the typical, $\rho_{{\rm typ}}$, and the mean, $\rho_{{\rm av}}$,
LDOS given by the expressions
\be\label{supp_eq:rhos}
\ln\rho_{{\rm typ}} = \la \ln \rho(x,E+i\eta)\ra, \quad
\rho_{{\rm av}} = \la \rho(x,E+i\eta)\ra,
\ee
with the LDOS before averaging written as
\be
\rho(x,E+i\eta) = \sum_{\mu} |\psi_\mu(x)|^2\frac{\eta/\pi}{(E-E_\mu)^2+\eta^2} \ .
\ee
The averaging in~\eqref{supp_eq:rhos} is taken over the disorder realizations,
over all coordinates $x$ (which are statistically equivalent in LN-RP) and over
$100$ energy values in the middle half of the spectrum.

As mentioned in the main text the ratio $ \rho_{{\rm typ}}/\rho_{{\rm av}}$ develops the
plateau $\sim N^{-1+D_1}$ in some range of bare level width parameter
$\eta\gg \delta$ large compared to the typical level spacing $\delta$.
However, at any finite sizes this plateau has a finite slope, especially
for the WE phase where $N^{D_1} = f N$ with a $N$-independent constant $f<1$
and, thus,
the plateau is also $N$-independent
\be
\phi(\eta\gg\delta) \propto 1-f = O(1) \
\ee
which is zero in the FE phase, $f=1$, and finite in the WE one, $f<1$.
\begin{figure}[hb!]
\centering
\includegraphics[width=0.48\linewidth]{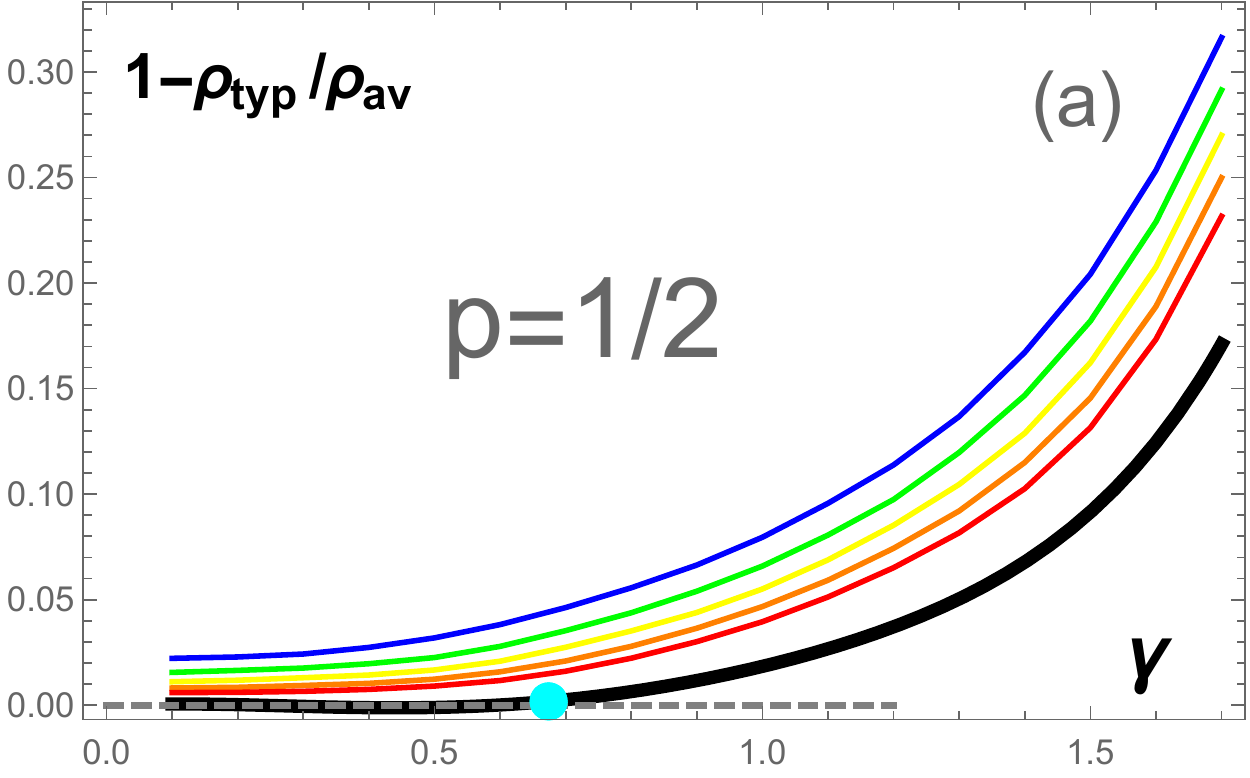}
\includegraphics[width=0.48\linewidth]{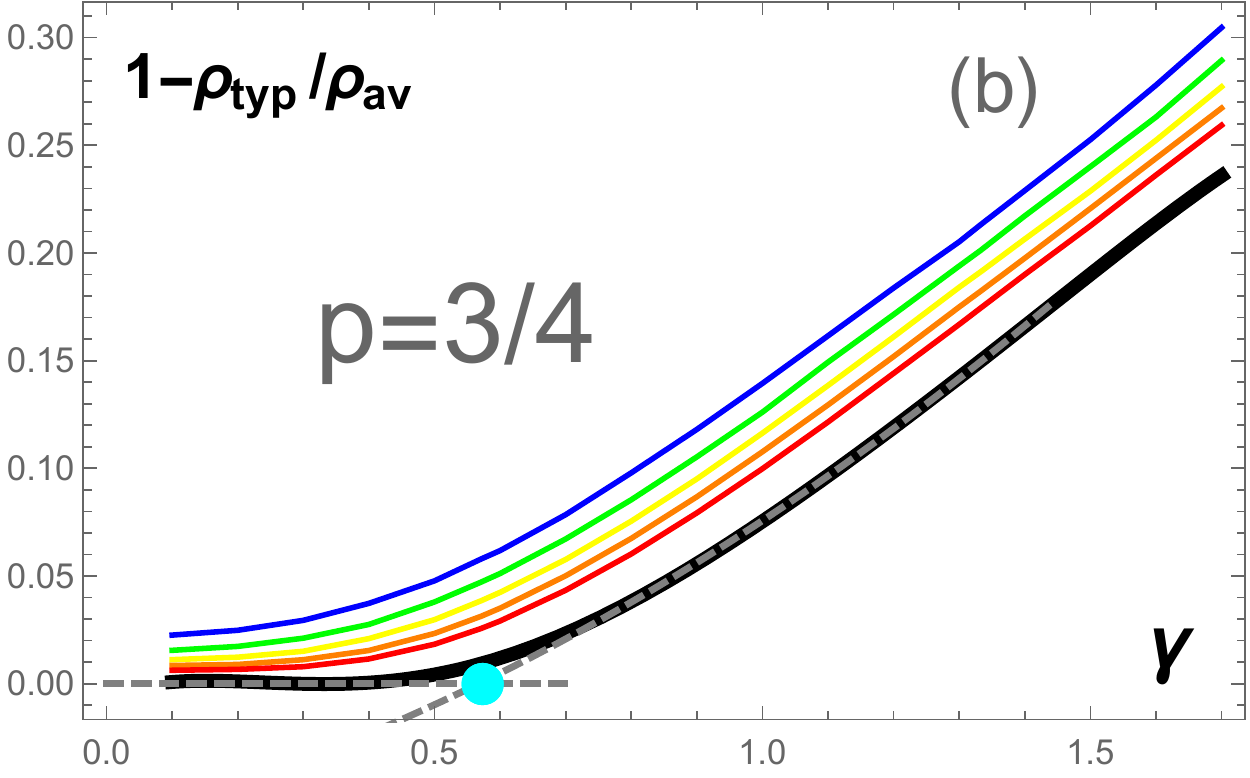}
\\
\includegraphics[width=0.485\linewidth]{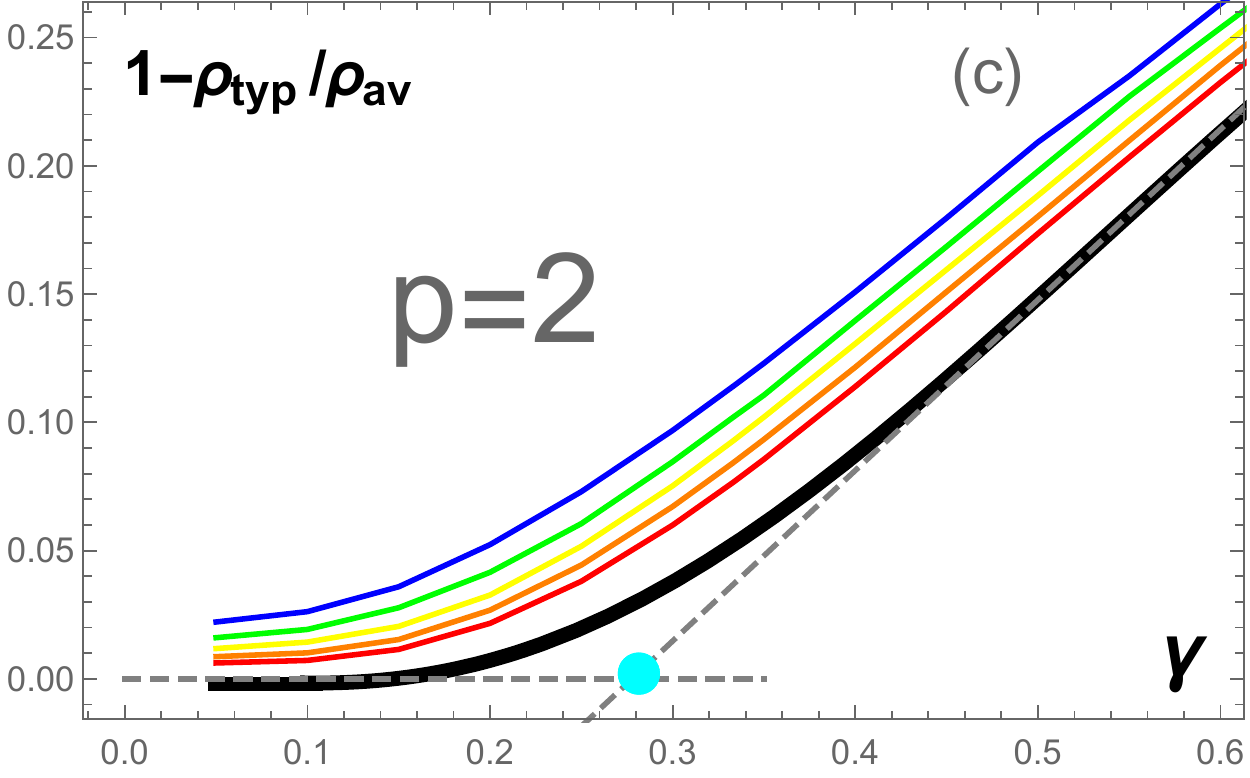}
\includegraphics[width=0.475\linewidth]{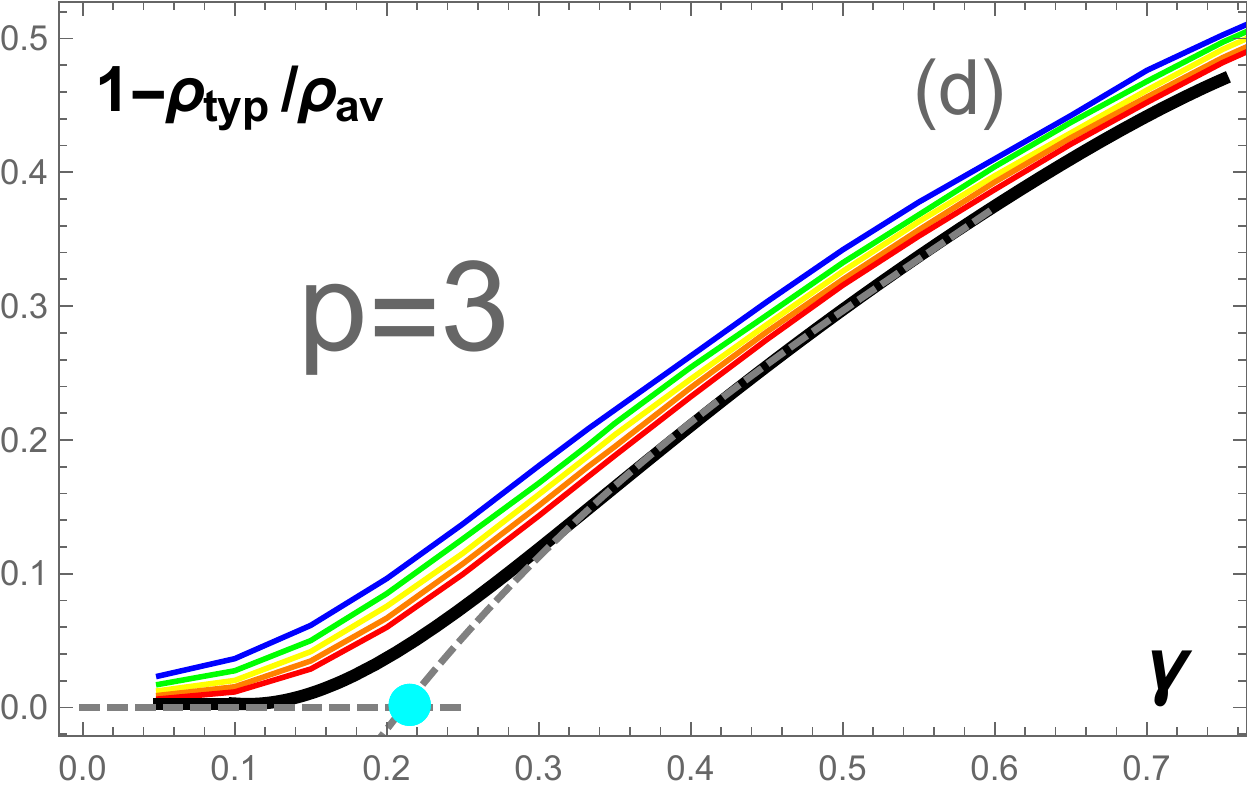}
\caption{
\textbf{Plots for $\phi=1-\rho_{{\rm typ}}/\rho_{{\rm av}}$ as a function of $\gamma$} at
$p=0.5$, $1$, $2$, $3$ for $N=512$, $1024$, $2048$, $4096$, $8192$ (from blue to red) and
extrapolated to $N=\infty$ (black). The gray dashed lines represent cubic polynomial
fits to the  points of extrapolation away from the transition. The
intersection of each of these lines with $\phi=0$ gives the numerical estimate of
$\gamma_{{\rm FWE}}$ (shown by bright blue point) which is compared in Table~\ref{Table:FWE} with the predicted
by~\eqref{FWT} in the main text.
 \label{Fig:FWE_trans} }
\end{figure}

In order to find the FWE transition accurately we develop the procedure
of the automatic selection of $\eta$ in the middle of the underdeveloped plateau.
For this purpose we take the second derivative of the ratio
$\rho_{{\rm typ}}/\rho_{{\rm av}}$ w.r.t. $\eta$ after the smoothening it with
the $5$-degree spline
and find the point of maximum of this derivative lying in between of two local
minima (see the lower panels in Fig.~\ref{Fig:rho_2der}).
Figure~\ref{Fig:rho_2der} shows several examples for $p=0.01$ and $p=1$
where the positions of the maxima of the second derivative are shown by crosses
of the corresponding color for all system sizes $N$.

\section{Location of FWE transition}\label{sec:FWE_trans}
The behavior of the order parameter~\eqref{OP} helps to
locate the FWE transition.
In Fig.~\ref{Fig:FWE_trans} we present the plots for $\phi=1-\rho_{{\rm typ}}/\rho_{{\rm av}}$
for different values of $p$ as a function of $\gamma$
calculated numerically by exact diagonalization for several values of $N$
\rvI{
and then extrapolated to $N=\infty$ as follows.
Similarly to the critical exponents
$\tau_q(N)$~\cite{Mirlin_Mildenberger_irrelevant_exp,luitz2019multifractality} or
the spectrum of fractal dimensions
$f(\alpha,N)$~\cite{DeLuca2014,gRP,Nosov2019correlations,Nosov2019mixtures,deng2020AnisBM}
we show that
  a linear in $1/\ln N$ function
\be\label{App_eq:extrap}
\phi(\gamma,p,N) = \phi(\gamma,p)+ \frac{c(\gamma,p)}{\ln N}
\ee
fits the data points at  fixed $(\gamma,p)$ for the available range of $N$
and take $\phi(\gamma,p)$ as an extrapolated value.
}
\newcolumntype{b}{>{\columncolor{blue!30!white!50}}c}
\begin{table}[h!]
\centering
\resizebox{0.7\columnwidth}{!}{
{
 \begin{tabular}{V{2.5} c V{2.5} b V{2.5} c V{3.0} b  V{2.5} c V{2.5} b V{2.5}}
\hline			
\Xhline{3\arrayrulewidth}
 p
& $0.50$ & $0.75$ & $1.00$ & $2.00$ & $3.00$
 \\
 \hline
\Xhline{3\arrayrulewidth}
\red{\gamma_{FWE}^{theor}}
& \red{0.66} & \red{0.57} & \red{0.5} & \red{0.33} & \red{0.25} \\
 \hline
\Xhline{3\arrayrulewidth}
\blue{\gamma_{FWE}^{num}}
& \blue{0.67\pm0.04} & \blue{0.57\pm0.04} & \blue{0.49\pm0.04} & \blue{0.28\pm0.07} & \blue{0.22\pm0.07}\\
\hline
\Xhline{3\arrayrulewidth}
 \end{tabular}
 }
}
\caption{ Comparison of analytical predictions (red),~\eqref{FWT},
and numerical data (blue) for the FWE transition points.
Numerical data is obtained by
exact diagonalization of LN-RP random matrices with $N=512-8192$ followed
by  extrapolation to $N\rightarrow\infty$ of the order parameter,~\eqref{OP},
given by the ratio of the typical $\rho_{{\rm typ}}$ and the mean $\rho_{{\rm av}}$
 local DOS.
}
\label{Table:FWE}
\end{table}
\begin{figure}[h!]
\center{
\includegraphics[width=0.46\linewidth]{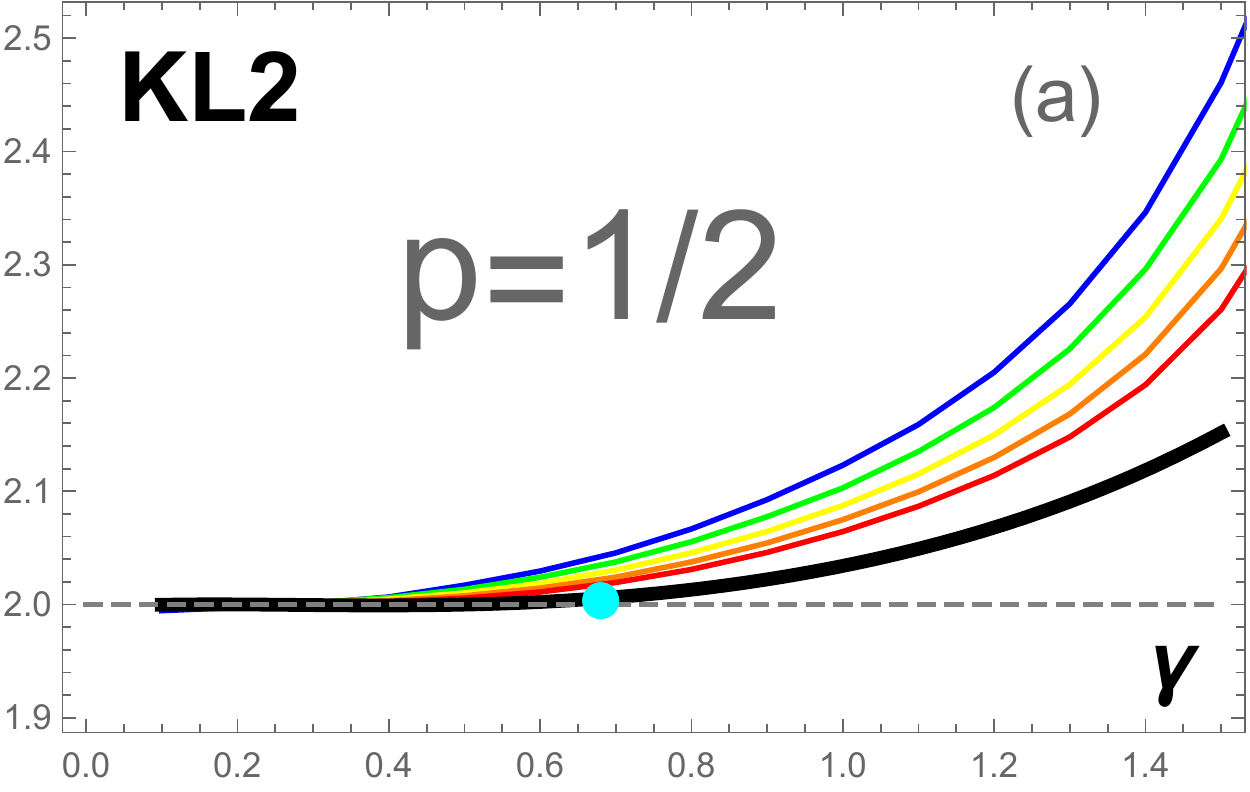}
\includegraphics[width=0.46\linewidth]{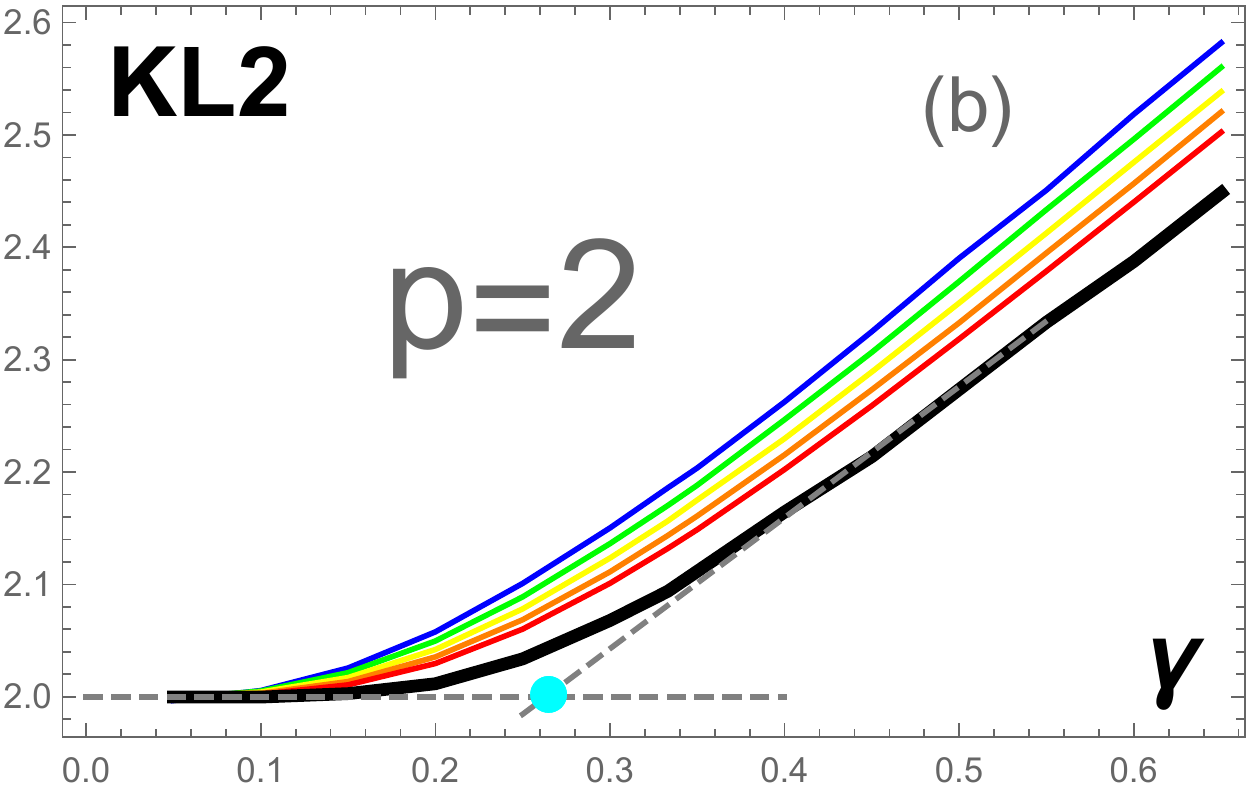}
 }
\caption{(Color online) {\bf Plots for ${\rm KL2}$ as a function of $\gamma$} at
$p=0.5$ (a) and $2$ (b) for $N=512$, $1024$, $2048$, $4096$, $8192$ (from blue to red) and
extrapolated to $N=\infty$ (black). The gray dashed lines represent cubic polynomial
fits to the  points of extrapolation away from the transition. The
intersection of each of these lines with the RMT value ${\rm KL2}=2$ gives the numerical estimate of
$\gamma_{{\rm FWE}}$ (shown by bright blue point).
  \label{Fig:KL_FWE_trans}  }
 \end{figure}

\begin{figure*}[t!]
\center{
\includegraphics[width=0.31\linewidth]{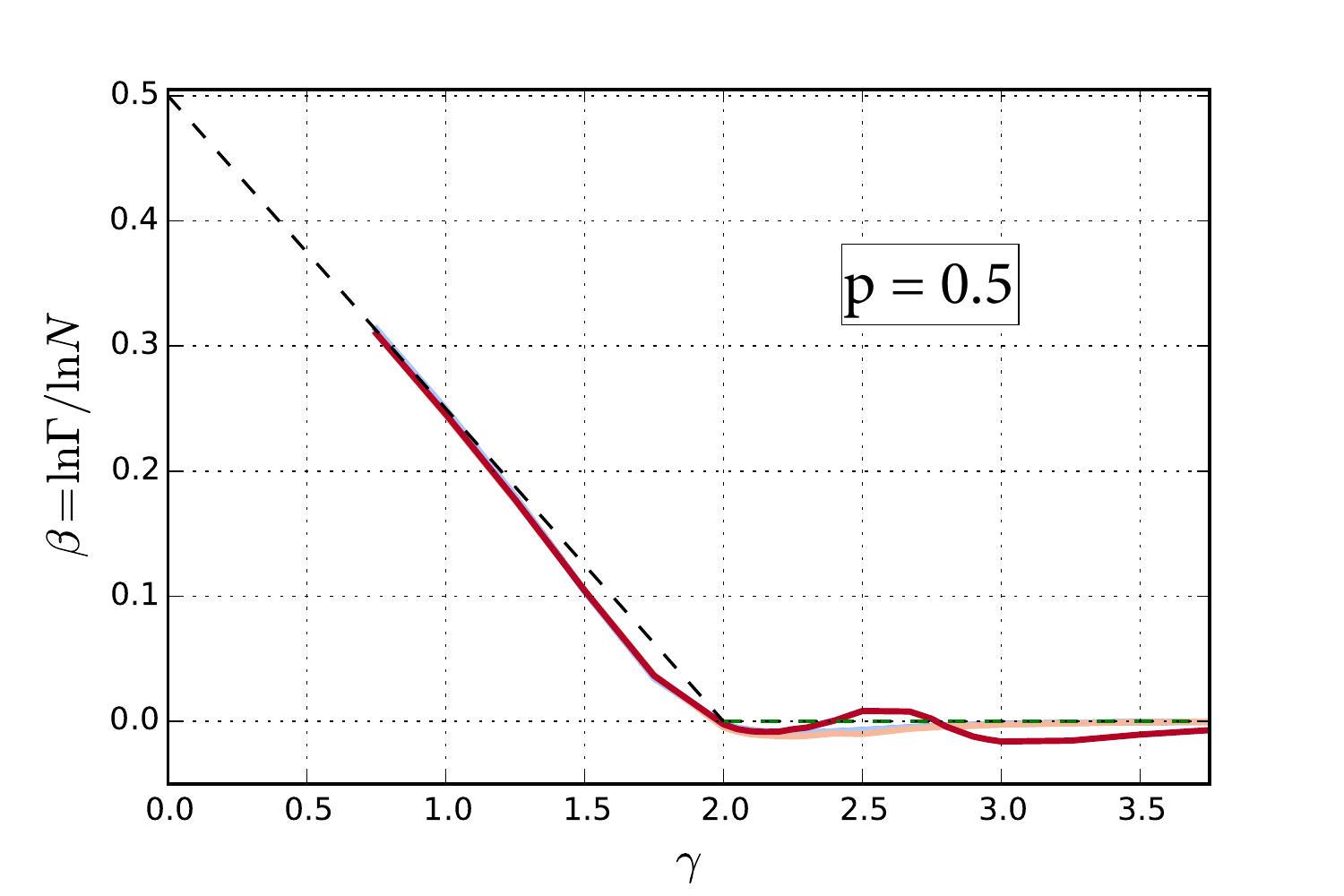}
\includegraphics[width=0.29\linewidth]{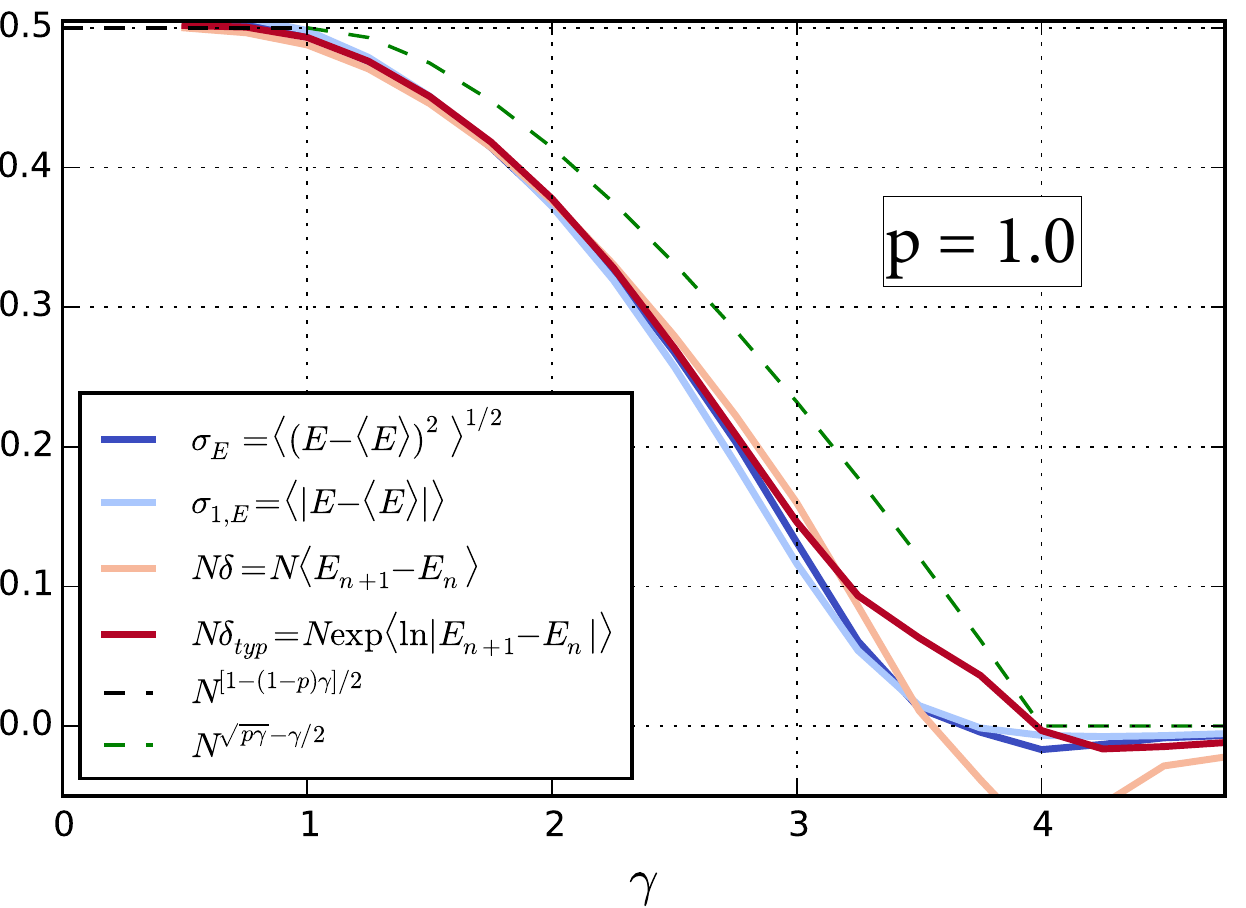}
\includegraphics[width=0.29\linewidth]{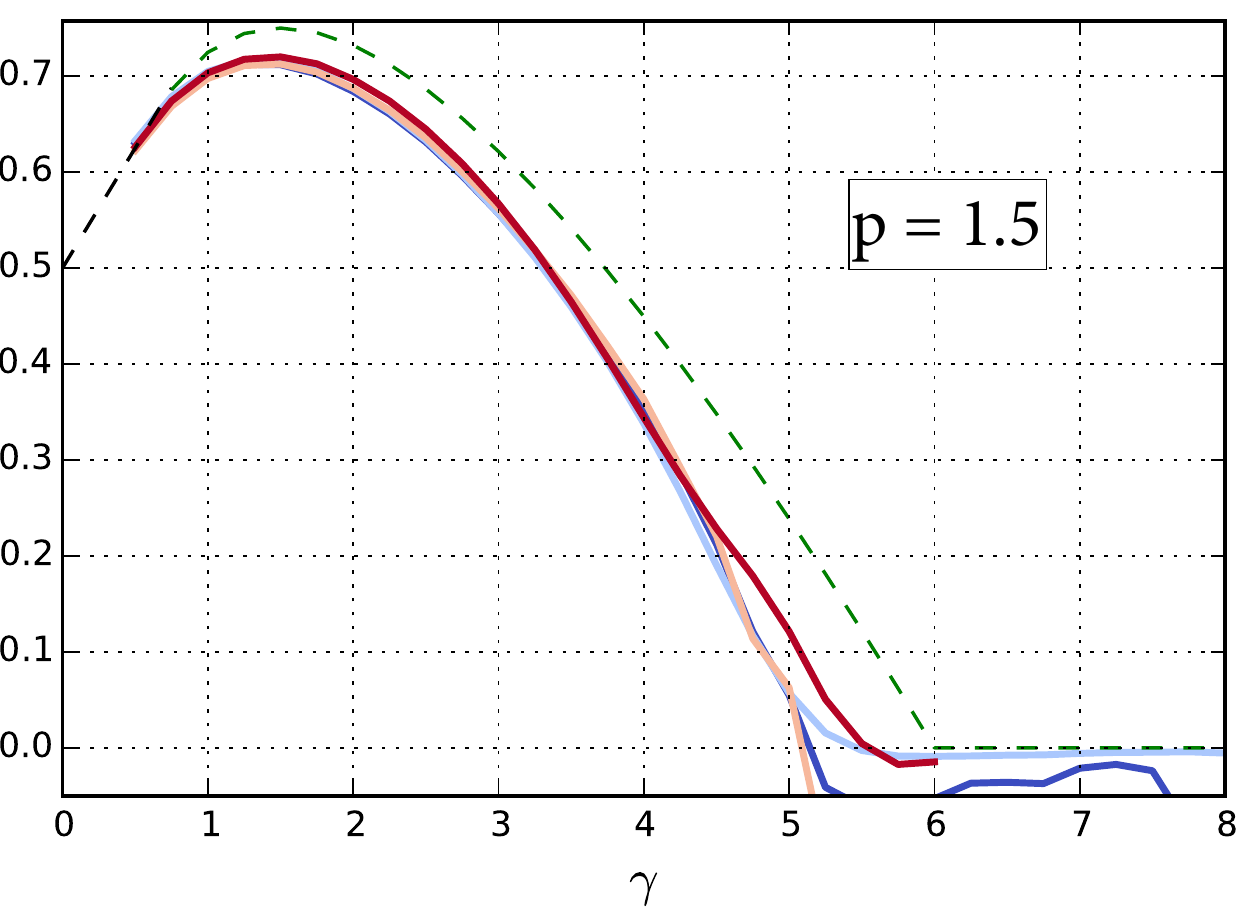}
}
\caption{(Color online) \textbf{Scaling of the spectral bandwidth
$E_{BW}$ with $N$ in different regions of $p$:}
(a) $p=1/2$; (b) $p=1$; (c) $p=1.5$ extracted numerically from the fitting to
$E_{BW} = c N^{\beta}$ of
the eigenvalue standard deviation $\sigma_E = \mean{(E - \mean{E})^2}^{1/2}$,
the  averaged absolute deviation of $E$ from its mean $\sigma_{1,E} = \mean{|E - \mean{E}|}$,
the mean $\delta = \mean{E_{n+1}-E_n}$ and typical
 $\delta_{typ} = \exp\mean{\ln\lrp{E_{n+1}-E_n}}$ global level spacings multiplied by $N$ for the system sizes $N=512-32768$.
All measures are calculated over the $90$~\% of the states in the middle of the spectrum.
The dashed lines represent the analytical prediction, Eq.~\eqref{App_eq:Gamma_res},
in the ergodic phases, $\gamma<\gamma_{ET}$,
while in the non-ergodic ones  $E_{BW}=W\sim N^0$.
\label{Fig:Bandwidth} }
\end{figure*}

While this extrapolation \rvI{(shown by black curves in the figures)} is reliable away from the transition, it is
not able to give the true singularity at the transition which would require an
extrapolation from much larger matrix sizes. Therefore for numerical location of the
transition $\gamma$ we
used the   cubic polynomial fit to the points of extrapolation
sufficiently remote from the transition (represented by gray dashed lines in
Fig.\ref{Fig:FWE_trans}).  Intersection of these lines with the
dashed line $\phi=0$ gives the numerical estimate of $\gamma_{{\rm FWE}}$.
Almost the same values of $\gamma_{{\rm FWE}}$ can be obtained by studying
${\rm KL2}$
statistics \rvI{(with the same extrapolation procedure given by Eq.~\eqref{App_eq:extrap})}. Some of the plots for ${\rm KL2}$ vs $\gamma$ are presented in
Fig.\ref{Fig:KL_FWE_trans}.

The results of this analysis are summarized and
compared with the prediction of Eq.(\ref{FWT}) in Table~\ref{Table:FWE}.
One can see that  \eqref{FWT} is well reproduced by our numerics.

\section{Mean and typical Breit-Wigner width of the mini-band}
According to the definition~\eqref{Mott} and the Fermi Golden rule the Breit-Wigner width is given by the following sum
\be
\Gamma_n \sim \rho_{av}\sum_{m=1 \atop |H_{mn}|<E_{BW}}^N |H_{mn}|^2
\ee
where $\rho_{av}\sim 1/E_{BW}$ is the mean global DOS and the spectral bandwidth $E_{BW} = \max(W,\Gamma)$ is given by the maximum of the bare on-site bandwidth $W$ and the mean Breit-Wigner miniband width
\be\label{App_eq:Gamma_av}
\Gamma = \mean{\Gamma_n} \sim \frac{N \mean{|H_{mn}|^2}_{E_{BW}}}{E_{BW}}
\ee
and should be found self-consistently.

The typical Breit-Wigner width
\be\label{App_eq:Gamma_typ}
\Gamma_{typ} = \exp\mean{\ln \Gamma_n} \sim S_3^{1/2} \sim \frac{N |H_{typ}|^2}{E_{BW}}
\ee
determines the FWE transition~\eqref{FW}.

Let's first calculate $\Gamma$ for different cases.

In the non-ergodic phases $\Gamma\ll E_{BW}\sim W$ and, thus, it is given by~\eqref{mom_q} with $q=2$ as
\be
\Gamma\sim S_2\sim N^{1-\gamma/\gamma_{ET}} \ ,
\ee
with $\gamma_{ET}$ given by~\eqref{ET}.

In the opposite limit of $\Gamma\gg W$ one should calculate the second moment of $H_{mn}$ self-consistently, taking into account in~\eqref{App_eq:Gamma_av}
$E_{BW} \simeq \Gamma$.
Parameterizing $\Gamma \sim S_2^{1/2} \sim N^{\beta}$ and $|H_{mn}|\sim N^{\alpha}$ with certain parameters $\beta>0$ and $\alpha<\beta$, one obtains self-consistency equation
\be\label{App_eq:Gamma_av_integral}
N^{2\beta-1} \sim \int\limits\limits_{-\infty}^{\beta} N^{2\alpha-\frac{(\alpha+\gamma/2)^2}{p\gamma}}d\alpha \ ,
\ee
where the integral can be calculated in the saddle-point approximation.

  For all $\beta> \alpha_{\max} = \gamma\lrp{p-\frac12}$ and $\beta>0$ one obtains:
\be
\beta = \frac12 - \frac{\gamma}2 \lrp{1 - p} \ .
\ee
The above conditions on $\beta$ restrict the validity range of this formula
to $\gamma<\min\lrp{\frac1p;\frac1{1-p}}$.

In the opposite limit of $0<\beta<\gamma\lrp{p-\frac12}$ the main contribution to the integral in~\eqref{App_eq:Gamma_av_integral} is given by $\alpha = \beta$ leading to
\be
2\beta - 1 = 2\beta - \frac{(\beta+\gamma/2)^2}{p\gamma} \Leftrightarrow \beta = \sqrt{p\gamma} - \frac{\gamma}{2} \ .
\ee
The above conditions on $\beta$ restrict the validity range of the latter to
$\frac1p <\gamma<4p$ which is achievable only for $p>\frac12$.

As a result  a new crossover parameter $\gamma_0 = \frac1p$ emerges in the scaling of
 $\Gamma$ with $N$:
\be\label{App_eq:Gamma_res}
\Gamma\sim \left\{
             \begin{array}{ll}
               N^{\frac{1-\gamma(1-p)}{2}}, & \gamma<\gamma_{ET},\gamma_0 \\
               N^{\sqrt{p\gamma} - \frac{\gamma}{2}}, & \gamma_0<\gamma<\gamma_{ET} \\
               N^{1-\gamma/\gamma_{ET}}, & \gamma>\gamma_{ET}
             \end{array}
           \right.
\ee
For the first two cases (corresponding to the ergodic phases) where $\Gamma$
determines the bandwidth $E_{BW}$ we also
check the above results numerically by calculating the scaling of different measures of
the total bandwidth, see Fig.~\ref{Fig:Bandwidth}. In all these cases a
semi-quantitative agreement is demonstrated with deviations for $p\geq 1$ caused probably by
finite-size effects.

 An important result of these numerics is  that for $90$~\% of the  states
(excluding $10$~\%  near the band edges)
the typical and the mean measures of $E_{BW}$
  scale in the same way.

Now we calculate the typical Breit-Wigner width given by~\eqref{App_eq:Gamma_typ} and show that the transition point $\gamma_{FWE}$,~\eqref{FWT}, is not affected by the presence of the crossover parameter $\gamma_0 = 1/p$.
Indeed, from~\eqref{App_eq:Gamma_typ} and~\eqref{App_eq:Gamma_res} one obtains
\be
\Gamma_{typ}\sim \left\{
             \begin{array}{ll}
               N^{\frac{1-\gamma(1+p)}{2}}, & \gamma<\gamma_{ET},\gamma_0 \\
               N^{1-\sqrt{p\gamma} - \frac{\gamma}{2}}, & \gamma_0<\gamma<\gamma_{ET} \\
               N^{1-\gamma}, & \gamma>\gamma_{ET}
             \end{array}
           \right.
\ee
and it is easy to check that there is the only solution of the equation $\Gamma_{typ} \sim N^0$ given by $\gamma = \gamma_{FWE}$.

\end{document}